\documentclass[11pt,a4paper]{article}
\pdfoutput=1
\usepackage{jheppub}

\newcommand{\be}{\begin{equation}}
\newcommand{\ee}{\end{equation}}
\newcommand{\bea}{\begin{eqnarray}}
\newcommand{\eea}{\end{eqnarray}}
\newcommand{\nn}{\nonumber}
\newcommand{\La}{\Lambda}
\newcommand{\e}{\mbox{e}}
\newcommand{\one}{\mbox{\bf 1}}

\def\hm{\hat{m}}

\def\hmu{\hat{\mu}}
\def\hg{\hat{g}}
\def\hh{\hat{h}}

\def\he{\eta}

\newcommand{\mcK}{\mathcal{K}}
\newcommand{\mcO}{\mathcal{O}}

\newcommand{\mm}{m}
\newcommand{\U}{\mathrm{U}}
\newcommand{\SU}{\mathrm{SU}}
\newcommand{\ab}[1]{\big|#1\big|}
\newcommand{\gl}[1]{\displaystyle \langle #1 \rangle}

\DeclareMathOperator\erfc{erfc}
\DeclareMathOperator\sgn{sgn}
\DeclareMathOperator\Pf{Pf}
\DeclareMathOperator\Tr{Tr}
\DeclareMathOperator\re{Re}
\DeclareMathOperator\im{Im}
\renewcommand\epsilon\varepsilon

\title{Random matrix theory of unquenched two-colour QCD with nonzero chemical potential}

\author[a,b]{G.~Akemann,}
\author[c]{T. Kanazawa,}
\author[a]{M.~J.~Phillips,}
\author[d]{and T.~Wettig}
\affiliation[a]{Department of Mathematical Sciences \& BURSt Research
  Centre, Brunel University West London, Uxbridge UB8 3PH, United Kingdom}
\affiliation[b]{Department of Physics, Bielefeld University, Postfach 100131 
  D-33615 Bielefeld, Germany}
\affiliation[c]{Department of Physics, University of Tokyo, Tokyo
  113-0033, Japan}
\affiliation[d]{Department of Physics, University of Regensburg, 93040
  Regensburg, Germany}
\emailAdd{akemann@physik.uni-bielefeld.de}
\emailAdd{tkanazawa@nt.phys.s.u-tokyo.ac.jp}
\emailAdd{michael.phillips@brunel.ac.uk}
\emailAdd{tilo.wettig@physik.uni-regensburg.de}

\abstract{
We solve a random two-matrix model with two real asymmetric matrices 
whose primary purpose is to describe certain aspects of 
quantum chromodynamics with two colours and dynamical fermions at nonzero
quark chemical potential $\mu$. 
In this symmetry class the determinant of the Dirac operator is real but not
necessarily positive. Despite this sign problem the unquenched matrix model
remains completely solvable and provides detailed predictions for the Dirac
operator spectrum in two different physical scenarios/limits:
(i) the $\epsilon$-regime of chiral perturbation theory at small $\mu$, where
$\mu^2$ multiplied by the volume remains fixed in the infinite-volume limit  and 
(ii) the high-density regime where a BCS gap is formed and $\mu$ is unscaled.
We give explicit examples for the complex, real, and imaginary
eigenvalue densities including $N_f=2$ non-degenerate flavours. Whilst the
limit of two degenerate masses has no sign problem and can be tested with
standard lattice techniques, we analyse the severity of the sign problem for
non-degenerate masses as a function of the mass split and of $\mu$.

On the mathematical side our new results include an analytical formula for
the spectral density of real Wishart eigenvalues in the limit (i)
of weak non-Hermiticity, thus completing the 
previous solution of the corresponding quenched model of two real asymmetric
Wishart matrices.
}


\begin{document}
\maketitle

\section{Introduction}
\label{sec:intro}

The phase diagram of quantum chromodynamics (QCD) as a function of
temperature $T$ and quark chemical potential $\mu$ is of great
importance for many phenomenological questions in, e.g., heavy-ion
collisions, the physics of neutron stars, and the early universe.
Clearly, nonperturbative methods are needed to study the phase
diagram.  Lattice QCD simulations have obtained solid quantitative
results at nonzero $T$, see, e.g., \cite{Aoki:2009sc}, but what
happens at nonzero $\mu$ is much less well established because lattice
simulations are hindered by the fermion sign problem, see
\cite{deForcrand:2010ys} for a review.  Some exact results could be
derived for asymptotically large $\mu$, see, e.g.,
\cite{Alford:2007xm} for a review, but otherwise studies of the
$\mu$-dependence of the QCD phase diagram had to rely on model
approaches.  Examples include the instanton liquid model
\cite{Rapp:1999qa}, the PNJL model \cite{Fukushima:2008wg}, and
universality arguments \cite{Halasz:1998qr}.  However, there are
QCD-like theories in which the fermion sign problem can be avoided for
certain choices of the parameters, and which can therefore be
simulated on the lattice.  An example is QCD with two colours, where
the single-flavour fermion determinant is real but not necessarily
nonnegative so that the sign problem is absent for an even number of
quark flavours whose masses are pairwise degenerate.  Such theories
are important testing grounds for effective theories or models since a
direct comparison of analytical results with lattice data is possible.
If the parameters of the theory are detuned from their ``sign-free''
values (e.g., if the degeneracy of the quark masses is lifted 
in two-colour QCD), a sign problem develops.  
As long as the sign problem is weak enough,
one can still obtain some information from the lattice data, e.g., by
reweighting, and therefore test the effective theory or model also in
the presence of the sign problem.  In fact, lattice studies of this
kind have already been done with two-colour adjoint staggered fermions
\cite{Hands2000,Hands:2001ee}, and we will comment on them at the end
of section~\ref{sec:sign_low}.

In this paper we will focus on two-colour QCD with $N_f$ quark flavours 
in the fundamental representation of the gauge group
at nonzero $\mu$.\footnote{Most of our results apply not only to
  two-colour QCD but also to any QCD-like theory with pseudo-real
  quarks.}  This theory has been investigated in the past in great
detail for both small and large $\mu$.  For small $\mu$, it was
studied using an effective chiral Lagrangian, see, e.g.,
\cite{Kogut:2000ek}, and in lattice simulations, see, e.g.,
\cite{Kogut:2001na}.  One of the main results from these studies is a
transition from a phase with a chiral condensate to a phase with a
diquark condensate at a critical value of $\mu$ equal to one half of
the Nambu-Goldstone boson mass.  For asymptotically large $\mu$, the
physics is completely different.  The ground state is believed to be a
BCS superfluid of diquark pairs, for which the BCS gap has been
computed in a weak-coupling approach in \cite{Son:1998uk,Schafer:2002yy}.  The
effective Lagrangian for this case has been derived in
\cite{Kanazawa:2009ks}.

A very useful analytical approach to QCD and QCD-like theories is
random matrix theory (RMT), which can be used either as an effective
theory that yields exact results in well-defined limits, or as a
schematic model to obtain qualitative results.  RMT has applications
in many areas of physics, see \cite{Beenakker2009,Zabrodin2009,Bouchaud2009,Verbaarschot:2009jz,Fyodorov:2010dt,Ferrari2010a,Majumdar2010,Zinn-Justin-Zuber2010} for reviews of some
of these areas.  The RMT approach to QCD was pioneered in
\cite{Shuryak:1992pi}, and reviews can be found in
\cite{Verbaarschot:2000dy,Akemann:2007rf,Verbaarschot:2009jz}.  Random matrix
models for 
QCD at nonzero $\mu$ have been proposed using a single random matrix
\cite{Stephanov:1996ki} or two random matrices \cite{Osborn:2004rf}.
Both models describe the same physics, but the two-matrix model turns
out to be easier to handle mathematically \cite{Akemann:2004dr}, due to the
fact that a representation in terms of complex eigenvalues exists.  
Random two-matrix models were proposed for adjoint QCD at nonzero $\mu$ 
\cite{Akemann:2005fd,Akemann:2006dj}
and most recently 
for two-colour QCD at nonzero $\mu$ 
\cite{Akemann:2008mp}, completing the set of all three 
chiral symmetry breaking patterns \cite{Halasz:1997fc}.  Each RMT yields exact results in
the so-called microscopic regime, corresponding to the lowest order in
the $\epsilon$-expansion \cite{Gasser:1987ah} of the chiral effective
theory.  For a comparison to lattice simulations (when these are possible) 
we refer to \cite{Akemann:2007rf}.

For the case of two-colour QCD, which is the topic of this paper, the
random two-matrix model
is explicitly given in eq.~\eqref{ZPQ} below.
It implicitly contains two parameters, see eq.~\eqref{weakscale}.  One
parameter relates the random-matrix quark masses to the physical quark
masses and can be identified with the low-energy constant $G$ (in the
notation of \cite{Kogut:2000ek}) in
the chiral Lagrangian.  The other parameter relates the random-matrix
chemical potential to the physical chemical potential and can be
identified with another low-energy constant, $F$.  Computing
analytical results from RMT and matching them to
lattice data then allows one to determine these two low-energy
constants.  It was realized in \cite{Kanazawa:2009en} that the model
for two colours proposed in \cite{Akemann:2008mp} also describes the completely
different physics at large $\mu$, provided that the random-matrix
parameter corresponding to $\mu$ is set to a particular value which
corresponds to maximal non-Hermiticity.  The remaining parameter can
then be identified with the BCS gap $\Delta$.  By matching the RMT
results to lattice data one could in principle determine $\Delta$, but
it is currently unclear what computational resources are needed to go
to large enough values of $\mu$ on the lattice.

The random two-matrix model proposed in \cite{Akemann:2008mp} was
solved in \cite{Akemann:2009fc} for $N_f=0$ only.  The eigenvalues of the
pseudo-real Dirac operator are either purely real, purely imaginary, or come in
complex conjugate pairs.  The spectral correlation functions of these
eigenvalues were computed in the quenched case (i.e., for $N_f=0$)
for finite $N$, where $N$ is related to the dimension of the random
matrix and assumed to be proportional to the four-volume $V_4$.  For
$N\to\infty$ two different mathematical limits were considered, that of weak
non-Hermiticity \cite{Fyodorov:1996sx,Fyodorov:1997zz} (corresponding to $\mu^2N$ finite) and that of
strong non-Hermiticity (corresponding to nonzero $\mu$ in the
$N\to\infty$ limit).  In the strong
non-Hermiticity limit, the microscopic spectral density was obtained
for the complex, real, and imaginary eigenvalues.  In the weak
non-Hermiticity limit, the microscopic spectral density could only be
obtained for the complex eigenvalues.  In this paper, we generalise
the results of \cite{Akemann:2009fc} to nonzero $N_f$ using the building block
for finite $N$ from  \cite{Akemann:2010mt}, and we
also obtain the microscopic spectral densities of the real and imaginary
eigenvalues in the weak non-Hermiticity limit.  Furthermore, we analyse the
sign problem that arises when the quark masses are detuned for even $N_f$ 
or when $\mu$ is increased for odd $N_f$.\footnote{%
The possibility of unequal chemical potentials as yet another source of 
the sign problem is interesting, see e.g. \cite{Klein:2004hv},  
but will not be considered in this paper.}  
It has been argued 
for three colours
that the sign problem and the resulting strong oscillations of the
spectral density are directly responsible for the breaking of chiral
symmetry in the unquenched theory \cite{Osborn:2005ss,Osborn:2008jp}.

In the remainder of this paper, we will sometimes refer to the two
mathematical limits of weak and strong non-Hermiticity as ``low
density'' and ``high density'', respectively, with the understanding
that by ``low density'' we mean the physical regime in which
$\mu^2F^2V_4$ is finite and by ``high density'' we mean the physical
regime in which a BCS gap is formed.  Our RMT results can be directly
applied to lattice simulations of two-colour QCD in both regimes.

The structure of this paper is as follows.  
In section \ref{ZandXPT} we define the partition
function in RMT as well as in the two limiting cases of low and high
density. New Pfaffian expressions for the latter are given.
Section \ref{sc:Dirac_Eigenvalues} first summarises the structure of
finite-$N$ results for all spectral densities, expressing the unquenched
quantities in terms of the quenched ones. Subsections \ref{lowmu} and
\ref{subsec_high_density} then give the final answers for the formulas and pictures of the
spectral densities at weak
and strong non-Hermiticity, respectively, including
the explicit examples of two degenerate and non-degenerate flavours. 
Section \ref{sc:sign-problem} is devoted to a detailed analysis of the
sign problem. In subsection \ref{phase} we first analyse the boundary, height,
and frequency of the highly oscillating region of the spectral density. In subsection 
\ref{sc:Avsign} we then define an observable which measures the severity of
the sign problem and compute this observable for low and high density. 
We conclude in section \ref{cons}.   The derivations of several technical details 
are given in three appendices: Appendix \ref{weakrealR} deals with the weak limit
of the quenched spectral density of real eigenvalues, 
in appendix \ref{heavy} we explicitly verify the decoupling of heavy flavours,
and 
appendix \ref{ZsignQ} expresses the
(partially) sign-quenched partition functions in terms of known
spectral densities.

\section{The partition function and two limiting theories}\label{ZandXPT}

In this section we briefly introduce the chiral RMT and 
discuss its two different regimes of applicability in the physics of
two-colour QCD. 

The RMT partition function at finite matrix size and fixed topology $\nu$ is given by%
\footnote{Note that our convention for the Gaussian weight is the 
same as in \cite{Akemann:2009fc,Akemann:2010mt} but different from \cite{Kanazawa:2009en}.
} 
\begin{multline}
  {\cal Z}_N^{(N_f,\,\nu)}(\mu;\{\mm\}) =
  \frac{1}{(2\pi)^{N(N+\nu)}}
  \!\!\!\!\!\!
  \int\limits_{\mathbb{R}^{N\times(N+\nu)}}\!\!\!\!\!\!dP\!\!
  \int\limits_{\mathbb{R}^{N\times(N+\nu)}}\!\!\!\!\!\!dQ~
  \exp\left[-\frac{1}{2}\Tr(PP^T+QQ^T)\right] \\
  \times \prod_{f=1}^{N_f} \det
  \begin{pmatrix}
  \mm_f\one_N& P+\mu_f Q\\
  P^T-\mu_f Q^T&\mm_f\one_{N+\nu}\\
  \end{pmatrix}
  .
  \label{ZPQ}
\end{multline}
Here, the 
rectangular matrices $P$ and $Q$ are each of size $N\times (N+\nu)$ with real
elements, without further symmetry restriction. 
The prefactor is chosen to ensure that ${\cal Z}_N^{(N_f=0,\,\nu)}=1$. 
The parameters $\mu_f$ 
and $\mm_f$ denote the chemical potential and quark mass, respectively, for
each flavour $f$ in the two-matrix model.  The mapping of these RMT
parameters to physical parameters is given in eqs.~\eqref{weakscale} and
\eqref{strongscale} below.
The RMT in eq.~\eqref{ZPQ} was introduced in \cite{Akemann:2008mp} as a model for two-colour QCD at
nonzero chemical potential. At $\mu=0$ the matrix $Q$ decouples, and the
remaining one-matrix model equals the known RMT for two-colour QCD \cite{Verbaarschot:1994ia}.
In eq.~(\ref{ZPQ}) the Dirac matrix is chosen to be 
Hermitian in the limit $\mu_f\to 0$, hence we must take the $\mm_f$ 
to be purely imaginary in applications to lattice QCD. 
In \cite{Akemann:2009fc} the above RMT was solved in the quenched case ($N_f=0$)
for finite $N$ and in two different large-$N$ limits, i.e.,
weak and strong non-Hermiticity as explained in section~\ref{sec:intro}.
In \cite{Kanazawa:2009en} the precise limits in which the above RMT matches two-colour QCD
with $\mu\neq0$ were identified. 

In the first large-$N$ limit, the microscopic limit 
at weak non-Hermiticity ($w$), all 
chemical potentials are set equal ($\mu_f=\mu$) and are taken to zero when $N\to\infty$ such that 
the product $N\mu^2$ remains finite. Likewise, 
the chiral limit 
is taken for all masses such that the product
$\sqrt{N}\mm_f$ remains finite. 
By integrating out the random matrices, followed by a Hubbard-Stratonovich
transformation and a saddle-point approximation, it was shown in \cite{Kanazawa:2009en} that 
the RMT partition function (\ref{ZPQ}) maps to 
the static part of the chiral Lagrangian, 
\begin{align}
  Z_w^{(N_f,\nu)}(\mu;\{\mm\})\sim
  \int\limits_{\U(1)} d\theta~\e^{i\nu\theta} 
  \!\!\!\!\!\!\!  \int\limits_{\SU(2N_f)}\!\!\!\!\!\! dU
  \exp\Big[&
  \frac{1}{2} N \mu^2 \Tr \Big(UIU^T \mathfrak{\cal B}^T (UIU^T)^\dagger 
  {\cal B}\Big)
  \notag\\
  &+ \sqrt{N} \re \Tr (\e^{i\theta/N_f} UIU^T i\mathfrak{T}^\dagger) \Big]\,,
  \label{Zweak}
\end{align}
which provides the dominant contribution
in the $\epsilon$-regime of chiral perturbation theory.
Here, we have introduced the $2N_f\times 2N_f$ matrices
\begin{equation}
  I\equiv\begin{pmatrix} 
    \mathbf{0} & \one \\
    -\one & \mathbf{0}
  \end{pmatrix}\,,\qquad
  {\cal B}\equiv\begin{pmatrix}
    \one&\mathbf{0}\\
    \mathbf{0}&-\one
  \end{pmatrix}\,,\qquad
  \mathfrak{T}\equiv \begin{pmatrix}
    \mathbf{0} & \mathbf{-t}^T\\
    \mathbf{t} & \mathbf{0}
  \end{pmatrix}\,,
\end{equation}
where $\mathbf{t}\equiv\text{diag}(\mm_1,\ldots,\mm_{N_f})$,
${\cal B}$ is the baryon charge matrix, and $\mathfrak{T}$ is
the mass matrix. 
Comparing eq.~\eqref{Zweak} with the static limit of the chiral
Lagrangian \cite{Kogut:2000ek} we obtain the following mapping of RMT
quantities to physical quantities,
\begin{subequations}
  \label{weakscale}
  \begin{alignat}{2}
    \hmu^2 & \equiv 2N\mu^2
    &&= 4 {\mu}_\text{phys}^2 F^2V_4\,, \\ 
    \label{eq:weakscale-m}
    \hm_f & \equiv 2\sqrt{N}\mm_f 
    &&= 2m_{f,\text{phys}}GV_4
  \end{alignat}
\end{subequations}
with the physical parameters ${\mu}_\text{phys}$ and
$m_{f,\text{phys}}$, the four-volume $V_4$, and the low-energy constant $G$
(in the notation of \cite{Kogut:2000ek}).
In particular, for equal masses 
these can be related to the pion mass
$m_\pi$
and the low-energy constant $F$ by $\hm=2m_\pi^2F^2V_4$,
where we have used the Gell-Mann--Oakes--Renner
relation.
The scaling of eq.~\eqref{eq:weakscale-m} 
at small chemical potential does
not come as a surprise since it also holds true for vanishing
chemical potential \cite{Halasz1995a}.

In the second large-$N$ limit, called strong non-Hermiticity ($s$)
in \cite{Akemann:2009fc}, a
map to a completely different regime at high density was found in \cite{Kanazawa:2009en}. In
this limit the RMT chemical potential is not scaled, i.e.,
$\mu=\mcO(1)$, and it was found for an even number $N_f$ of
flavours that
\bea
  Z_s^{(N_f,\nu=0)}(\mu=1;\{\mm\}) \sim \int\limits_{\U(1)}\! d\phi 
  \!\!\int\limits_{\SU(N_f)} 
  \!\!\! d\tilde{U}\,d\tilde{V}~
  \exp\left[-\frac{1}{2} \re\Tr (
  \e^{2i\phi} \mathbf{t} \tilde{U}I\tilde{U}^T
  \mathbf{t}^T \tilde{V}^*I\tilde{V}^\dagger)\right]
  \label{Zstrong}
\eea
in the sector of zero topology. Here we have to identify 
\be
  \hm^2_f \equiv \mm^2_f =
  \frac{3}{\pi^2}V_4\Delta^2m_{f,\text{phys}}^2\,,
  \label{strongscale}
\ee
where $\Delta$ is the BCS gap. Note that we have $\hm_f=\mm_f$, i.e., no
rescaling by a power of $N$ is necessary
in this regime \cite{Akemann:2009fc}.

Whilst in \cite{Kanazawa:2009en} this mapping was found at maximal non-Hermiticity
($\mu=1$), we will show below that the same result holds for $0<\mu\leq 1$, both
for the partition function and for the eigenvalue density correlation functions,
subject to a trivial rescaling. 
We also note here that the two different limits in $\mu$ can be formally
related. By taking the limit $\hmu\to\infty$ of the results in the weak large-$N$ limit
(and suitably rescaling the eigenvalues and masses), the
results in the strong limit can be recovered.

The group integrals in eqs.~(\ref{Zweak}) and (\ref{Zstrong}) are in general very
difficult to evaluate explicitly. However, 
in \cite{Akemann:2010mt} a Pfaffian expression was computed for 
the RMT partition function (\ref{ZPQ}) for finite $N$ and any number of
flavours. Combined with the large-$N$ limit of the building blocks of the
Pfaffian expression, the kernel from \cite{Akemann:2009fc},
and the limiting individual skew-orthogonal polynomials of even degree, 
we find
\be
  {Z}^{(N_f,\nu)}_{w/s}(\mu;\{\hm\})
  \sim\frac{(-)^{[N_f/2]}
  }{\Delta_{N_f}(\{\hm^2\})}
  \begin{cases}
    \underset{1\leq f,g\leq N_f}{\Pf}
    \left[{\cal K}^{}_{w/s}(\hm_f^2,\hm_g^2)\right] 
    &N_f \text{ even}\,,\\[5mm]
    \underset{1\leq f,g\leq N_f}{\Pf}
    \begin{bmatrix}
      0 & q_{w}(\hm_g^2)\\
      -q_{w}(\hm_f^2)  & {\cal K}^{}_{w}(\hm_f^2,\hm_g^2)\\ 
    \end{bmatrix}
    &{N_f\text{ odd }(w)\,,}\\
  \end{cases}
  \label{ZNF}
\ee
where $[x]$ denotes the integer part of $x$ and the Vandermonde determinant 
is defined as 
$\Delta_{N}(\{ z \})\equiv \prod_{k>l}^{N}(z_k-z_l)$. 
Note that the limiting partition function for the strong case only
exists when there is an even number of flavours.
The building blocks in the weak limit ($w$) are the quenched 
limiting kernel 
\be
  \label{weakK}
  {\cal K}^{}_{w}(u,v) = \frac{1}{256\pi\hat{\mu}^2}
  \int_0^1 \,ds\, s^2\,\e^{-2\hmu^2 s^2}\,
  \Big\{ \sqrt{u}\,J_{\nu+1}(s{\sqrt{u}})J_{\nu}(s{\sqrt{v}}) - 
  {\sqrt{v}}\,J_{\nu+1}(s{\sqrt{v}})J_{\nu}(s{\sqrt{u}})
  \Big\}
\ee
and the limit of the quenched even skew-orthogonal polynomial 
$q_N$ given in eq.~\eqref{q2N},
\begin{align}
  q_{w}(z) & \equiv \lim_{N\to\infty} \frac{1}{(4N)^{\nu/2}N!}
  \left(\frac{z}{4N}\right)^{\nu/2}	q^{}_{N}
  \left( \frac{z}{4N} \right)
  \notag\\
  & = 2^{-\nu} \e^{-\hmu^2/2}J_{\nu}(\sqrt{z})\,.
  \label{weakQ}
\end{align}
In the strong limit ($s$) we have\footnote{In order to emphasise the
  similarities between the weak and strong case, we have incorporated
  (by convention) an extra factor of $(uv)^{\nu/2}$ into the limiting
  strong kernel ${\cal K}^{}_{s}$ compared with
  \cite{Akemann:2009fc}.  Corresponding factors have been removed from
  the limiting weight functions in eqs.~\eqref{ghats} and
  \eqref{eq:def_h_s} below. All observables remain 
  unaffected by this choice. \label{factor_removal}}
\be
  {\cal K}^{}_s(u,v)= \frac{\eta_+^3}{8\pi}
  (u - v)\,\e^{-\eta_-(u+v)}
  \,I_{\nu}\left(
  2 \eta_+\sqrt{uv} \right),
  \label{strongK}
\ee
where we have defined
\be
  \label{eta_hat_def}
  \eta_{\pm} \equiv \frac{1 \pm \mu^2}{4\mu^2}\,.
\ee
These objects will also appear as building blocks in
expressions for the densities of Dirac eigenvalues. 
They are related to the quenched kernel ${\cal K}^{}_N$ 
(given explicitly in eq.~(\ref{kernel2N})) as
\begin{align}
  {\cal K}^{}_w(u,v) & = 
  \lim_{N\to\infty}\frac{1}{(4N)^2}
  \left(\frac{uv}{(4N)^2}\right)^{\nu/2}{\cal K}^{}_N
  \left(\frac{u}{4N},\frac{v}{4N}\right)\Big|_{\hmu =
    \sqrt{2N}\mu\text{ fixed}}\,,
  \\
  {\cal K}^{}_s(u,v) & =
  (uv)^{\nu/2}\lim_{N\to\infty}{\cal K}^{}_N(u,v)\,.
  \label{eq:K_s-and-K_N}
\end{align}

We also note that the building blocks are directly related to the one- and two-flavour
partition functions,
\begin{align}  
  {Z}^{(N_f=2,\nu)}_{w/s}(\hat{\mu};\hm_1,\hm_2)
  &\sim
  \frac{{\cal K}^{}_{w/s}(\hm_1^2,\hm_2^2)}{\hm_1^2-\hm_2^2}\,,
  \label{ZNf2}
  \\
  {Z}^{(N_f=1,\nu)}_{w}(\hat{\mu};\hm)
  &\sim q_{w}(\hm^2)\,.
  \label{ZNf1}
\end{align}

From the explicit knowledge of the partition functions above one could in
principle derive detailed 
sum rules for the Dirac operator eigenvalues, as was
suggested in \cite{Kanazawa:2009ks,Kanazawa:2009pc}. 
However, we will be able to give more detailed
unquenched density correlation functions in the following section, from which
such sum rules also follow.

\section{Dirac eigenvalue correlation functions}
\label{sc:Dirac_Eigenvalues}

\subsection[Results for finite $N$]{\boldmath Results for finite $N$}

In the following 
we will be interested in finding the correlations of the eigenvalues of the 
Dirac operator. 
Within RMT this operator is given by the matrix
\be
  {\cal D}(\mu)\equiv \begin{pmatrix}
  \mbox{\bf 0}_N& P+\mu Q\\
  P^T-\mu Q^T&\mbox{\bf 0}_{N+\nu}\\
  \end{pmatrix}
  \equiv
  \begin{pmatrix}
  \mbox{\bf 0}_N& A\\
  B^T&\mbox{\bf 0}_{N+\nu}\\
  \end{pmatrix},
  \label{Mdef}
\ee
with the Gaussian weights specified in  eq.~(\ref{ZPQ}). 
In our convention, the Dirac operator is symmetric at $\mu=0$,
with the eigenvalues lying on the real axis.  Since the Dirac operator in
Euclidean field theory is actually anti-symmetric, it is necessary to multiply
all eigenvalues by $i$ when mapping from RMT to QCD.

Finding the spectrum is equivalent to finding the eigenvalues $\Lambda_j^2$ of 
the Wishart matrix $AB^T$, 
as can be seen from the characteristic equation 
\be
  0=\det[\Lambda \one_{2N+\nu}-{\cal D}(\mu)]=\La^\nu\det[\La^2 \one_N-AB^T]=
  \La^\nu\prod_{j=1}^N(\La^2-\La_j^2)\,.
  \label{evrel}
\ee
Because the matrix elements of $A$ and $B$ are real the solutions of the
characteristic equation 
in $\La^2$ are either real or come in complex conjugate pairs. 
Consequently the determinant of the Dirac operator is always real, but not
necessarily positive. Hence, for two-colour QCD there is no phase
problem but a true sign
problem. 

Switching back from Wishart 
to Dirac eigenvalues $\La_j$ we therefore have three 
possibilities: The $\La_j$ can be real ($\La_j^2>0$), 
imaginary ($\La_j^2<0$), or come in complex
quadruplets ($\pm\La_j,\pm\La_j^*$). In addition there are generically $\nu$ 
eigenvalues equal to zero.

For simplicity we will assume that $N=2n$ is even. Because we will subsequently take the
large-$N$ limit this is not important.\footnote{Results for odd $N$ are also
known, and it is expected that they will lead to the same large-$N$ spectral
densities.}  The partition function (\ref{ZPQ}) can be expressed in terms of the Wishart
eigenvalues $z_j \equiv \La_j^2$, 
and the most compact form derived in \cite{Akemann:2010mt} reads
\begin{align}
  {\cal Z}_{N}^{(N_f,\nu)}(\mu;\{\mm\}) 
  & \equiv 
  \prod_{k=1}^{N}  \int\limits_{\mathbb{C}} d^{\,2}z_k~
  P^{(N_f)}_N(z_1,\ldots,z_N)
  \notag\\
  & \sim
  \prod_{f=1}^{N_f}\mm_f^\nu 
  \prod_{k=1}^{N}  \int\limits_{\mathbb{C}} d^{\,2}z_k
  \prod\limits_{j=1}^{N/2}F^{(N_f)}(z_{2j-1},z_{2j}) 
  \,\Delta_{N}(\{z\})\,,
  \label{Zgen}
\end{align}
where in the second line we have ordered the eigenvalues. This is achieved 
by the anti-symmetric bi-variate weight function defined as
\begin{multline}
  F^{(N_f)}(z_{1},z_{2})
  \equiv \prod_{f=1}^{N_f}(z_{1}-\mm_f^2)(z_{2}-\mm_f^2)\Big\{
  {2i} \, g(z_{1},z_{2})[\Theta(\im z_{1})-\Theta(\im z_{2})]\,
  \delta^2(z_{2}-z_{1}^*)
  \\
  +  h(z_{1})h(z_{2})\delta(\im z_{1})\delta(\im z_{2})
  \sgn(\re\,z_{2}-\re\,z_{1})\Big\}\, .
  \label{Fdef}
\end{multline}
In contrast to \cite{Akemann:2010mt} we have made the mass terms explicit in $F^{(N_f)}$.
It contains the quenched weight functions
\begin{align}
  \label{wch}
  h(z) &\equiv  2|z|^{\nu/2} \e^{\eta_-z} K_{\frac{\nu}{2}}(\eta_+|z|)
  \,,\\
  g(z_1,z_2) & \equiv
  2|z_1z_2|^{\nu/2}\e^{\eta_-(z_1+z_2)}\!
  \int_0^\infty\! \frac{dt}{t}\e^{-\eta_+^2 t(z_1^2+z_2^2)-{1}/{4t}}
  K_{\frac{\nu}{2}}\left(2\eta_+^2t z_1 z_2\right)
  \erfc\left(\eta_+\sqrt{t}\,|z_2-z_1|\right),
  \label{eq:def_g}
\end{align}
where $K$ and $\erfc$ denote the modified Bessel function of the
second kind and the complementary error function, respectively.
We note that in the limit of real arguments $x$,
$\lim_{z_{1,2}\to x}g(z_1,z_2)=h(x)^2$.

We are now able to give the result for all density correlation functions. The
$k$-point density correlation function for Wishart eigenvalues is defined in the standard way. 
For finite (and even) $N$, it reads
\begin{align}
  & R_k^{(N_f)}(z_1,\ldots,z_k)
  \equiv 
  \frac{N!}{(N-k)!} 
  \frac{1}{{\cal Z}_{N}^{(N_f,\nu)}(\mu;\{\mm\})} 
  \int\limits_{\mathbb{C}}
  d^2z_{k+1}\cdots \int_{\mathbb{C}} d^2z_N~ P^{(N_f)}_N(\{z\}) 
  \notag\\
  &
  =\underset{1 \leq i,j \leq k}{\Pf}
  \begin{bmatrix}
    {\cal K}_N^{(N_f)}(z_i,z_j)
    & \displaystyle \int\limits_{\mathbb{C}} d^2z \, 
    {\cal K}_N^{(N_f)}(z_i,z)F^{(N_f)}(z,z_j)\\
    \\[-10pt]
    \displaystyle 
    -\int\limits_{\mathbb{C}} d^2z \, 
   {\cal K}_N^{(N_f)}(z_j,z)F^{(N_f)}(z,z_i)
    &\quad
    \begin{matrix}
      F^{(N_f)}(z_i,z_j)
      \displaystyle 
      -\int\limits_{\mathbb{C}} d^2z F^{(N_f)}(z_i,z) \\
       \displaystyle \times \int\limits_{\mathbb{C}} d^2z^\prime\,{\cal
	K}_N^{(N_f)}(z,z^\prime) F^{(N_f)}(z^\prime,z_j)
    \end{matrix}
  \end{bmatrix}
  . 
  \label{RnPf}
\end{align}
It is given by a Pfaffian of the ordinary, $2k \times 2k$ matrix composed of
the matrices inside the square brackets, 
containing the unquenched kernel ${\cal K}_N^{(N_f)}(u,v)$
and single or double integrals folded with the anti-symmetric weight function 
$F^{(N_f)}(u,v)$ from eq.~(\ref{Fdef}). 
Thus for fixed $k$ there will be various contributions to 
$R_k^{(N_f)}$, beginning with one having only $k$ complex eigenvalues, then ranging over all mixed
possibilities, to one having only $k$ real eigenvalues. 

The unquenched kernel can be expressed as follows \cite{Akemann:2010mt}
in terms of the quenched kernel ${\cal K}_N$ (given explicitly in eq.~(\ref{kernel2N}))
at finite (and even) $N$,
\be 
  {{\cal K }^{(N_f)}_{N} (u,z)}
  = 
  \frac{\Pf
  \begin{bmatrix}
    0 & {\cal K }^{}_{N+N_f}(u,z) & {\cal K }^{}_{N+N_f}(u,\mm_g^2)
    \\
    {\cal K }^{}_{N+N_f}(z,u) & 0  & {\cal K }^{}_{N+N_f}(z,\mm_g^2)
    \\
    {\cal K }^{}_{N+N_f}(\mm_f^2,u) & {\cal K }^{}_{N+N_f}(\mm_f^2,z) 
    & {\cal K }^{}_{N+N_f}(\mm_f^2,\mm_g^2)\\
  \end{bmatrix}
  }{\prod_{f=1}^{N_f}(u-\mm_f^2)(z-\mm_f^2) \,
  \Pf
  \left[{\cal K
    }^{}_{N+N_f}(\mm_f^2,\mm_g^2)\right]}\,.
  \label{KNFe}
\ee
This expression is valid for an even number $N_f$ of flavours, and even $N$. 
A similar expression holds for odd
$N_f$ as given in eq.~(5.32) in \cite{Akemann:2010mt} and 
also includes skew-orthogonal polynomials of even degree, as given in 
eq.~(\ref{q2N}) (cf.\ eqs.~(\ref{rhoCweakNf1}) and (\ref{rhoRweakNf1})
for $N_f=1$ as an example). 
Because of eq.~(\ref{ZNF}) the unquenched kernel 
can also be written as the ratio of two partition functions, for both even and
odd $N_f$. 

Let us also comment on the two different large-$N$ limits that can be obtained
from eqs.~(\ref{RnPf}) and (\ref{KNFe}). In the strong limit for an even
number of flavours, all $k$-point correlation functions follow unambiguously by
replacing the kernels in eq.~(\ref{KNFe}) with the one  
in the strong limit, eq.~(\ref{strongK}). 
At weak non-Hermiticity the large-$N$ limit is more involved, as was already
noted in \cite{Akemann:2009fc}, due to the non-interchangeability of the limit
and the integration.\footnote{This fact has nothing to do with the sign problem
as it also occurs at $\mu=0$ \cite{Verbaarschot:1994ia}.}
Considering only complex eigenvalues this is not
an issue as all integrations are ``killed'' by the delta functions in
$F^{(N_f)}$ from eq.~(\ref{Fdef}), 
and the $k$-point densities follow by using the kernel in the weak limit,
eq.~(\ref{weakK}), in eq.~(\ref{KNFe}). 
Once correlations of real Wishart eigenvalues are computed the  
 non-interchangeability becomes significant, as is explained for the quenched building
 blocks in appendix \ref{weakrealR}.   
For the spectral density of real eigenvalues we explicitly evaluate the
limiting expressions for $N_f=0,1,2$, 
and then a higher number of flavours readily follows. 
Higher $k$-point correlation functions in the weak limit that contain real
eigenvalues will also involve double integrals (see the lower right block in the
Pfaffian (\ref{RnPf})) which may require further analysis.

Although the structure of the most general $k$-point
function with arbitrary $N_f$ is clear from the results above, in the following we will focus 
mainly on the most useful examples, that is, the spectral densities of complex, real, and imaginary Dirac
eigenvalues with few (or no) flavours. 
They can be obtained from the spectral density (i.e., the one-point function)
of the Wishart eigenvalues, which follows from a single kernel and is given by 
\be
  R_1^{(N_f)}(z_1)= \int_{\mathbb{C}}  d^2z \, {\cal K}^{(N_f)}_N(z_1,z)
  {F}^{(N_f)}(z,z_1)
  \equiv R_1^{(N_f,\mathbb{C})}(z_1)+\delta(y_1)R_1^{(N_f,\mathbb{R})}(x_1)\,,
  \label{density}
\ee
where we denoted $z_1=x_1+iy_1$.
The delta functions in the first and second line of ${F}^{(N_f)}$ in
eq.~(\ref{Fdef}) lead to the split into separate densities of
complex and real Wishart eigenvalues. 
The densities of complex, real, and imaginary Dirac eigenvalues are
then obtained from
\begin{align}
  \rho_1^{(N_f,\mathbb{C})}(z)
  &= 4 |z|^2 R_1^{(N_f,\mathbb{C})}(z^2)\,, \nn \\ 
  \rho_1^{(N_f,\mathbb{R})}(x)
  &= 2 |x| R_1^{(N_f,\mathbb{R})}(x^2)\,,\nn \\ 
  \rho_1^{(N_f,i\,\mathbb{R})}(iy)
  &= 2 |y| R_1^{(N_f,\mathbb{R})}(-y^2)\,.
  \label{eq:weak_mappings}
\end{align}
To clarify the difference to the non-Hermitian RMT for 
three-colour QCD with $\mu\neq0$ \cite{Osborn:2004rf,Akemann:2004dr} 
let us briefly summarise the distinctive features of the spectral density 
for complex Dirac eigenvalues in the present case:
\begin{enumerate}
  \item[(A)] $\rho_{1}^{(N_f,\mathbb{C})}(z)$ has no imaginary part.
  \item[(B)] $\rho_{1}^{(N_f,\mathbb{C})}(z)=\rho_{1}^{(N_f,\,\mathbb{C})}(z^*)=\rho_{1}^{(N_f,\,\mathbb{C})}(-z^*)$.
  \item[(C)] $\rho_{1}^{(N_f,\mathbb{C})}(z)$ becomes zero for real and
    imaginary $z$.
\end{enumerate}
(A) and (B) follow from the reality and chiral symmetry of the Dirac determinant, while 
(C) is a result of the eigenvalue repulsion: 
If $z$ approaches the real axis, $z$ and $z^*$ come close to each
other, but the corresponding probability 
is highly suppressed by the Vandermonde determinant 
$\Delta_N(\{z^2\})$ of Dirac eigenvalues,
and the same statement holds for   $z$ and $-z^*$.
(A)--(C) hold for any finite $N$,
and as we will see 
also in the large-$N$ limit.

We now turn directly to the results in the two different large-$N$
limits.  For
some details in the quenched case, see also \cite{Akemann:2010mt}.

\subsection{Eigenvalue densities at low density}
\label{lowmu}

In this subsection we give the results valid for the microscopic limit at weak
non-Hermiticity ($w$).
In this limit we rescale the chemical potential, the masses, and the Dirac eigenvalues in the
following way (see eq.~(\ref{weakscale})),
\bea
  \hmu^2&\equiv&2N\mu^2\,, \nn\\
  \hm_{f}&\equiv& 2\sqrt{N}\mm_f\,, \nn\\
  \xi &\equiv& 2\sqrt{N}\La \,.
  \label{weaklim}
\eea
Under these scalings, and at large $N$, the microscopic densities for Wishart (squared) eigenvalues are given by
\bea
  R_w^{(N_f,\mathbb{C})}(z) &\equiv& \lim_{N\to\infty} 
  \frac{1}{(4N)^2} R_1^{(N_f,\mathbb{C})}\left( \frac{z}{4N} \right), \nn\\
  R_w^{(N_f,\mathbb{R})}(x) &\equiv& \lim_{N\to\infty} 
  \frac{1}{4N} R_1^{(N_f,\mathbb{R})}\left(\frac{x}{4N}\right).
  \label{rhoweakdef}
\eea
We can then switch back to Dirac eigenvalues using the mappings in
eq.~\eqref{eq:weak_mappings}.
Because $R_1^{(N_f,\mathbb{R})}(-x^2)\neq 
R_1^{(N_f,\mathbb{R})}(x^2)$, i.e., the function is not symmetric in general, the densities of real
and imaginary Dirac eigenvalues will also differ in general: $\rho_{w}^{(N_f,\mathbb{R})}(\xi)\neq
\rho_{w}^{(N_f,i\mathbb{R})}(i\xi)$. 
Later in the strongly non-Hermitian limit a 
symmetry between the real and imaginary densities will emerge.

\begin{figure}[t]
  \unitlength1.0cm
  \centering
  \includegraphics[clip=,width=5.8cm]{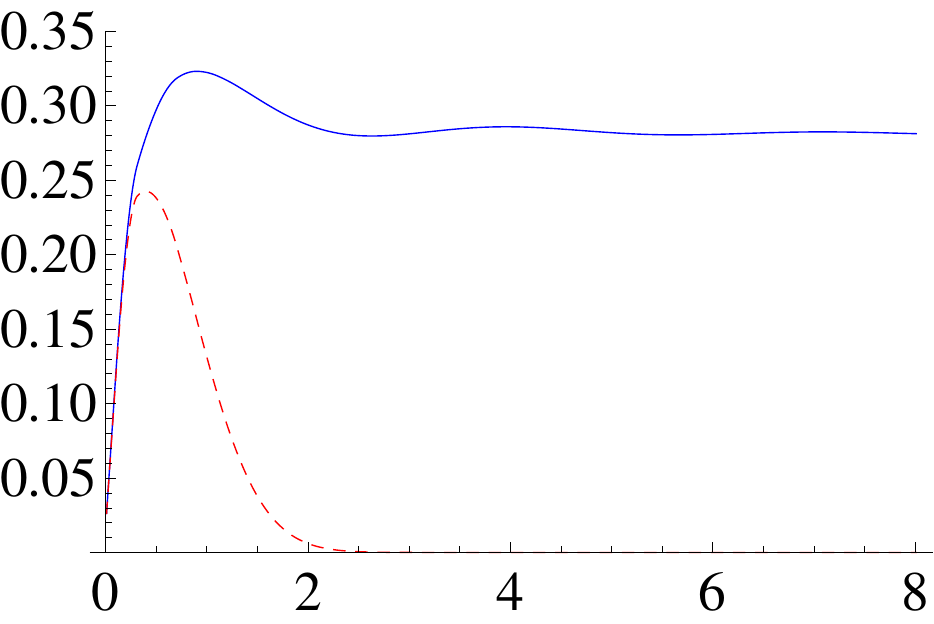}
  \qquad \quad   
  \includegraphics[clip=,width=5.8cm]{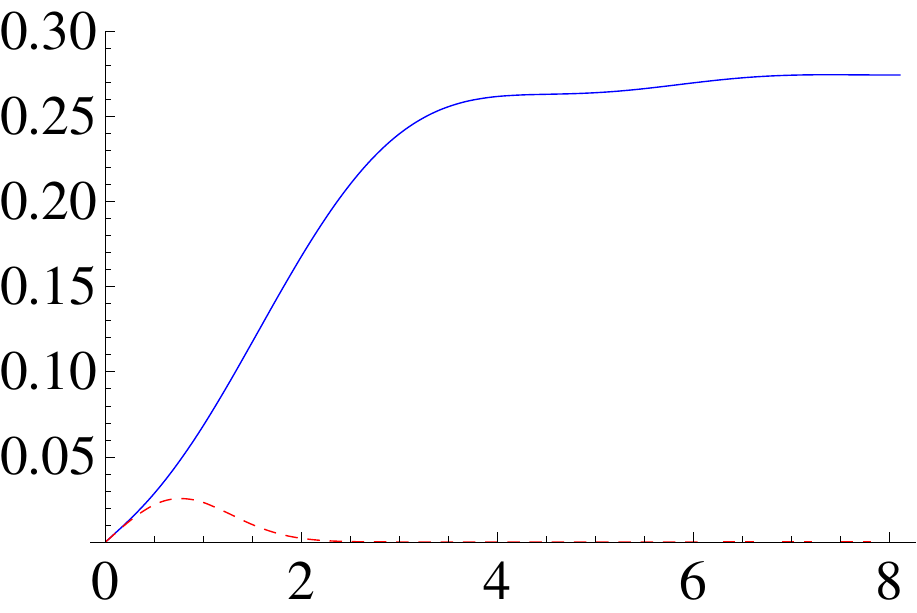}
  \put(0.3,0.3){$\xi$}
  \put(-6.7,0.3){$\xi$}
  \put(-6.4,4.2){$\rho_{w}^{(N_f=0,\,(i)\mathbb{R})}\big((i)\xi\big)$}
  \put(-13.5,4.2){$\rho_{w}^{(N_f=0,\,(i)\mathbb{R})}\big((i)\xi\big)$}
  \caption{\label{fg:weak_real_density-1}Quenched spectral density of
    real (blue solid line) and imaginary (red dashed line) Dirac
    eigenvalues on the positive half-line in the microscopic limit at
    weak non-Hermiticity for $\hmu=\sqrt{0.2}$ at $\nu=0$ (left) and
    $\nu=2$ (right).}
\end{figure}

\begin{figure}[t]
  \unitlength1.0cm
  \centering
  \includegraphics[clip=,width=5.8cm]{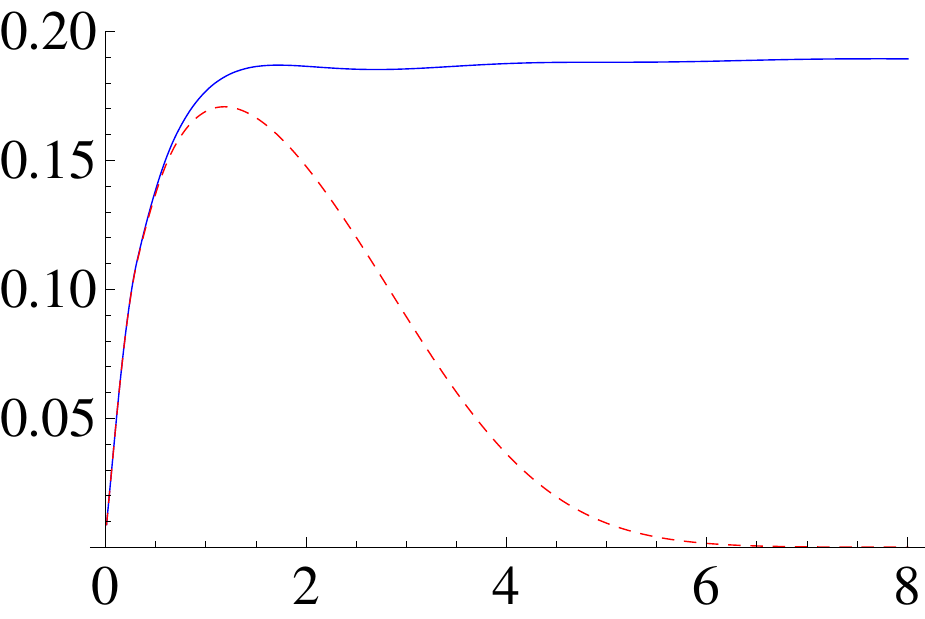}
  \qquad\qquad
  \includegraphics[clip=,width=5.8cm]{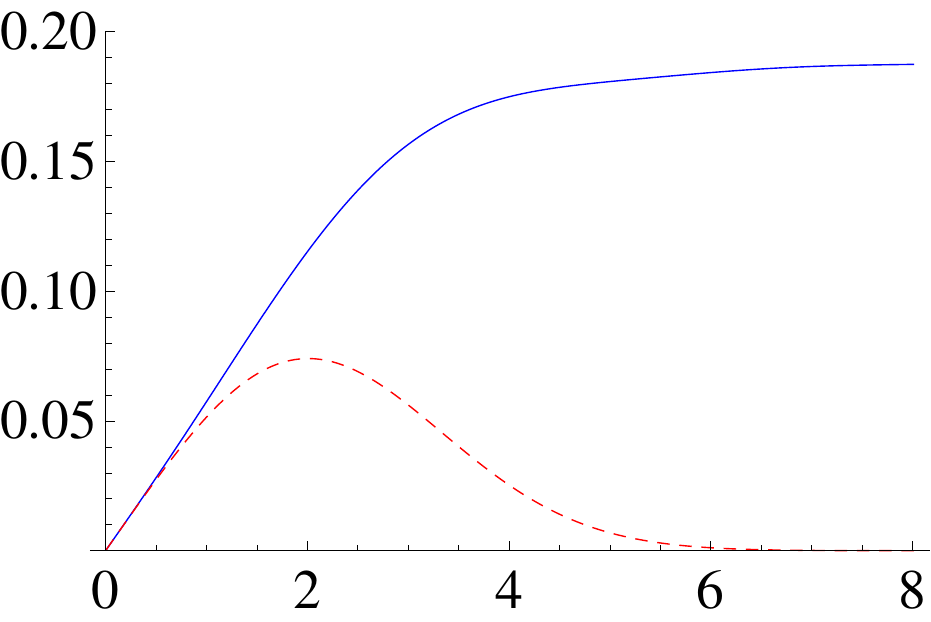}
  \put(0.3,0.3){$\xi$}
  \put(-6.7,0.3){$\xi$}
  \put(-6.4,4.1){$\rho_{w}^{(N_f=0,\,(i)\mathbb{R})}\big((i)\xi\big)$}
  \put(-13.5,4.2){$\rho_{w}^{(N_f=0,\,(i)\mathbb{R})}\big((i)\xi\big)$}
  \caption{\label{fg:weak_real_density-2}Same as
    figure~\ref{fg:weak_real_density-1} but for $\hmu=1$ at $\nu=0$
    (left) and $\nu=2$ (right).}
\end{figure}

We will first give the quenched quantities. As was shown in \cite{Akemann:2009fc}
from eqs.~(\ref{density})
and (\ref{Fdef}) the density of complex eigenvalues is given simply by the
rescaled weight function $\hg_w(\xi^{*2},\xi^2)$ multiplied by the 
weak limiting kernel in eq.~(\ref{weakK}),
\be
  \label{rhoCweakQ}
  \rho_{w}^{(N_f=0,\mathbb{C})}(\xi)=
  4 |\xi|^2 \hg_w(\xi^{*2},\xi^2){\cal K}_w(\xi^2,\xi^{*2}) \,,
\ee
where
\begin{align}
  \label{ghatw}
  \hg_w(z,z^*) & =  - \hg_w(z^*,z) 
  \equiv  2i\,\sgn(\im z)\,
\lim_{N\to\infty}\frac1{(4N)^2} 
  \,
\left(\frac{4N}{|z|}\right)^\nu
g\left(\frac{z}{4N},\frac{z^*}{4N}\right)\\
  & = 4i \, \sgn(\im\,z)\,\e^{\frac{1}{4 \hmu^2}\re\,z}\! 
  \int_0^\infty\!
  \frac{dt}{t}\e^{-\frac{t}{64\hmu^4}
  (z^2+z^{*\,2})-\frac{1}{4t}}
  K_{\frac{\nu}{2}}\Big(\frac{t}{32\hmu^4}|z|^2\Big)
  \erfc\Big(\frac{\sqrt{t}}{4\hmu^2}|\im z|\Big)\,.\notag
\end{align}
The weak limit of the density of real and imaginary eigenvalues is
not quite so straightforward. From eqs.~(\ref{density})
and (\ref{Fdef}) the density of Wishart eigenvalues $R_1^{(N_f,\mathbb{R})}(x)$ is given by a real
integral over the finite-$N$ kernel. It turns out that it is not possible to
exchange the limit $N\to\infty$ with this integration, a fact that
already occurs for $\mu=0$ and $N_f=0$ in \cite{Verbaarschot:1994ia}.
In appendix \ref{weakrealR} we derive the new quenched result 
\be
  \label{rhoRweakQ}
  \rho_{w}^{(N_f=0,(i)\mathbb{R})}(\xi) = -2 |\xi| G_w(\xi^2,\xi^2)\,,
\ee
where
\begin{align}
\!  G_w(x,x') &\equiv -
  \frac{\hh_w(x')}{[\sgn(x')]^{\nu/2}} 
  \Bigg\{\!\! \left( (-i)^{\nu} \int_{-\infty}^0 dy +
  \frac{2}{[\sgn(x')]^{\nu/2}} \int_{0}^{x'} dy \right) 
  {\cal K}_w(x,y) \hh_w(y) 
  \label{Gdef}
  \\
  & - \frac{1}{32\sqrt{\pi}} 
  \Bigg[ - \, \frac{1}{\hmu}\,\e^{-\hmu^2}\,J_{\nu}(\sqrt{x})
  + \frac{2 \hmu^{\nu}}{\Gamma\left(\frac{\nu+1}{2}\right)}\,
  \int_0^1 ds\,\e^{-\hmu^2 s^2} s^{\nu+2} 
  \nn
  \\
  & \times\! \left(
  \frac{\sqrt{x}}{2}\,E_{\frac{1-\nu}{2}}(\hmu^2 s^2)\,
  J_{\nu+1}(s \sqrt{x}) - \hmu^2s\left(
  E_{\frac{-1-\nu}{2}}(\hmu^2 s^2) - E_{\frac{1-\nu}{2}}(\hmu^2 s^2) \right)
  J_{\nu}(s \sqrt{x}) \right)\!\! \Bigg]\!
  \Bigg\}\,.
  \nn
\end{align}
Here we have defined the rescaled real weight function
\be
  \hh_w(x)\equiv
\lim_{N\to\infty}\frac1{4N}
\left(\frac{4N}{|x|}\right)^{\nu/2}
h\left(\frac x{4N}\right)
  = \e^{x/8\hmu^2}\,
  2K_{\frac{\nu}{2}}\left( \frac{|x|}{8\hmu^2} \right)\label{hhdef}
\ee
and used the exponential integral, which is defined as
\begin{equation}
  E_n(x)\equiv\int_1^\infty dt \, t^{-n}\,e^{-x t}\,.
  \label{Edef}
\end{equation}
Equation~\eqref{rhoRweakQ} is plotted in figures~\ref{fg:weak_real_density-1} 
and \ref{fg:weak_real_density-2} 
for $\hmu=\sqrt{0.2}$ and $1$, each at $\nu=0$ and $2$.  We observe
that the presence of exact zero modes leads to a depletion of the spectral 
densities near zero.

We now turn to the unquenched densities, where we will give two examples
which are both new:
two generic flavours ($N_f=2$), including the degenerate mass limit, and a single
flavour ($N_f=1$). 
We start with a general remark: when expanding the Pfaffian in the numerator
of the unquenched kernel in eq.~(\ref{KNFe}), we find that
it always contains the quenched kernel ${\cal
  K}_{N+N_f}(u,v)$ times a Pfaffian $\Pf[{\cal K}_{N+N_f}(m_f^2,m_g^2)]$
that is the same as the  Pfaffian in the denominator and hence cancels; 
the remaining product in the denominator is cancelled by factors in the
weight function. Therefore, on inserting this into eq.~(\ref{density}), the unquenched density
is seen always to be of the form of the quenched density plus a correction term,
\be
  R_1^{(N_f)}(z)\equiv R_1^{(N_f=0)}(z)
  +\Delta R_1^{(N_f)}(z)\,,
\label{correction}
\ee
for both the complex and the real Wishart eigenvalues.
This argument holds for both an even and an odd number of flavours.

The two-flavour density of complex eigenvalues is therefore obtained in a straightforward
manner,
\begin{align}
  \rho_{w}^{(N_f=2,\mathbb{C})}(\xi;\hm_1,\hm_2)&=
  \rho_{w}^{(N_f=0,\mathbb{C})}(\xi) 
  \label{rhoCweakNf2}
\\
  &+4 |\xi|^2 \hg_w(\xi^{*2},\xi^2) 
  \frac{{\cal K}_w(\xi^2,\hm_2^2){\cal K}_w(\xi^{*2},\hm_1^2)-
  {\cal K}_w(\xi^2,\hm_1^2){\cal K}_w(\xi^{*2},\hm_2^2)}{{\cal K}_w(\hm_1^2,\hm_2^2)}\,, 
\nn
\end{align}
with the quenched density from eq.~(\ref{rhoCweakQ}) and the limiting 
weight function  (\ref{ghatw}). It is of the same form as for the
symplectic ensemble relevant for adjoint QCD \cite{Akemann:2006dj}. 
In the limit of
degenerate masses we simply have to Taylor-expand the ratio of quenched
kernels in the second line, where the denominator vanishes like
$\hm_1-\hm_2$. For convenience we also give the 
limiting expression,
\begin{align}
  \lim_{\hm_2\to\hm_1} &
  \frac{{\cal K}_w(\xi^2,\hm_2^2){\cal K}_w(\xi^{*2},\hm_1^2)-
    {\cal K}_w(\xi^2,\hm_1^2){\cal K}_w(\xi^{*2},\hm_2^2)}{{\cal K}_w(\hm_1^2,\hm_2^2)} \nn \\
  &\qquad=\frac{{\cal D}_w(\xi^2,\hm_1^2){\cal K}_w(\xi^{*2},\hm_1^2)-
  {\cal K}_w(\xi^2,\hm_1^2){\cal D}_w(\xi^{*2},\hm_1^2)}{{\cal D}_w(\hm_1^2,\hm_1^2)}\,,
\end{align}
where
\begin{align}
  {\cal D}_w(x,y)   \equiv   
  \int_0^1 ds\, s^2\,\e^{-2\hmu^2 s^2}\,
  \Big[&\sqrt{x}J_{\nu+1}(s\sqrt{x})\Big(\frac{\nu}{\sqrt{y}}J_\nu(s\sqrt{y})-sJ_{\nu+1}(s\sqrt{y})\Big)
  \notag\\
  &+ J_\nu(s\sqrt{x})\Big(\nu J_{\nu+1}(s\sqrt{y})-s\sqrt{y}J_{\nu}(s\sqrt{y})\Big) \Big]\,.
  \label{eq:Dw}
\end{align}
Looking at eq.~(\ref{ZNf2}) and noting that ${\cal D}_w$ is the
derivative of ${\cal K}_w$ with respect to the second argument,
eq.~\eqref{eq:Dw} with $x=y$ also provides the mass-degenerate
two-flavour partition function, 
\be
{Z}^{(N_f=2,\nu)}_{w}(\hmu;\hm,\hm) \sim \int_0^1 \,ds\,
s^3\,\e^{-2\hmu^2 s^2}\,
\hm\Big(J_{\nu+1}(s\hm)J_{\nu-1}(s\hm)-J_{\nu}(s\hm)^2\Big)\,,
  \label{ZNf2deg}
\ee
after using an identity for Bessel functions. 
The two-flavour density of the real and imaginary eigenvalues can readily be
obtained from the above building blocks, using eqs.~(\ref{KNFe}) and
(\ref{density}), 
\be
  \rho_{w}^{(N_f=2,(i)\mathbb{R})}(\xi;\hm_1,\hm_2)=
  \rho_{w}^{(0,(i)\mathbb{R})}(\xi)
  \ +\ 2|\xi| 
  \frac{{\cal K}_w(\xi^2,\hm_2^2)G_w(\hm_1^2,\xi^2)-
  {\cal K}_w(\xi^2,\hm_1^2)G_w(\hm_2^2,\xi^2)}{{\cal K}_w(\hm_1^2,\hm_2^2)}
  \label{rhoRweakNf2}
\ee
with $G_w$ defined in eq.~(\ref{Gdef}). However, there is no simple expression for the degenerate-mass limit
in this case.

Our second unquenched 
example is the case of a single flavour ($N_f=1$). The density of
complex eigenvalues reads
\be
  \rho_{w}^{(N_f=1,\mathbb{C})}(\xi;\hm)=
  \rho_{w}^{(0,\mathbb{C})}(\xi)
  \, +\, 4 |\xi|^2 \hg_w(\xi^{*2},\xi^2) 
  \frac{q_w(\xi^2){\cal K}_w(\xi^{*2},\hm^2)-
  q_w(\xi^{*2}){\cal K}_w(\xi^2,\hm^2)}{q_w(\hm^2)}\,, 
  \label{rhoCweakNf1}
\ee
which includes the second building block eq.~(\ref{weakQ}),
\be
\nn
  q_{w}(z)= 2^{-\nu} \e^{-\hmu^2/2}J_{\nu}(\sqrt{z})\,.
\ee
For the density of real and imaginary eigenvalues we again encounter 
the problem of noncommutativity of the large-$N$ limit and integration. By handling 
this issue similarly to the quenched case (see above), we obtain
\be
  \rho_{w}^{(N_f=1,(i)\mathbb{R})}(\xi;\hm) 
  = \rho_{w}^{(N_f=0,(i)\mathbb{R})}(\xi) +  2|\xi| 
  \frac{q_w(\xi^2)G_w(\hm^2,\xi^2)-Q_w(\xi^2){\cal K}_w(\xi^2,\hm^2)}
  {q_w(\hm^2)}\,,
  \label{rhoRweakNf1}
\ee
where
\begin{multline}
  Q_w(x) \equiv 
  \frac{\hh_w(x)}{[\sgn(x)]^{\nu/2}}
  \Biggl\{\left((-i)^{\nu}\int_{-\infty}^{0}\!\!\! dy+\frac{2}{[\sgn(x)]^{\nu/2}}\int_{0}^{x}\!dy\right)q_w(y)\hh_w(y)
  \\
  -~2^{3-\nu} \sqrt{\pi}\,\e^{\hmu^2/2}\left(  \frac{\hmu^{\nu+2}E_{\frac{1-\nu}{2}}(\hmu^2)}{\Gamma\big(\frac{\nu+1}{2}\big)}+\hmu  \right)  \Biggr\}\,.
\end{multline}
The proof follows along similar lines to that of $G_w(u,v)$.

\begin{figure}[t]
    \unitlength1.0cm
    \centering
       \includegraphics[clip=,width=6.7cm]{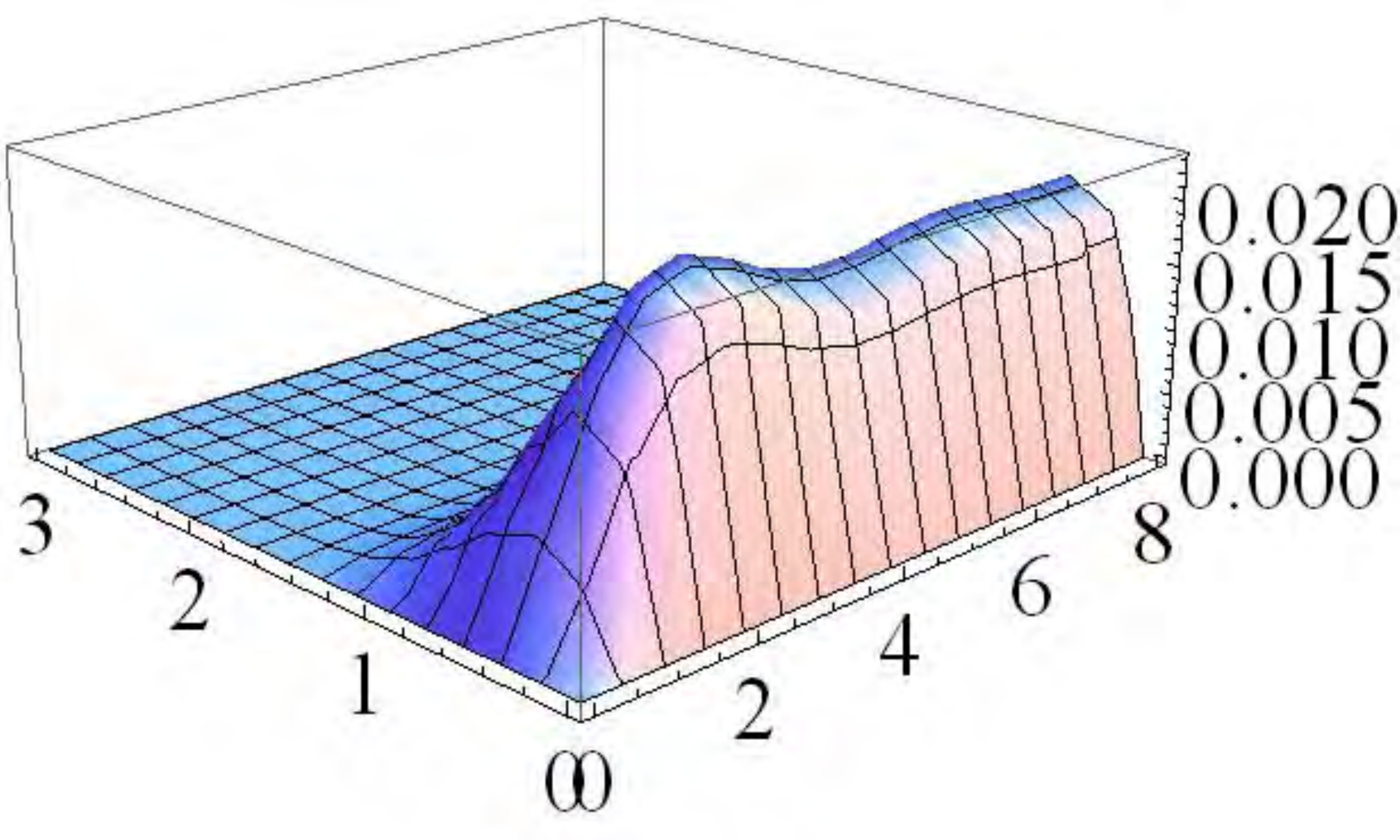} \qquad 
       \includegraphics[clip=,width=6.7cm]{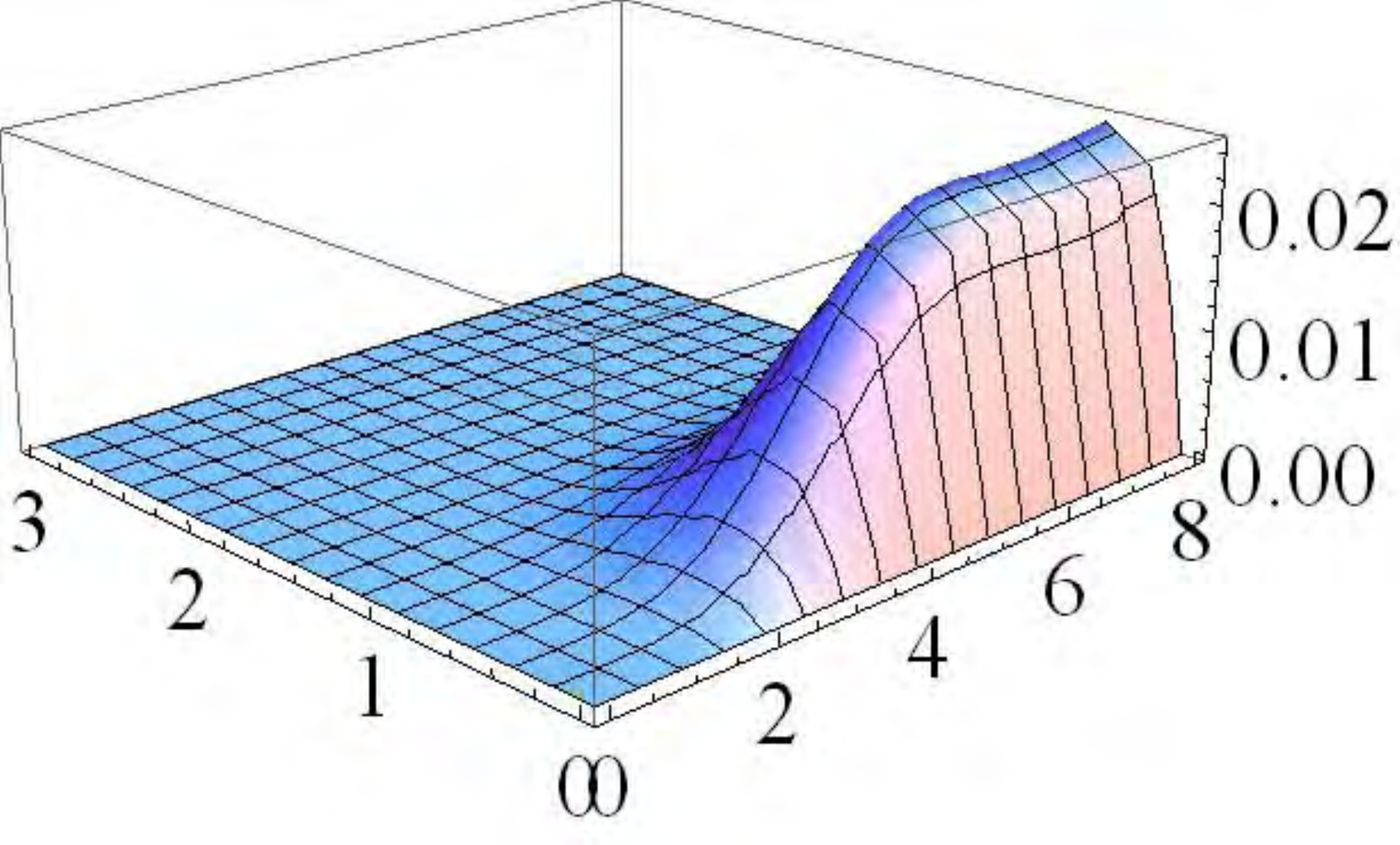}
      \put(-1.8,0.4){$\re[\xi]$}
      \put(-6.4,0.4){$\im[\xi]$}
      \put(-9.5,0.4){$\re[\xi]$}
      \put(-13.9,0.4){$\im[\xi]$}
      \put(-6.8,4.3){$\rho_{w}^{(N_f=1,\mathbb{C})}(\xi;\hm)$}
      \put(-14.4,4.3){$\rho_{w}^{(N_f=0,\mathbb{C})}(\xi)$}
    \caption{\label{fg:Complex-weak-density_Nf=1___No-1}
      Quenched (left) and $N_f=1$ (right) spectral density of 
      complex Dirac eigenvalues at weak non-Hermiticity, both for $\hmu=\sqrt{0.2}$ at $\nu=0$. 
      The right figure is in the chiral limit ($\hm=0$).
    }
\end{figure}
\begin{figure}[t]
    \unitlength1.0cm
    \centering
      \includegraphics[clip=,width=6.7cm]{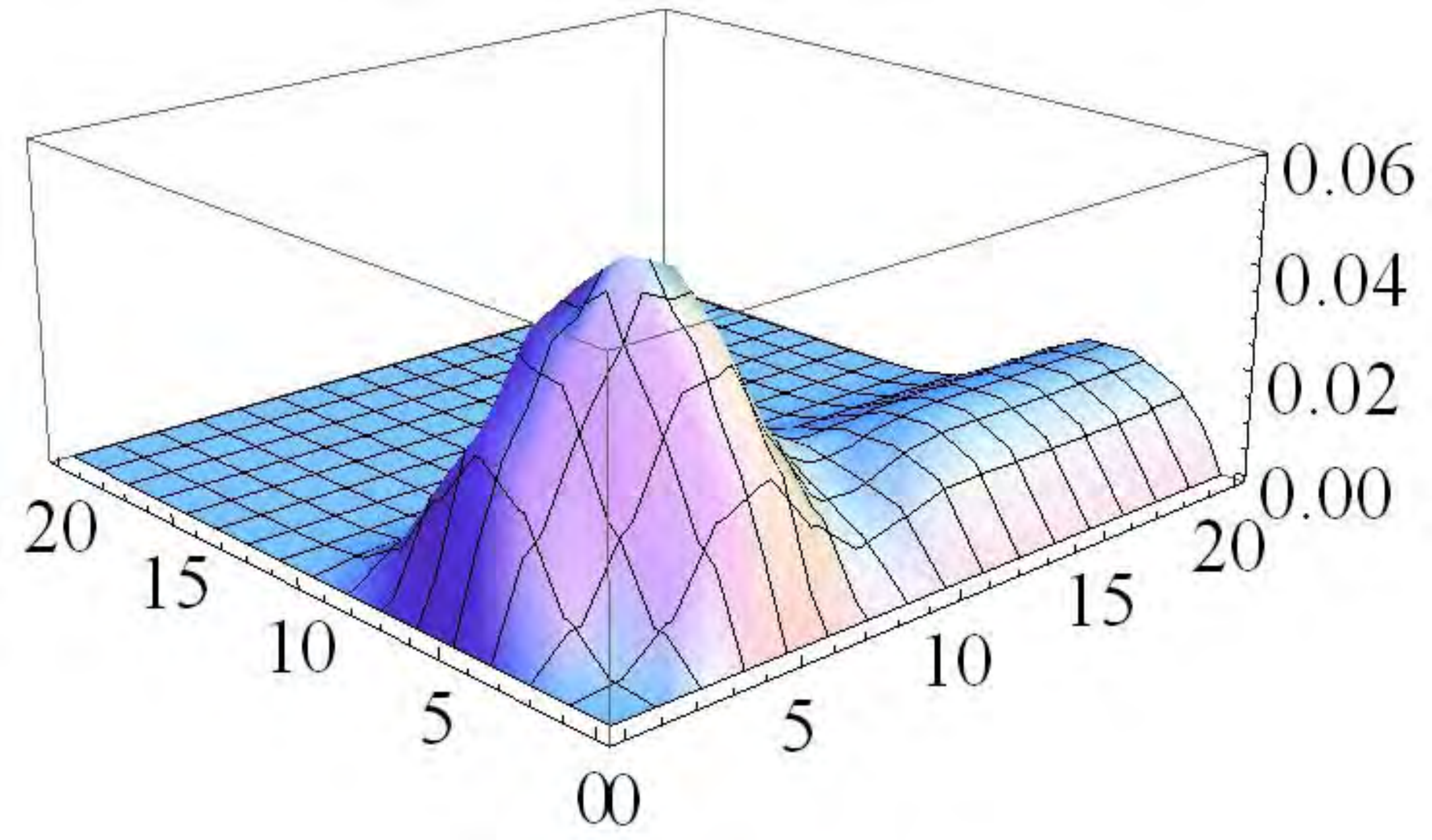} \qquad 
      \includegraphics[clip=,width=6.7cm]{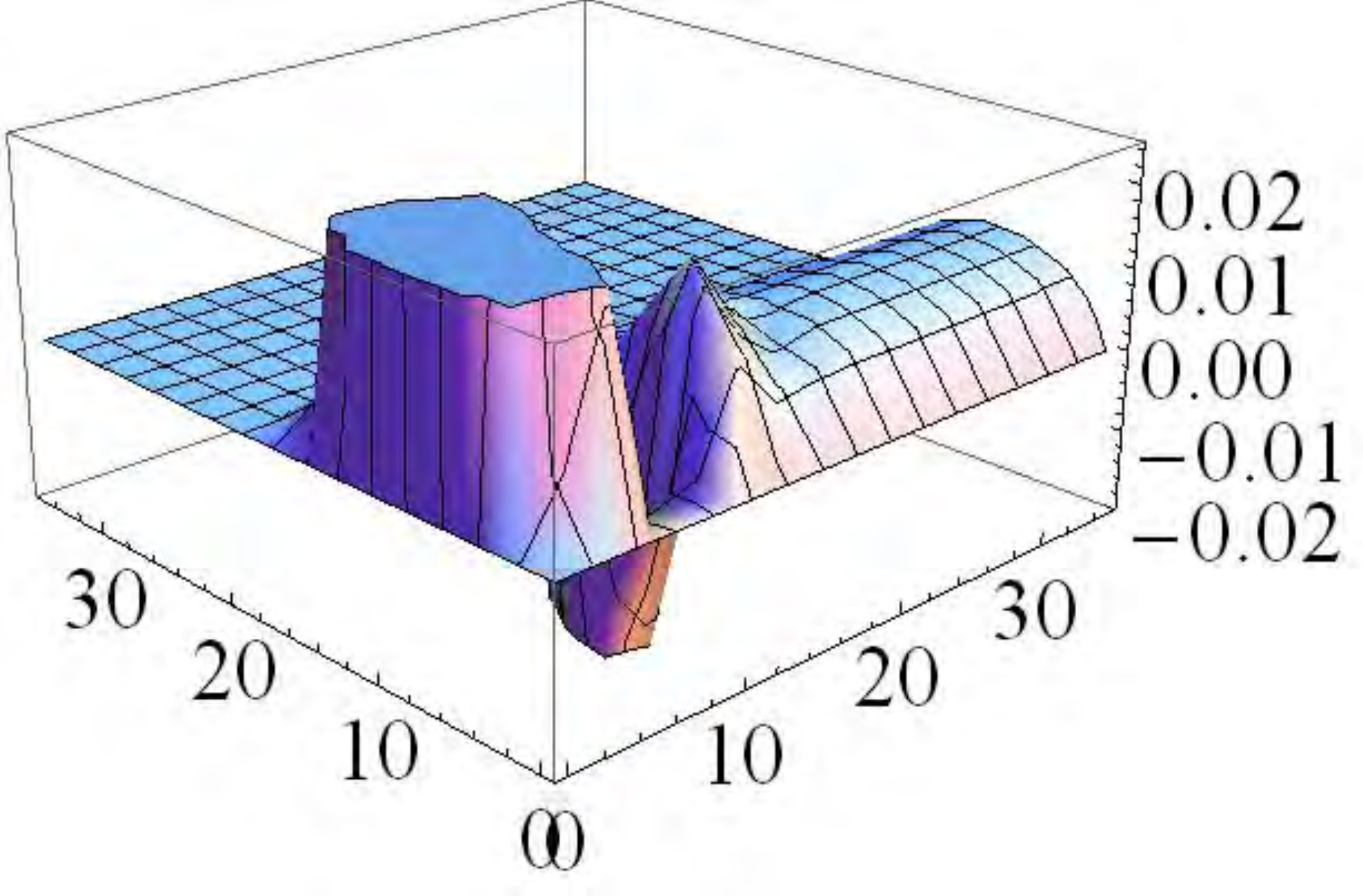}
      \put(-1.8,0.4){$\re[\xi]$}
      \put(-6.4,0.4){$\im[\xi]$}
      \put(-9.4,0.3){$\re[\xi]$}
      \put(-14.0,0.3){$\im[\xi]$}
      \put(-6.8,4.7){$\rho_{w}^{(N_f=1,\mathbb{C})}(\xi;\hm)$}
      \put(-14.4,4.2){$\rho_{w}^{(N_f=1,\mathbb{C})}(\xi;\hm)$}
    \caption{\label{fg:Complex-weak-density_Nf=1___No-2}
      The $N_f=1$ spectral density of complex Dirac eigenvalues 
      at weak non-Hermiticity for $\hmu=1.8$ (left) and $\hmu=2.5$ (right) at $\nu=0$ 
      in the chiral limit ($\hm=0$). Here and in the following the peaks of
      the oscillations are clipped.
    }
\end{figure}
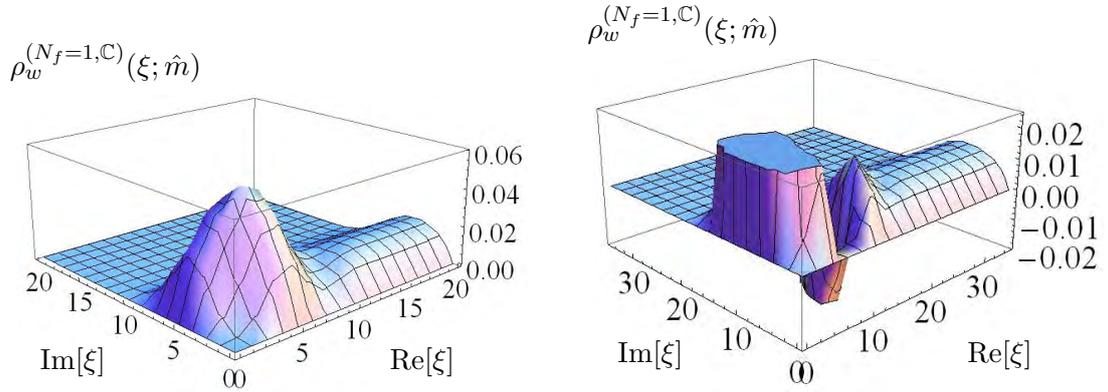

In
figures~\ref{fg:Complex-weak-density_Nf=1___No-1}--\ref{fg:Complex-weak-density_Nf=1___No-3}
we show plots of $\rho_{w}^{(N_f=0,\mathbb{C})}(\xi)$ from eq.~\eqref{rhoCweakQ}
and $\rho_{w}^{(N_f=1,\mathbb{C})}(\xi;\hm)$ from eq.~\eqref{rhoCweakNf1}
for several values of $\hmu$ and $\hm$. 
(Due to the symmetry $\rho_w^{(N_f,\mathbb{C})}(\xi)=\rho_w^{(N_f,\mathbb{C})}(\xi^*)
=\rho_w^{(N_f,\mathbb{C})}(-\xi)$ we only present the result in the first quadrant. 
The quark masses are taken to be imaginary as our Dirac operator is symmetric at $\mu=0$.) 
For small $\hmu$ (figure~\ref{fg:Complex-weak-density_Nf=1___No-1}), the effect of the massless 
flavour is to deplete the spectral density near the origin. As $\hmu$ increases 
(figures~\ref{fg:Complex-weak-density_Nf=1___No-2} and \ref{fg:Complex-weak-density_Nf=1___No-3}), 
qualitatively different effects emerge. In the region between the location of the mass 
(in this case, the origin) and the edge of the spectrum, there appears a domain of 
strong oscillations with the shape of an ellipse 
that grows in its peak size and in the number of ripples as $\hmu$
gets larger (see section~\ref{phase} for a detailed analysis of these
oscillations).  

\begin{figure}[t]
    \unitlength1.0cm
    \centering
      \qquad \quad 
      \includegraphics[clip=,width=6.5cm]{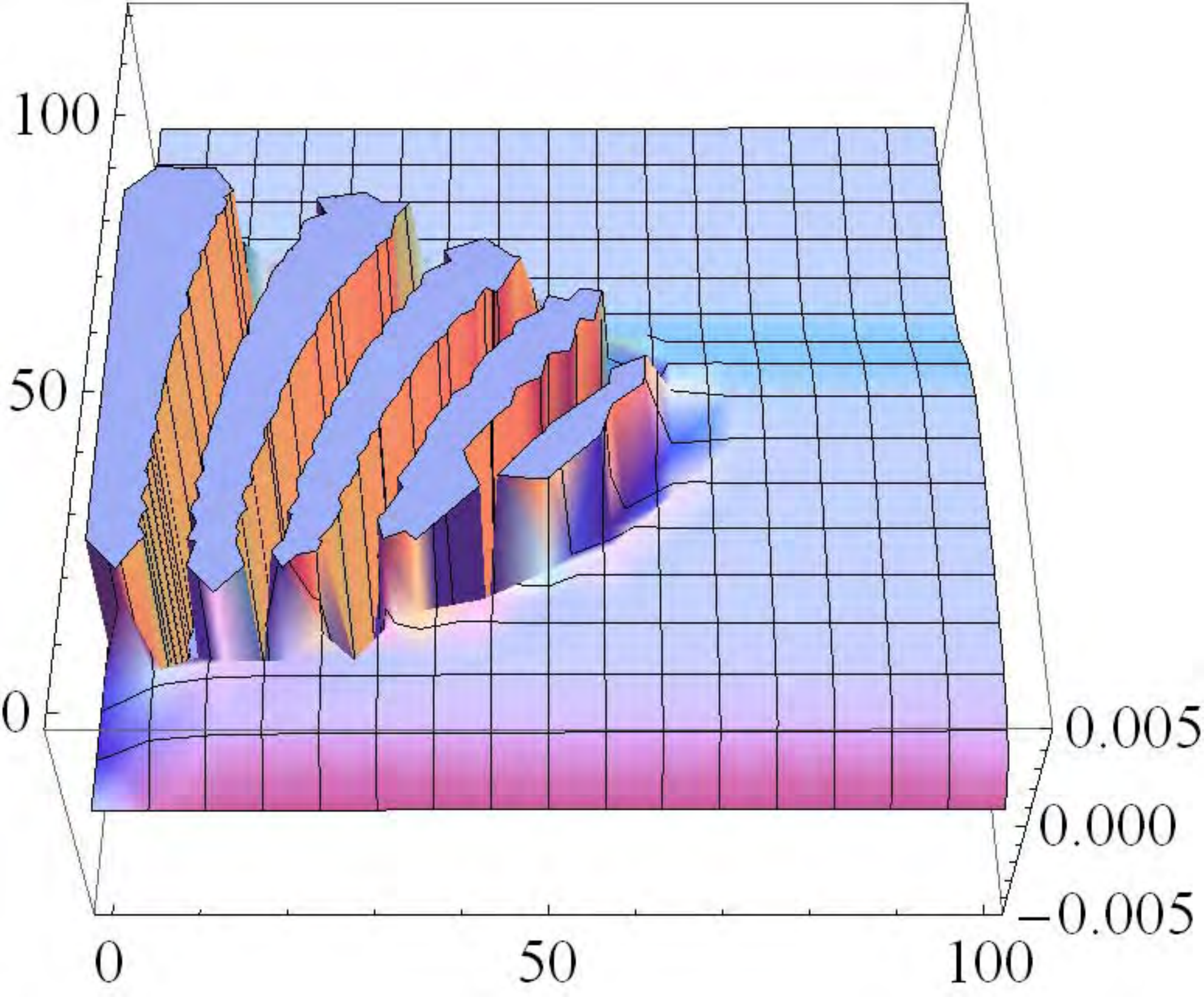}
      \qquad 
      \includegraphics[clip=,width=6.5cm]{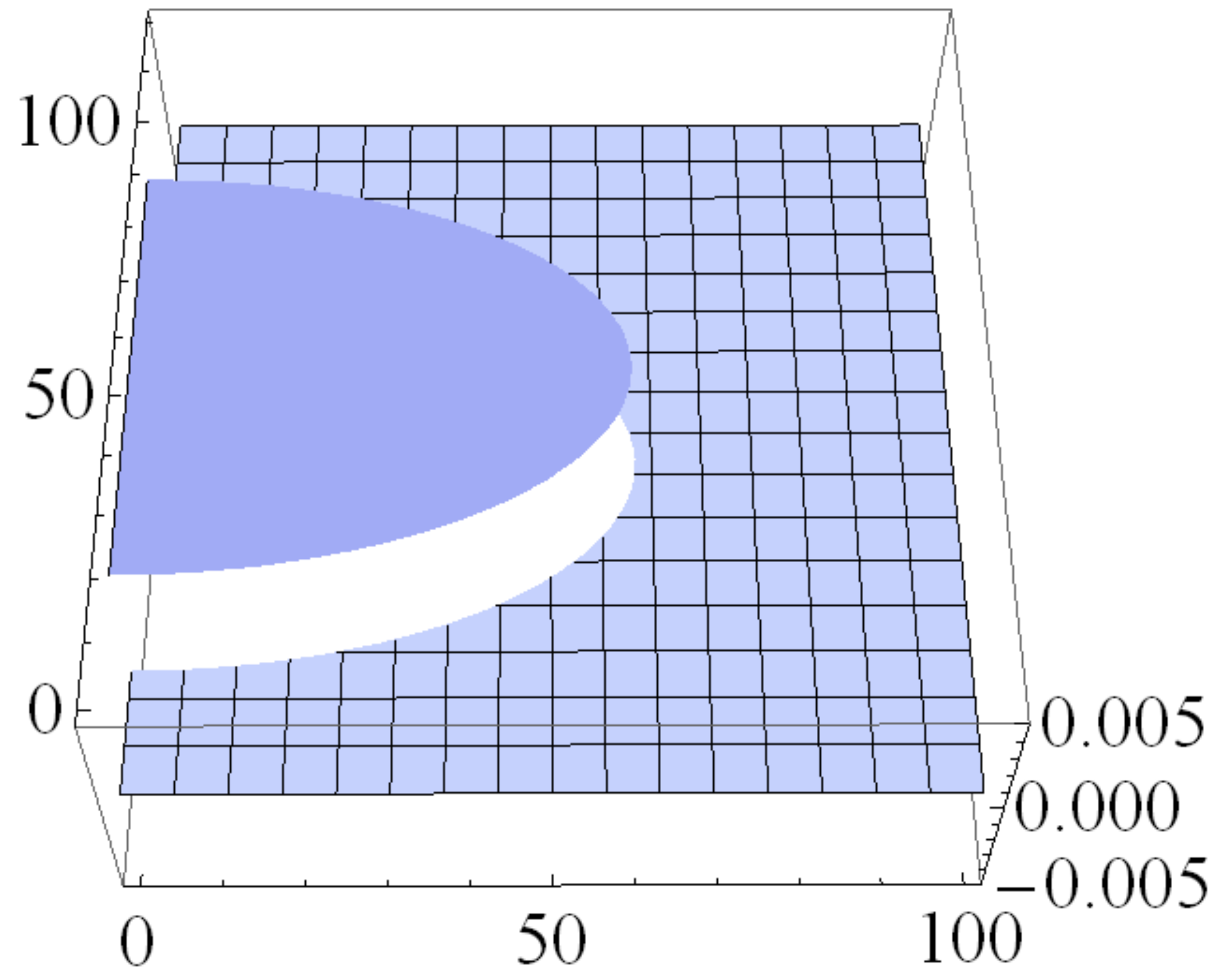}
      \put(-11.3,-0.5){$\re[\xi]$}
      \put(-15,3.1){$\im[\xi]$}
      \put(-3.9,-0.5){$\re[\xi]$}
      \put(-7.6,3){$\im[\xi]$}
      \put(-12.3,6){$\rho_{w}^{(N_f=1,\mathbb{C})}(\xi;\hm)$}
    \caption{\label{fg:Complex-weak-density_Nf=1___No-3}
      Left: $N_f=1$ spectral density of complex Dirac eigenvalues 
      at weak non-Hermiticity for $\hmu=6$ and $\hm=20i$ at $\nu=0$. 
      Right: Envelope of the oscillating region given by 
      eq.~\eqref{eq:envelope_weak_limit}, see section~\ref{phase} for more
      details. 
    }
\end{figure}
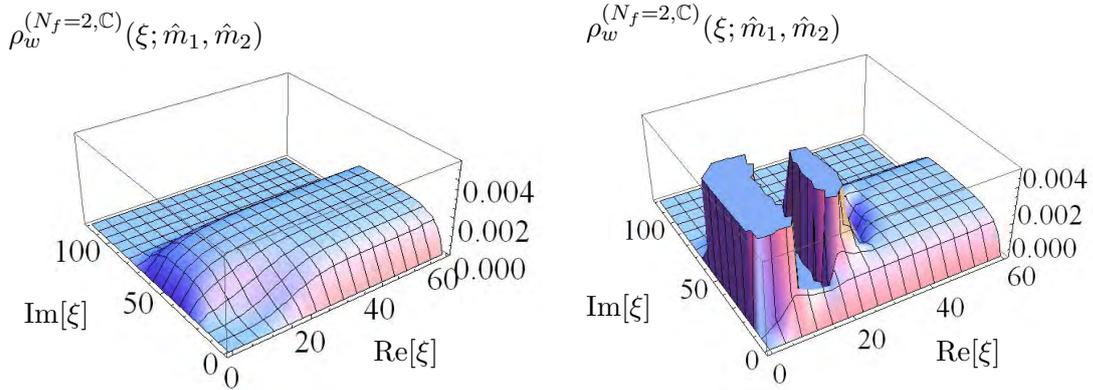
\begin{figure}[t]
    \unitlength1.0cm
    \centering
      \includegraphics[clip=,width=6.5cm]{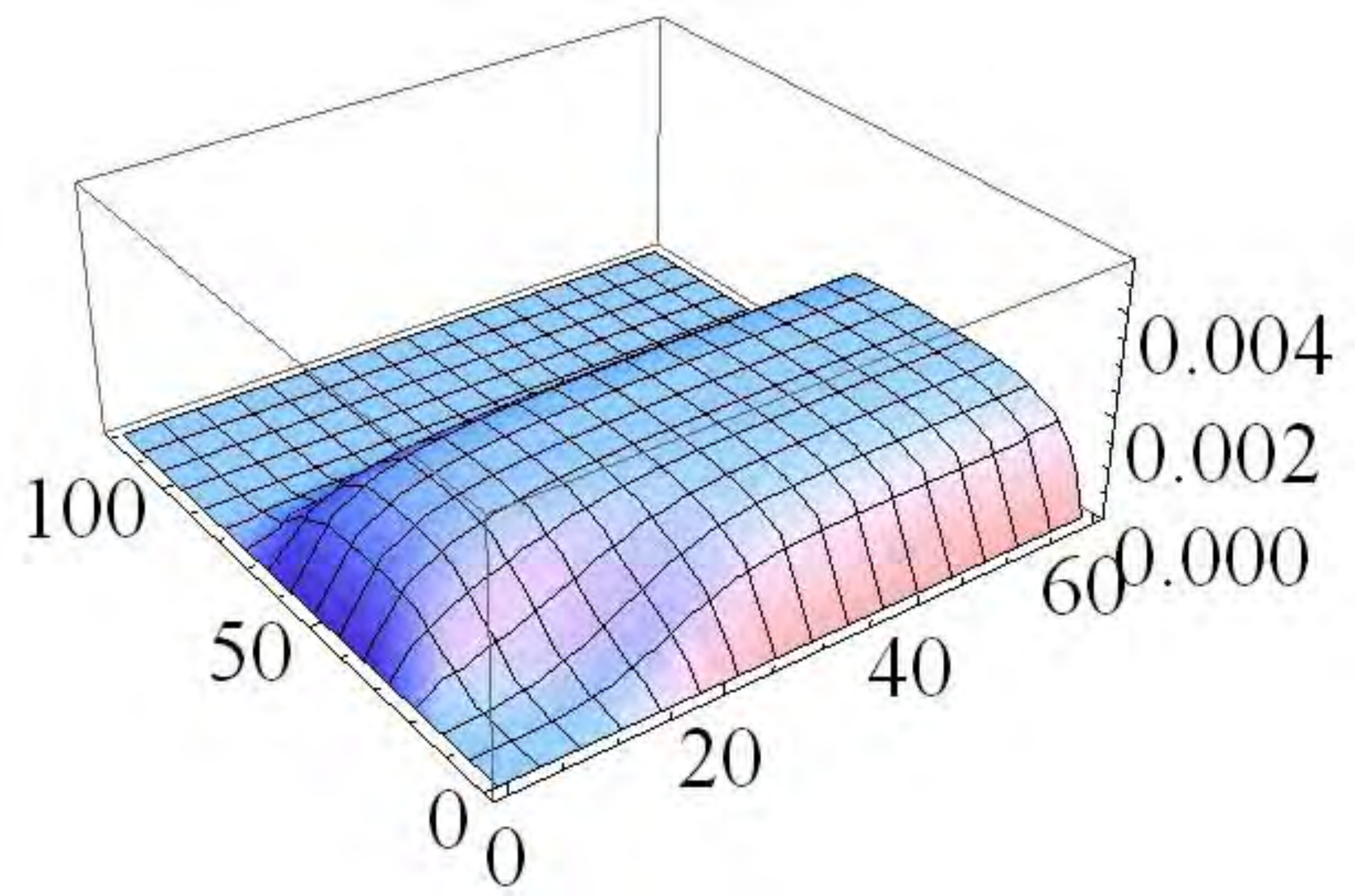} \qquad 
      \includegraphics[clip=,width=6.5cm]{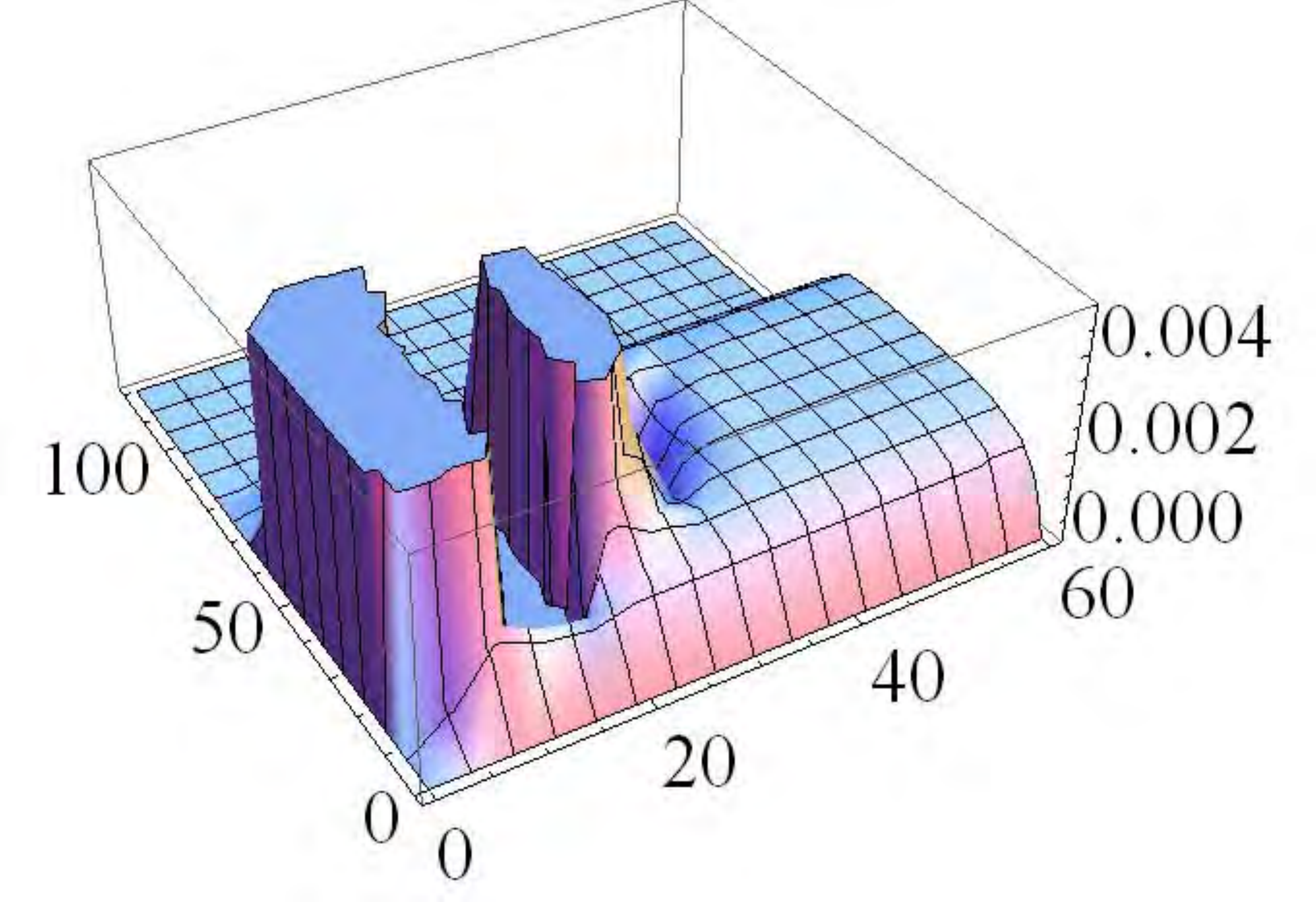}
      \put(-2.2,0.4){$\re[\xi]$}
      \put(-6.8,1.0){$\im[\xi]$}
      \put(-9.6,0.4){$\re[\xi]$}
      \put(-14.2,0.9){$\im[\xi]$}
      \put(-6.8,4.7){$\rho_{w}^{(N_f=2,\mathbb{C})}(\xi;\hm_1,\hm_2)$}
      \put(-14.4,4.6){$\rho_{w}^{(N_f=2,\mathbb{C})}(\xi;\hm_1,\hm_2)$}
      \vspace{0.2\baselineskip}
    \caption{\label{fg:Complex-weak-density_Nf=2}
      The $N_f=2$ spectral density of complex Dirac eigenvalues 
      at weak non-Hermiticity for $\hmu=6$ and $\nu=0$, for $\hm_1=\hm_2=10i$ (left) 
      and $\hm_1=10i$ and $\hm_2=70i$ (right). In the right figure only the positive 
      part of $\rho_{w}^{(N_f=2,\mathbb{C})}$ is shown for better readability.
    }
\end{figure}

This change in the spectrum signals the emergence of a severe sign problem. For $N_f=1$ 
the Dirac determinant is real but not necessarily positive, 
hence the spectral ``density'' could take negative values. The structure of the
oscillating domain, to be analysed in more detail in section~\ref{phase}, is 
quite similar to the corresponding one in the non-Hermitian RMT for three-colour QCD with $\mu\neq0$
\cite{Akemann:2004dr}, 
despite the difference in symmetries between two and three colours. 

Another important remark is in order. When $\hm$ becomes so large that it is
located outside the support  
of the quenched spectrum, the spectrum shows no change as compared to the quenched case. 
This is consistent with the physical expectation that a heavy flavour should
effectively decouple from  
the rest of the system. More importantly, it suggests that the sign problem will be milder 
when the quark mass is larger than a certain critical value, set by the
width of the quenched support.  
An essentially identical conclusion was reached for the three-colour RMT
(see, e.g., \cite{Splittorff:2008sw} for a review), 
again despite the difference from our two-colour model.

Next, we turn to the plot of
$\rho_{w}^{(N_f=2,\mathbb{C})}(\xi;\hm_1,\hm_2)$
from eq.~\eqref{rhoCweakNf2}
for two sets of masses in figure~\ref{fg:Complex-weak-density_Nf=2}.
For degenerate masses (left), there is no sign problem and 
no change in the spectrum except for
a depletion of eigenvalues  near the location of the masses. This situation is
qualitatively very similar to the symmetry class for adjoint QCD (which has no
real or imaginary eigenvalues), see the figures in \cite{Akemann:2005fd}.
For widely different masses, the density develops a domain of oscillations, 
which resembles the $N_f=1$ case. This implies the decoupling of one 
flavour to become heavy. 
In section~\ref{subsec_high_density} it will be shown that  
a similar oscillating spectrum emerges at high density for
well-separated masses for $N_f=2$.  
This was to be anticipated from the connection between the weak and strong large-$N$ limits 
(see the comment in the paragraph following eq.~\eqref{strongscale}).

\begin{figure}[t]
    \unitlength1.0cm
    \centering
      \includegraphics[clip=,width=6.6cm]{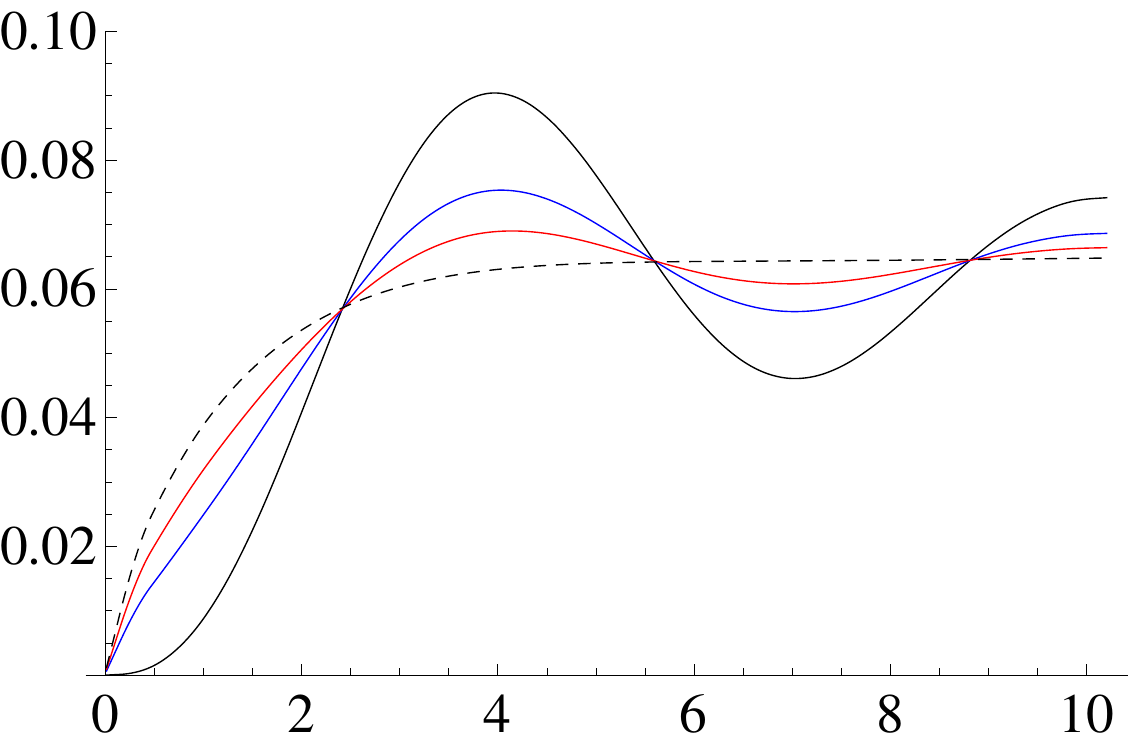} \qquad ~~
      \includegraphics[clip=,width=6.6cm]{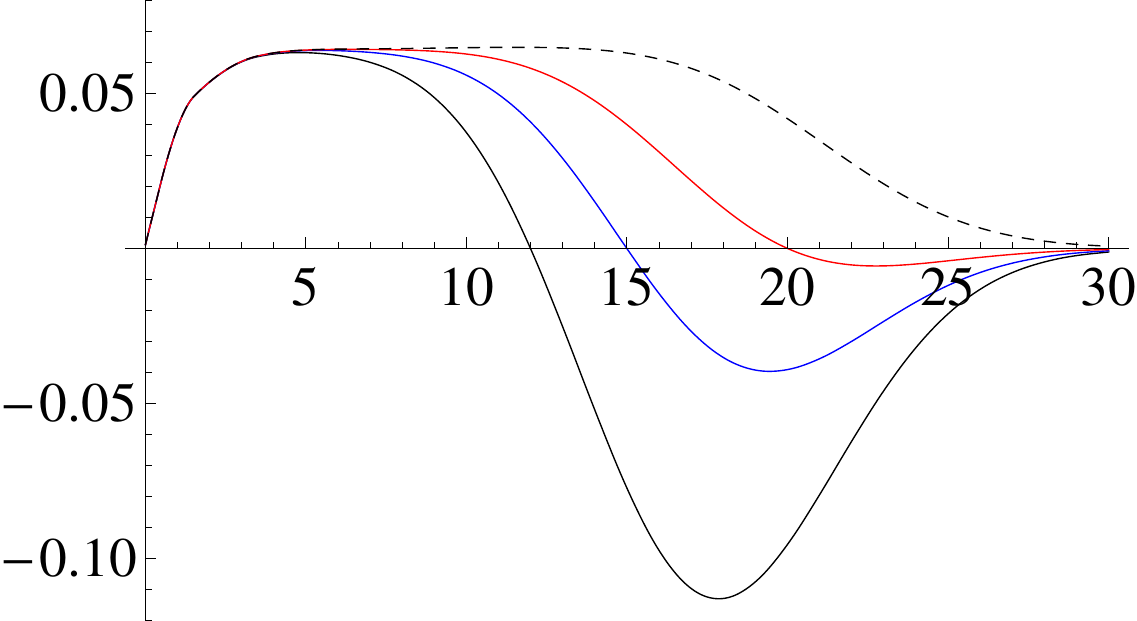}
      \put(0.2,2.1){$\xi$}
      \put(-7.6,0.35){$\xi$}
      \put(-6.8,4){$\rho_{w}^{(N_f=1,i\mathbb{R})}(i\xi;\hm)$}
      \put(-14.8,4.7){$\rho_{w}^{(N_f=1,\mathbb{R})}(\xi;\hm)$}
    \caption{\label{fg:Weak-density_Nf=1_for_real_and_imaginary}
      The $N_f=1$ spectral density of real (left) and imaginary (right) Dirac eigenvalues 
      at weak non-Hermiticity for $\hmu=3$ and $\nu=0$. 
      Left: $\hm=0.1i$ (black), $\hm=2i$ (blue), $\hm=3i$ (red), and
      quenched case (dashed).  
      Right: $\hm=12i$ (black), $\hm=15i$ (blue), $\hm=20i$ (red), and
      quenched case (dashed). Note that the zeros coincide with the mass values. 
    }
\end{figure}
\begin{figure}[t]
    \unitlength1.0cm
    \centering
      \includegraphics[clip=,width=6.6cm]{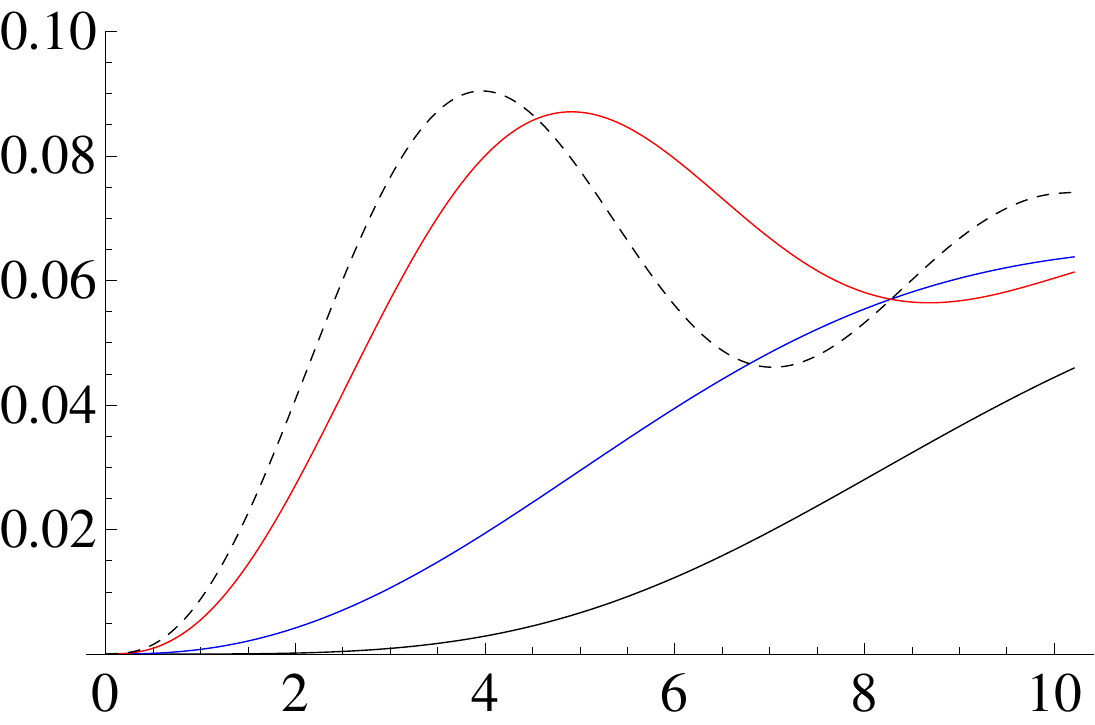} \qquad ~~
      \includegraphics[clip=,width=6.6cm]{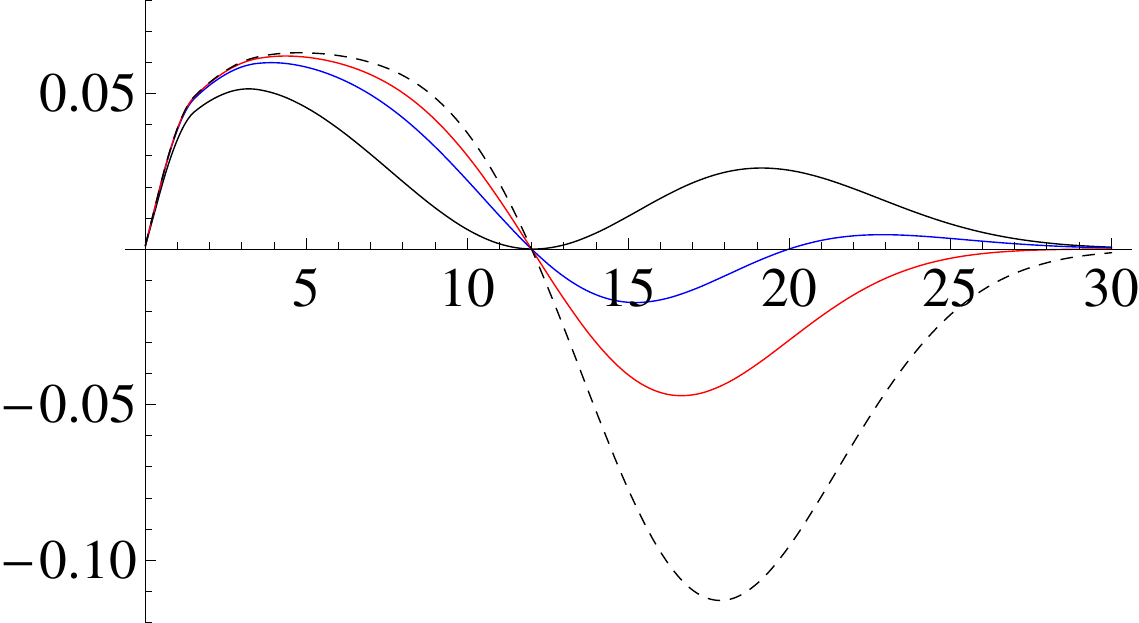}
      \put(0.2,2.1){$\xi$}
      \put(-7.6,0.3){$\xi$}
      \put(-6.8,4){$\rho_{w}^{(N_f=2,i\mathbb{R})}(i\xi;\hm_1,\hm_2)$}
      \put(-14.8,4.7){$\rho_{w}^{(N_f=2,\mathbb{R})}(\xi;\hm_1,\hm_2)$}
    \caption{\label{fg:Weak-density_Nf=2_for_real_and_imaginary}
      The $N_f=2$ spectral density of real (left) and imaginary (right) Dirac eigenvalues 
      at weak non-Hermiticity for $\hmu=3$ and $\nu=0$. 
      Left: $(\hm_1,\hm_2)=(0.1i,0.1i)$ (black), $(0.1i,10i)$ (blue),
      $(0.1i,30i)$ (red), and $N_f=1$ case with $\hm=0.1i$ (dashed).  
      Right: $(\hm_1,\hm_2)=(12i,12i)$ (black), $(12i,20i)$ (blue),
      $(12i,30i)$ (red), and $N_f=1$ case with $\hm=12i$ (dashed).  
Note that the zeros coincide with the mass values. 
    }
\end{figure}

In figure~\ref{fg:Weak-density_Nf=1_for_real_and_imaginary} 
we show plots of $\rho_{w}^{(N_f=1,\,(i)\mathbb{R})}(\xi;\hm)$ from eq.~\eqref{rhoRweakNf1}. 
The effect of a dynamical flavour is always stronger for the imaginary eigenvalues than for 
the real eigenvalues. The reason is that the partition function and densities
vanish when the imaginary eigenvalues equal the (in our case) imaginary masses,
see eqs.~\eqref{Fdef} and \eqref{Zgen}. Thus the imaginary density
crosses the horizontal axis at the value of the mass.%
\footnote{Using eq. \eqref{eq:R_Z_sign_k=1} in the appendix and the positivity of the partition 
function for imaginary masses, one can prove for general $N_f$ that the imaginary density cannot cross 
the horizontal axis at values away from the masses. This explains why a rapid oscillation as in figure 
\ref{fg:Complex-weak-density_Nf=1___No-3} does not occur in the imaginary density.} 
Interestingly the dynamical flavour brings about an oscillatory behaviour in the real density. 
For both real and imaginary densities, one can clearly observe a convergence
to the quenched density as the mass is increased, which leads to an effective
quenching (see appendix \ref{heavy} for an analytic confirmation of this fact).  

In figure~\ref{fg:Weak-density_Nf=2_for_real_and_imaginary} 
we show plots of $\rho_{w}^{(N_f=2,\,(i)\mathbb{R})}(\xi;\hm_1,\hm_2)$
from eq.~\eqref{rhoRweakNf2}.  
We illustrate the effect of dynamical flavours by changing one of the masses
while keeping the other one fixed.  
For both real and imaginary densities, one can clearly observe a convergence to the $N_f=1$ density 
as one of the masses is increased. Note also that the spectral density for
degenerate masses is positive,  
as it should be, because there is no sign problem there.

Because of the issue of non-interchangeability of integration and large-$N$
limit in the weak limit, as well as in order to better understand explicitly the relation 
between theories with different flavour content, we have checked in appendix
\ref{heavy} that sending a subset of masses to infinity leads to the decoupling of these.

\subsection{Eigenvalue densities at high density}
\label{subsec_high_density}

In this subsection we will give the results for the spectral density in 
the large-$N$ limit at strong non-Hermiticity, which corresponds physically to the
high-density regime.
In this large-$N$ limit the rescaling of the parameters is different
from the one in eq.~(\ref{weaklim}) and
is given by (see also eq.~(\ref{strongscale}))
\bea
  \hm_{f} & \equiv  & \mm_f\,, \nn\\
  \xi & \equiv  &  \La\,,
  \label{stronglim}
\eea
i.e., the masses and the Dirac eigenvalues, as well as $\mu$, 
need no rescaling at all. 
For the microscopic densities of Wishart eigenvalues we have
\bea
  R_s^{(N_f,\mathbb{C})}(z) &\equiv& \lim_{N\to\infty} 
  R_1^{(N_f,\mathbb{C})}(z)\,, \nn\\
  R_s^{(N_f,(i)\mathbb{R})}(x) &\equiv& \lim_{N\to\infty} 
  R_1^{(N_f,\mathbb{R})}(x)\,,
  \label{rhostrongdef}
\eea
and the densities of Dirac eigenvalues are again obtained using the
mappings in eq.~\eqref{eq:weak_mappings}.
We again begin by quoting the quenched densities from \cite{Akemann:2009fc}, 
which will constitute the building blocks for the new unquenched
results.  We have
\be
  \rho_s^{(N_f=0,\mathbb{C})}(\xi) =  4 |\xi|^2 \hg_s(\xi^{*2},\xi^2) \mcK_s(\xi^2,\xi^{*2})\,,
  \label{rhoCstr}
\ee
where the complex weight in the strong limit is defined as
\begin{align}
  \label{ghats}
  \hg_s(z,z^*) & = - \hg_s(z^*,z)
  \equiv 2i \, \sgn(\im z) \, \frac{1}{|z|^\nu}\,g(z^*,z) \\
  & = 4i \sgn(\im z)\, \e^{2\eta_-\re z} \int_0^\infty\!
  \frac{dt}{t}\e^{-\eta_+^2 t(z^2+z^{*2})-\frac{1}{4t}}
  K_{\frac{\nu}{2}}\left(2\eta_+^2t |z|^2\right)
  \erfc\big(2\eta_+\sqrt{t}\,|\im z|\big)\notag
\end{align}
with $g$ from eq.~\eqref{eq:def_g}.
It can be seen from that equation that the cases of general $\mu\leq1$ and of 
maximal non-Hermiticity with $\mu=1$ (and $\he_+=\tfrac{1}{2}$) are related simply by a
rescaling of the complex eigenvalue $\he_+\xi^2\to\xi^2/2$. The same is true
for the densities of the real and imaginary eigenvalues,
\begin{align}  
  \rho_s^{(N_f=0,(i)\mathbb{R})}(\xi) = 2|\xi|\frac{\he_+^3}{8\pi}
  \, 2K_{\frac{\nu}{2}}(\he_+|\xi|^2) & \left(
  \int_{0}^{\infty}dx'\,|\xi^2 -x'|\
  2K_{\frac{\nu}{2}}(\he_+|x'|)\
  I_{\nu}( 2\he_+ \xi\sqrt{x'})\right.
  \label{rhoRstr}
  \\
  &\left.
  +\int_{-\infty}^{0}dx'\,|\xi^2 -x'|\
  2K_{\frac{\nu}{2}}(\he_+|x'|)\
  J_{\nu}( 2\he_+ \xi\sqrt{|x'|})
  \right).
  \notag
\end{align}
Unlike in the weak limit, in the strong limit no complications arise
from an interchange of the large-$N$ limit and the integration.

The example of two non-degenerate flavours is now straightforward, and we
only quote the result which includes eq.~(\ref{rhoCstr}) as the first term,
\begin{align}
  &\rho_s^{(N_f=2,\mathbb{C})}(\xi;\hm_1,\hm_2) \notag\\
  &\quad= \rho_s^{(0,\mathbb{C})}(\xi)
  +4 |\xi|^2 \hg_s(\xi^{*2},\xi^2)
  \frac{{\cal K}_s(\xi^2,\hm_2^2){\cal K}_s(\xi^{*2},\hm_1^2)-
  {\cal K}_s(\xi^2,\hm_1^2){\cal K}_s(\xi^{*2},\hm_2^2)}
  {{\cal K}_s(\hm_1^2,\hm_2^2)}\,.
  \label{rhoCstrongNf2}
\end{align}
It is identical to the form of the weak result in eq.~(\ref{rhoCweakNf2})
apart from the kernel in the strong limit from eq.~(\ref{strongK}),
\be
  {\cal K}^{}_s(u,v)\ =\ \frac{\eta_+^3}{8\pi}
  (u-v)\,\e^{-\eta_{-} (u+v)}
  \,I_{\nu}\left(
  2 \eta_+ \sqrt{uv} \right).
  \nn
\ee
It is easy to obtain the limiting result for two degenerate masses which we
also give here for completeness,
\begin{align}
  & \rho_s^{(N_f=2,\mathbb{C})}(\xi;\hm_1,\hm_2) 
  \notag
  \\
  & \quad= \rho_s^{(N_f=0,\mathbb{C})}(\xi)   +4 |\xi|^2 \hg_s(\xi^{*2},\xi^2)
  \frac{\eta_+^3\,\e^{-\eta_{-} (\xi^2+\xi^{*\,2})}}{8\pi I_\nu(2\eta_+\hm_1^2)}
  \Big[(\xi^2-\xi^{*\,2})\big|I_\nu(2\eta_+\hm_1\xi)\big|^2 \notag\\
  & \qquad+
  (\xi^2-\hm_1^2)(\xi^{*\,2}-\hm_1^2)\frac{\eta_+}{\hm_1}
  \Big( I_\nu(2\eta_+\hm_1\xi^*) \xi I_{\nu+1}(2\eta_+\hm_1\xi)
  -(\xi\leftrightarrow\xi^*)
  \Big) 
  \Big].
\end{align}
Similarly, the extension of eq.~\eqref{rhoRstr} to $N_f=2$ reads
\begin{multline}
  \rho_s^{(N_f=2,(i)\mathbb{R})}(\xi;\hm_1,\hm_2) = \rho_s^{(N_f=0,(i)\mathbb{R})}(\xi)
  + 2|\xi| \hh_s(\xi^2)
  \int_{\mathbb{R}}dx~ \hh_s(x) \sgn(\xi^2-x)
  \\
  \times \frac{{\cal K}_s(\xi^2,\hm_2^2){\cal K}_s(x,\hm_1^2)
  -{\cal K}_s(\xi^2,\hm_1^2){\cal K}_s(x,\hm_2^2)}
  {{\cal K}_s(\hm_1^2,\hm_2^2)}\,,
\end{multline}
where the strong limit of the real weight is defined as 
\be
  \hh_s(z) \equiv \frac{1}{(-z)^{\nu/2}}h(z)
  = 2 \frac{|z|^{\nu/2}}{(-z)^{\nu/2}} \e^{\eta_-z}K_{\frac\nu2}(\eta_+|z|)\,.
  \label{eq:def_h_s}
\ee
\begin{figure}[t]
    \unitlength1.0cm
    \centering
      \includegraphics[clip=,width=6cm]{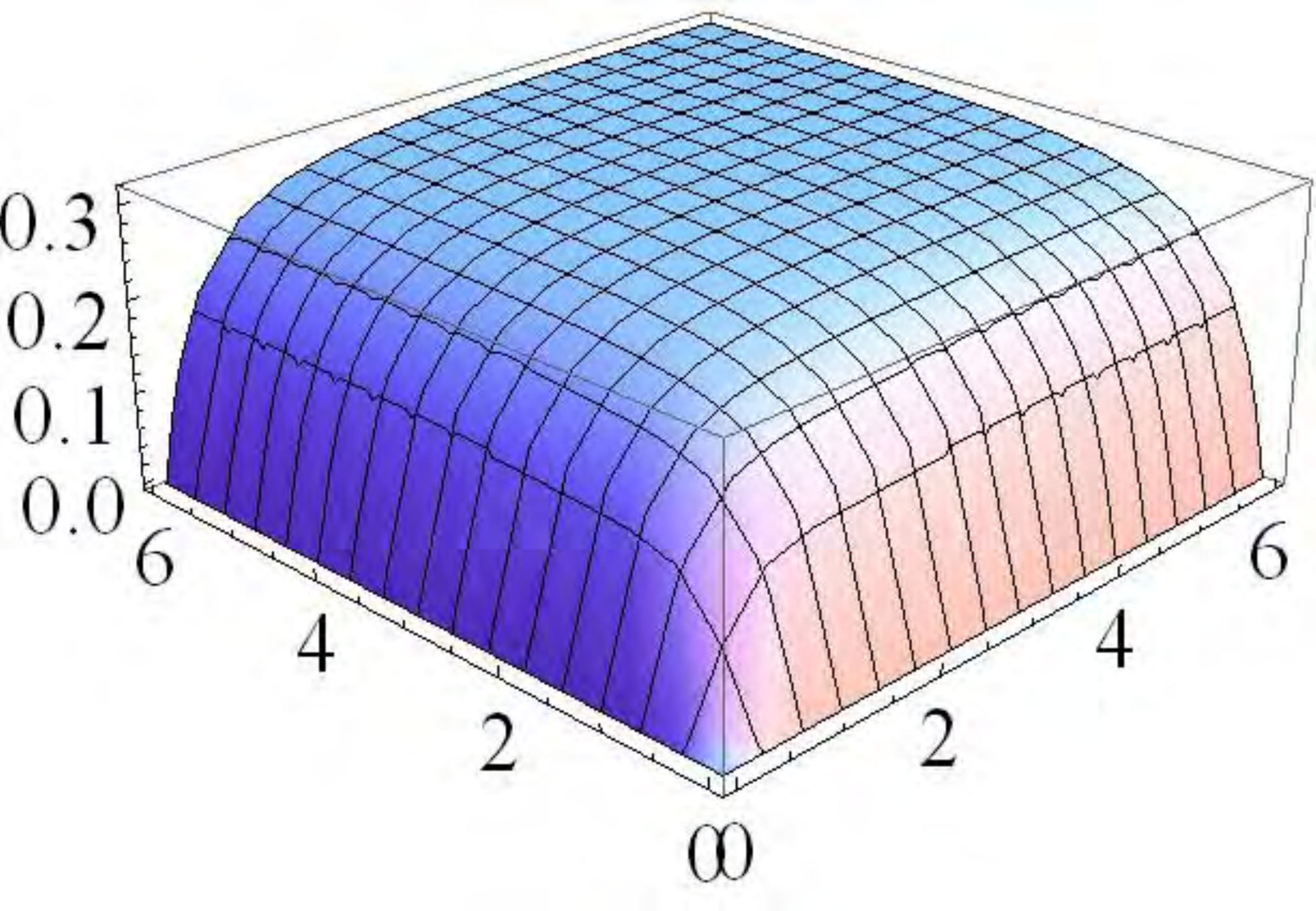} \qquad \quad  
      \includegraphics[clip=,width=6cm]{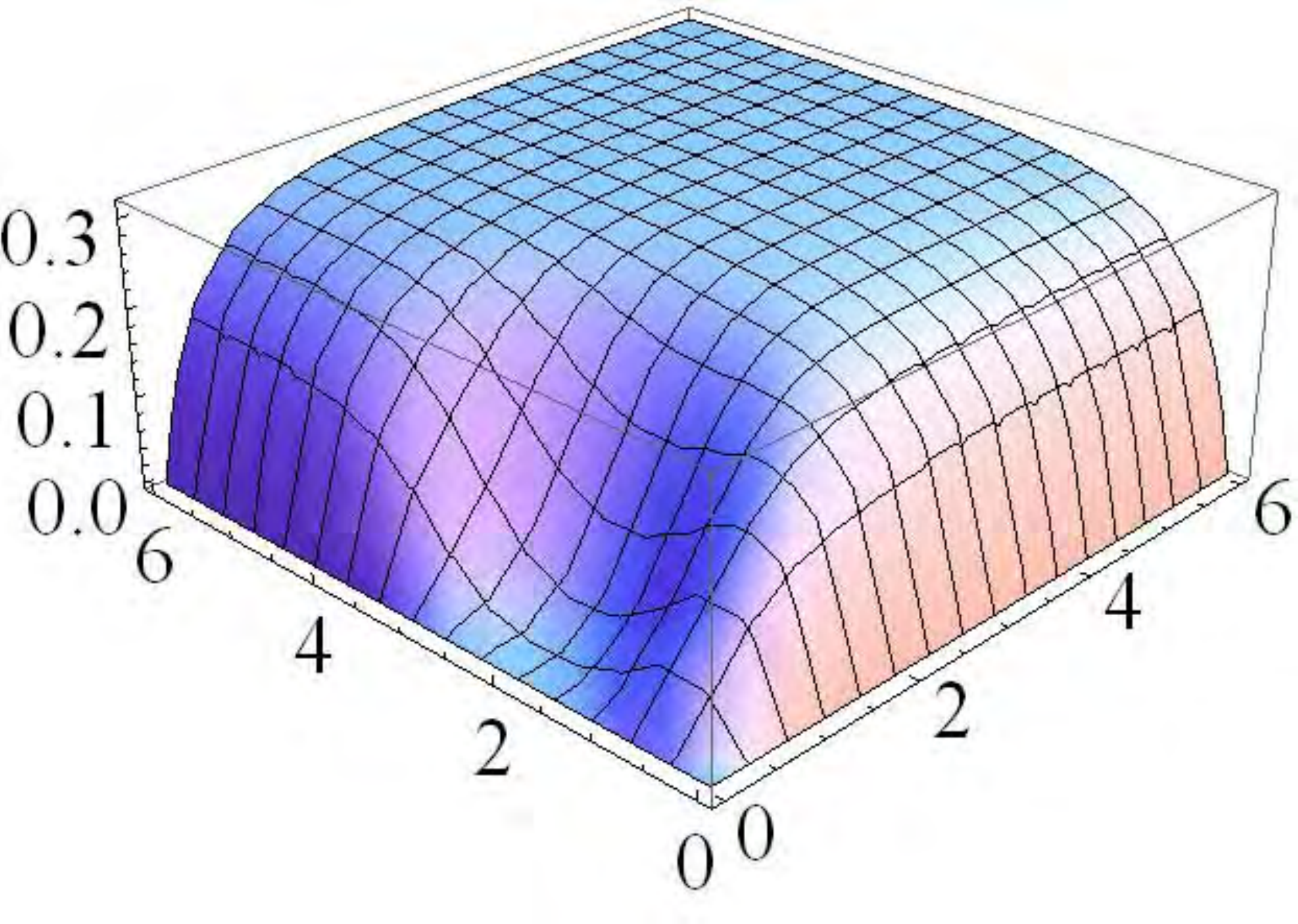}
      \put(-0.9,0.5){$\re[\xi]$}
      \put(-5.7,0.5){$\im[\xi]$}
      \put(-8.3,0.6){$\re[\xi]$}
      \put(-13.0,0.6){$\im[\xi]$}
      \put(-6.4,4.0){$\rho_{s}^{(N_f=2,\mathbb{C})}(\xi)$}
      \put(-13.5,4.1){$\rho_{s}^{(N_f=0,\mathbb{C})}(\xi)$}
    \caption{\label{fg:Complex-strong-density_1}
      Quenched (left) and $N_f=2$ (right) spectral density of complex Dirac eigenvalues 
      at maximal non-Hermiticity $(\mu = 1)$ for $\nu=0$, 
      with equal $\hm_1=\hm_2=2i$ in the right figure. Note again the
      similarity with adjoint QCD \cite{Akemann:2005fd}.
    }
\end{figure}%
\begin{figure}[t]
    \unitlength1.0cm
    \centering
      \includegraphics[clip=,width=6cm]{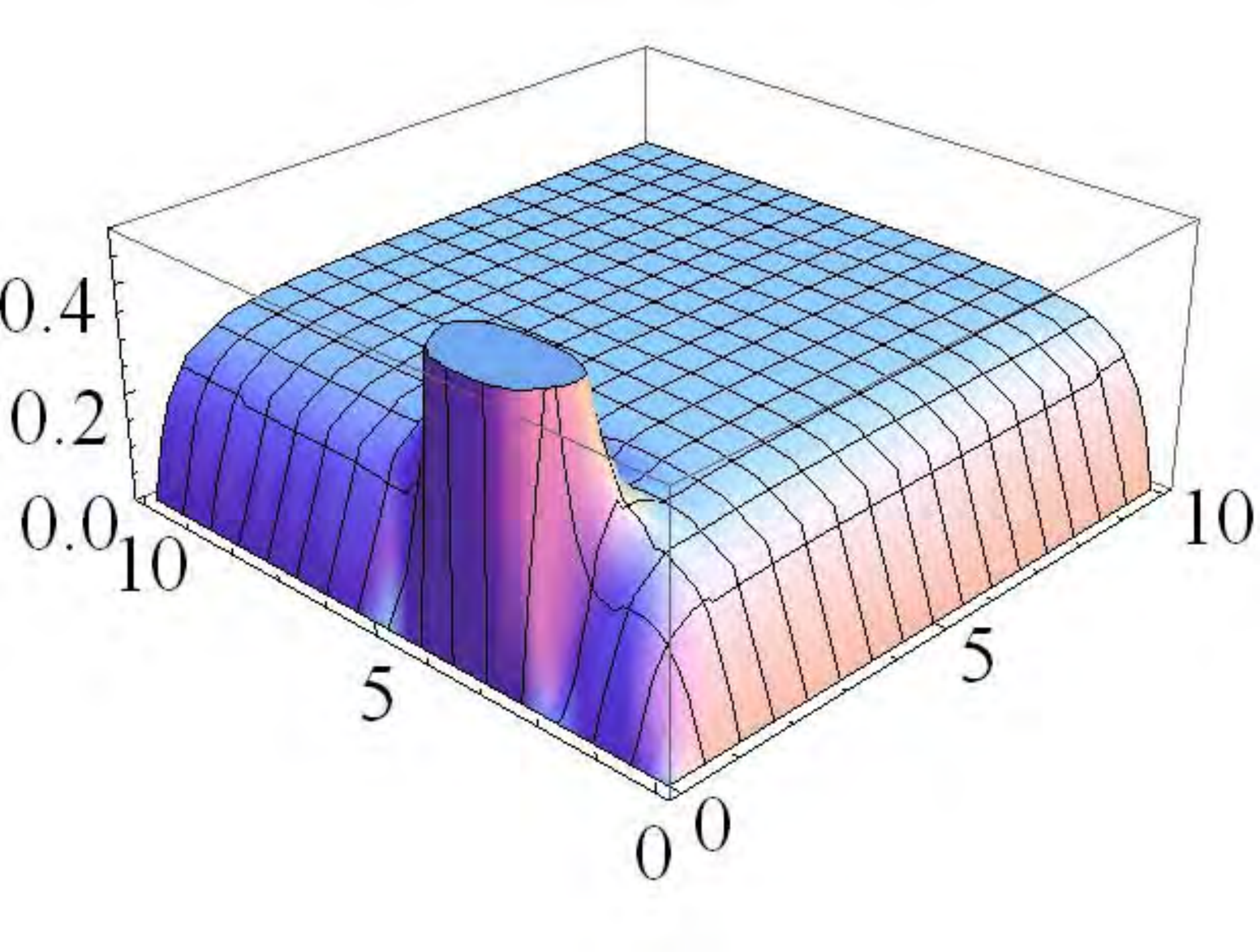} \qquad \quad  
      \includegraphics[clip=,width=6cm]{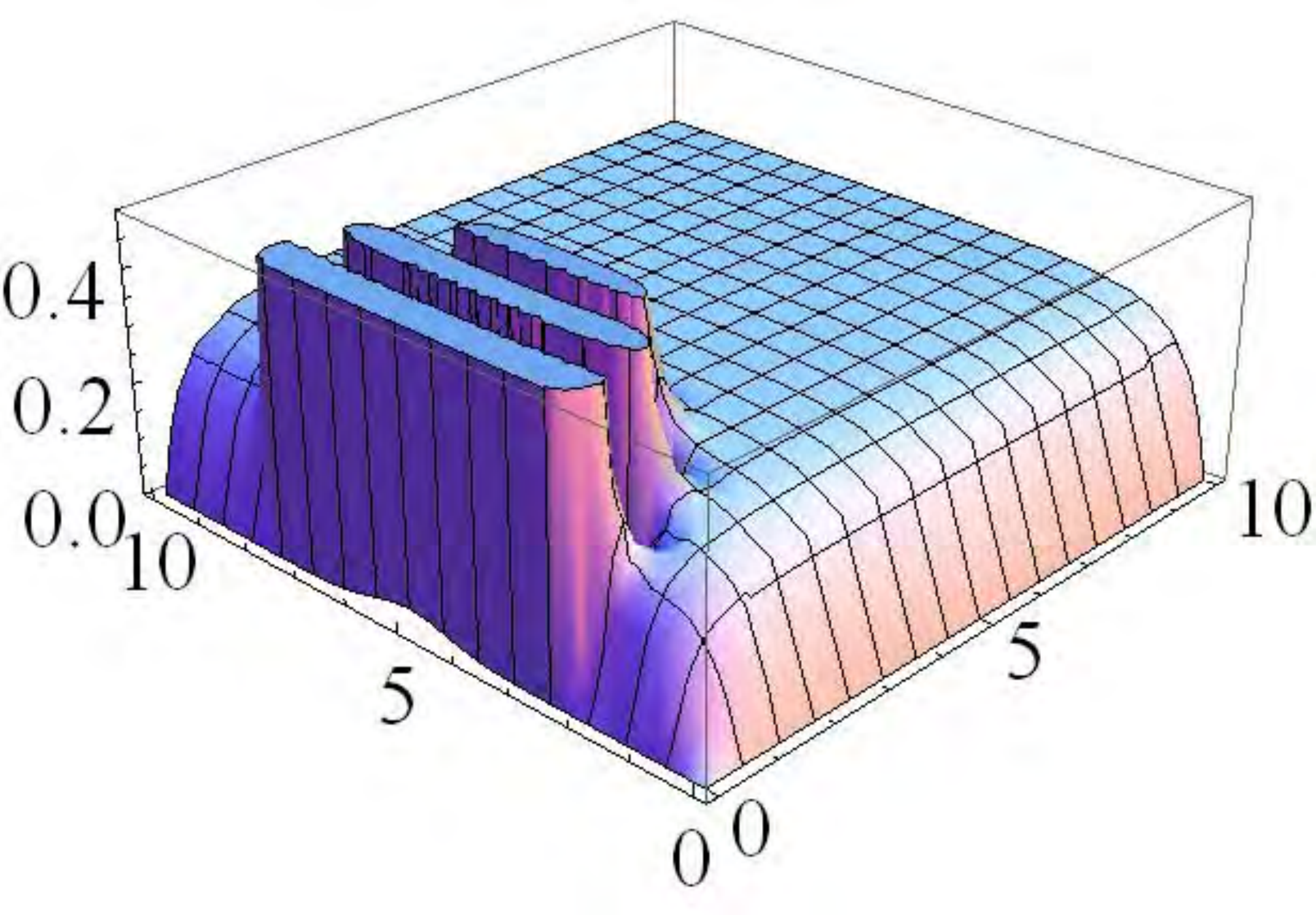}
      \put(-1.2,0.6){$\re[\xi]$}
      \put(-5.5,0.5){$\im[\xi]$}
      \put(-8.5,0.5){$\re[\xi]$}
      \put(-12.9,0.5){$\im[\xi]$}
      \put(-6.7,4.2){$\rho_{s}^{(N_f=2,\mathbb{C})}(\xi)$}
      \put(-13.9,4.2){$\rho_{s}^{(N_f=2,\mathbb{C})}(\xi)$}
    \caption{\label{fg:Complex-strong-density_2}
      The $N_f=2$ spectral density of complex Dirac eigenvalues 
      at maximal non-Hermiticity $(\mu = 1)$ for $\nu=0$ 
      with $(\hm_1,~\hm_2)=(2i,~5i)$ (left) and $(\hm_1,~\hm_2)=(2i,~8i)$ (right). 
      In the right figure the negative part of $\rho_s$ is cut for better readability.
    }
\end{figure}
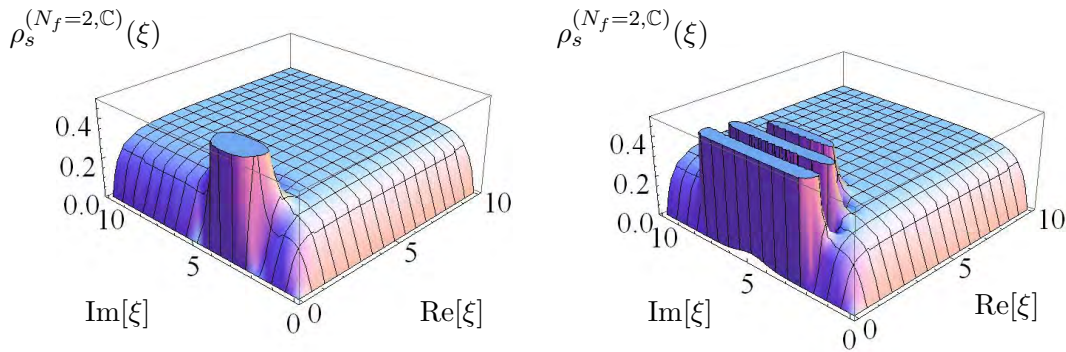%
The density $\rho_s^{(N_f=2,\mathbb{C})}(\xi;\hm_1,\hm_2)$ is plotted in 
figures~\ref{fg:Complex-strong-density_1} and
\ref{fg:Complex-strong-density_2}.  
In the former the quenched result is also included for comparison. 
(Again, we only present the result in the first quadrant
because of the symmetry $\rho_s^{(N_f,\mathbb{C})}(\xi)=\rho_s^{(N_f,\mathbb{C})}(\xi^*)
=\rho_s^{(N_f,\mathbb{C})}(-\xi)$, and we take 
the quark masses to be imaginary as our Dirac operator is symmetric at $\mu=0$. 
Since the difference between $0<\mu<1$ and $\mu=1$ is trivial we set $\mu=1$ in the figures.) 
We observe that the effect of dynamical flavours with equal masses is to simply 
generate a dip in the quenched spectrum. The effect of unequal masses is qualitatively different.  
For a sufficiently large mass difference, the spectrum exhibits a domain of strong oscillations, 
taking both positive and negative values, which signals the appearance of the sign problem. 
We caution that sending one of the masses to infinity does not yield 
$\rho_{s}^{(N_f=1,\mathbb{C})}(\xi;\hm)$ but merely makes the oscillating
domain extend to the entire spectrum.\footnote{This is not in
  contradiction with the usual decoupling of heavy flavours, see
  appendix~\ref{heavy}, since in the strong limit we are restricted to
  even $N_f$.}

\begin{figure}[t]
    \unitlength1.0cm
    \centering
      \includegraphics[clip=,width=5.8cm]{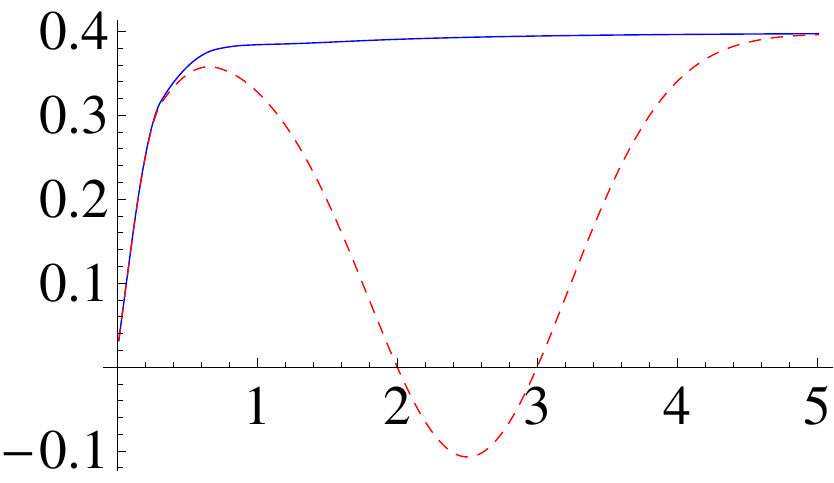} \qquad \quad  
      \includegraphics[clip=,width=5.8cm]{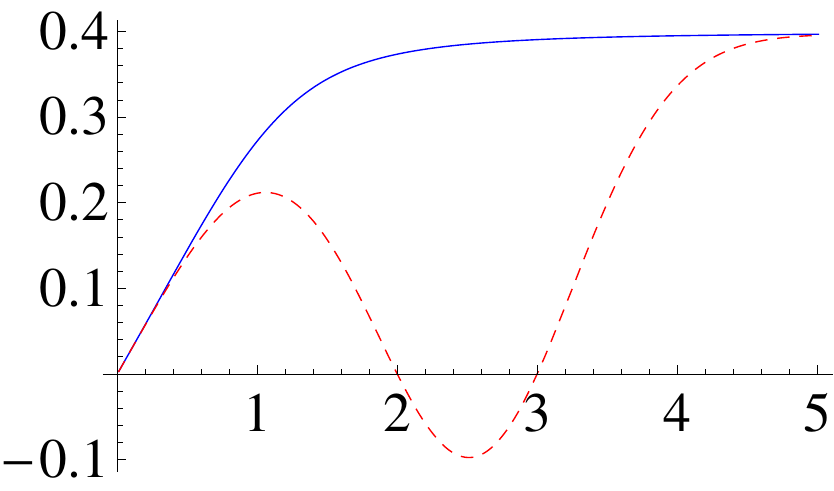}
      \put(0.3,0.7){$\xi$}
      \put(-6.7,0.7){$\xi$}
      \put(-6.4,3.9){$\rho_{s}^{(N_f=2,\,(i)\mathbb{R})}\big((i)\xi\big)$}
      \put(-13.5,3.9){$\rho_{s}^{(N_f=2,\,(i)\mathbb{R})}\big((i)\xi\big)$}
    \caption{\label{fg:Real-strong-density___m1=2i___m2=3i}
      The $N_f=2$ spectral density of Dirac eigenvalues 
      at maximum non-Hermiticity 
      ($\mu = 1$) for $(\hm_1,~\hm_2)=(2i,~3i)$ 
      at $\nu=0$ (left) and $\nu=2$ (right). 
      The densities of both real
      (blue full line) and imaginary (red dashed line) eigenvalues
      are displayed in the same plot for comparison. The latter again has
      zeros at the locations of the masses.
    }
\end{figure}

The density $\rho_{s}^{(N_f=2,\,(i)\mathbb{R})}\big((i)\xi;\hm_1,\hm_2\big)$ is plotted in 
figure~\ref{fg:Real-strong-density___m1=2i___m2=3i}.
As in the case of weak non-Hermiticity, the density of imaginary
eigenvalues goes through zero at $\xi=i\hm_1$ and $\xi=i\hm_2$.
However, unlike in the case of weak non-Hermiticity, no oscillation of
the density of real eigenvalues is observed. 
A more detailed analysis of the sign problem will be presented in section \ref{sc:sign-problem}.

We now consider the case of four quark flavours, which is
interesting since at high density $N_f=2$ and $N_f\geq 4$ (even) have
distinct chiral symmetry breaking patterns and a new class of
Nambu-Goldstone bosons emerges for $N_f\ge4$ \cite{Kanazawa:2009ks}.
With non-degenerate masses, 
we have to expand the Pfaffian of a $6\times6$ matrix in order to 
write down the complex density explicitly. The result is
\begin{align}
  \label{rhoCstrongNf4}
  \rho_s^{(N_f=4,\mathbb{C})}&(\xi;\{\hm\}_4)
  =\rho_s^{(N_f=0,\mathbb{C})}(\xi)\\
  &+ \frac{4 |\xi|^2 \hg_s(\xi^{*2},\xi^{2})}{{\cal K}_s(\hm_1^2,\hm_2^2){\cal K}_s(\hm_3^2,\hm_4^2) 
  - {\cal K}_s(\hm_1^2,\hm_3^2){\cal K}_s(\hm_2^2,\hm_4^2) + {\cal K}_s(\hm_1^2,\hm_4^2){\cal K}_s(\hm_2^2,\hm_3^2)} \nn  \\
  &\;  \times \Big\{ 
  {\cal K}_s(\hm_1^2,\xi^{*2}){\cal K}_s(\hm_2^2,\xi^2){\cal K}_s(\hm_3^2,\hm_4^2) - 
  {\cal K}_s(\hm_1^2,\xi^2){\cal K}_s(\hm_2^2,\xi^{*2}){\cal K}_s(     \hm_3^2,\hm_4^2) \nn \\
  &\quad- {\cal K}_s(\hm_1^2,\xi^{*2}){\cal K}_s(\hm_2^2,\hm_4^2){\cal K}_s(\hm_3^2,\xi^2) 
  + {\cal K}_s(\hm_1^2,\hm_4^2){\cal K}_s(\hm_2^2,\xi^{*2}){\cal K}_s(\hm_3^2,\xi^2) \nn \\
   &\quad+ {\cal K}_s(\hm_1^2,\xi^2){\cal K}_s(\hm_2^2,\hm_4^2){\cal K}_s(\hm_3^2,\xi^{*2}) 
  - {\cal K}_s(\hm_1^2,\hm_4^2){\cal K}_s(\hm_2^2,\xi^2){\cal K}_s(\hm_3^2,\xi^{*2}) \nn \\
  &\quad+ {\cal K}_s(\hm_1^2,\xi^{*2}){\cal K}_s(\hm_2^2,\hm_3^2){\cal K}_s(\hm_4^2,\xi^2) 
  - {\cal K}_s(\hm_1^2,\hm_3^2){\cal K}_s(\hm_2^2,\xi^{*2}){\cal K}_s(\hm_4^2,\xi^2) \nn \\
  &\quad+ {\cal K}_s(\hm_1^2,\hm_2^2){\cal K}_s(\hm_3^2,\xi^{*2}){\cal K}_s(\hm_4^2,\xi^2) 
  - {\cal K}_s(\hm_1^2,\xi^2){\cal K}_s(\hm_2^2,\hm_3^2){\cal K}_s(\hm_4^2,\xi^{*2}) \nn \\
  &\quad+ {\cal K}_s(\hm_1^2,\hm_3^2){\cal K}_s(\hm_2^2,\xi^2){\cal K}_s(\hm_4^2,\xi^{*2}) 
  - {\cal K}_s(\hm_1^2,\hm_2^2){\cal K}_s(\hm_3^2,\xi^2){\cal K}_s(\hm_4^2,\xi^{*2})
  \Big\}\,.
  \nn
\end{align}

\begin{figure}[t]
    \unitlength1.0cm
    \centering
      \includegraphics[clip=,width=6cm]{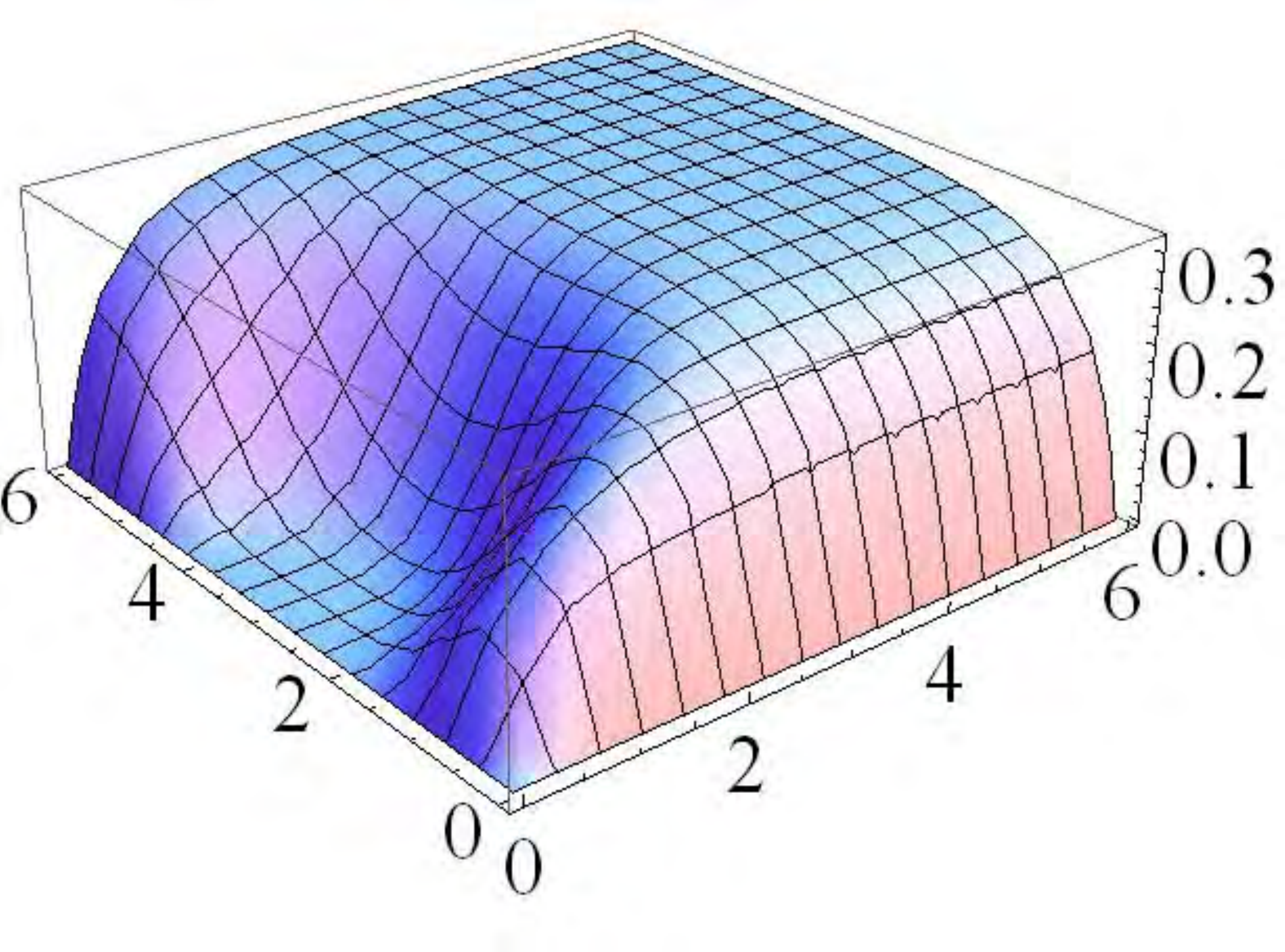} \qquad \quad  
      \includegraphics[clip=,width=6cm]{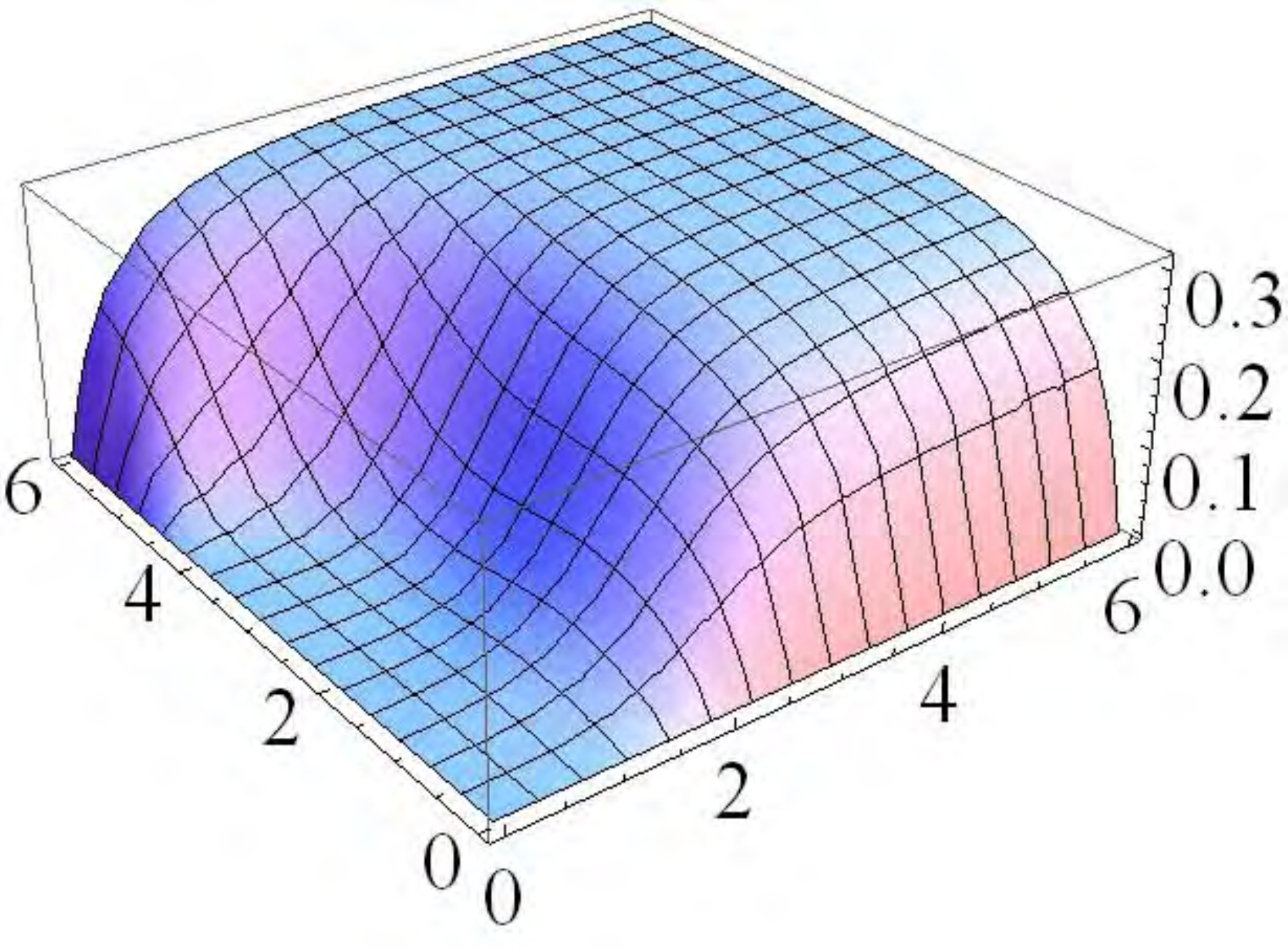}
      \put(-1.6,0.7){$\re[\xi]$}
      \put(-6.2,0.9){$\im[\xi]$}
      \put(-8.9,0.6){$\re[\xi]$}
      \put(-13.5,0.8){$\im[\xi]$}
      \put(-6.7,4.6){$\rho_{s}^{(N_f=4,\mathbb{C})}(\xi)$}
      \put(-13.6,4.6){$\rho_{s}^{(N_f=4,\mathbb{C})}(\xi)$}
    \caption{\label{fg:Complex-strong-density_Nf=4_1}
      The $N_f=4$ spectral density of complex Dirac eigenvalues 
      at maximal non-Hermiticity $(\mu = 1)$ 
      for $\hm_1=\hm_2=\hm_3=\hm_4=3i$ 
      at $\nu=0$ (left) and $\nu=6$ (right). 
    }
\end{figure}
\begin{figure}[t]
    \unitlength1.0cm
    \centering
      \includegraphics[clip=,width=6cm]{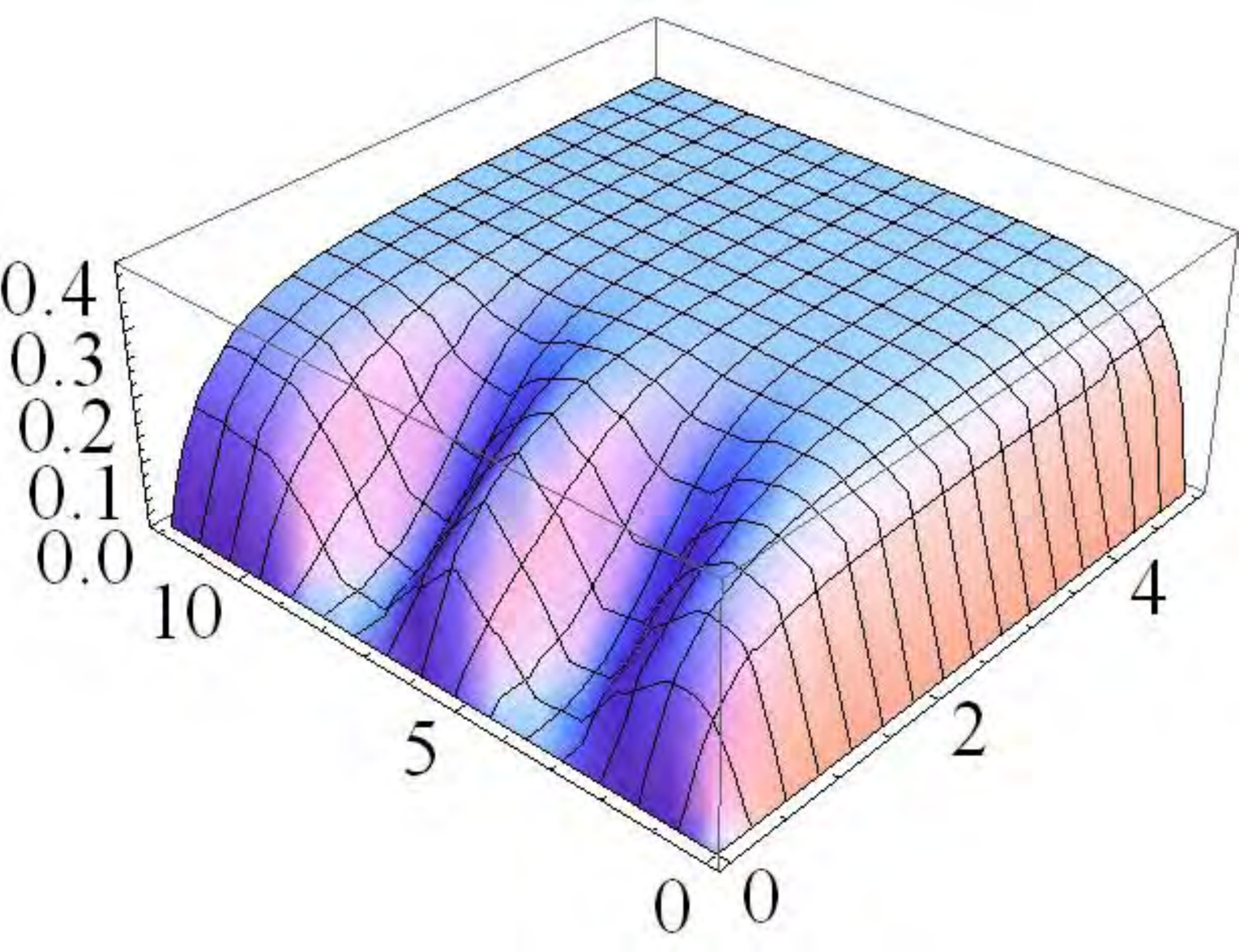} \qquad \quad  
      \includegraphics[clip=,width=6cm]{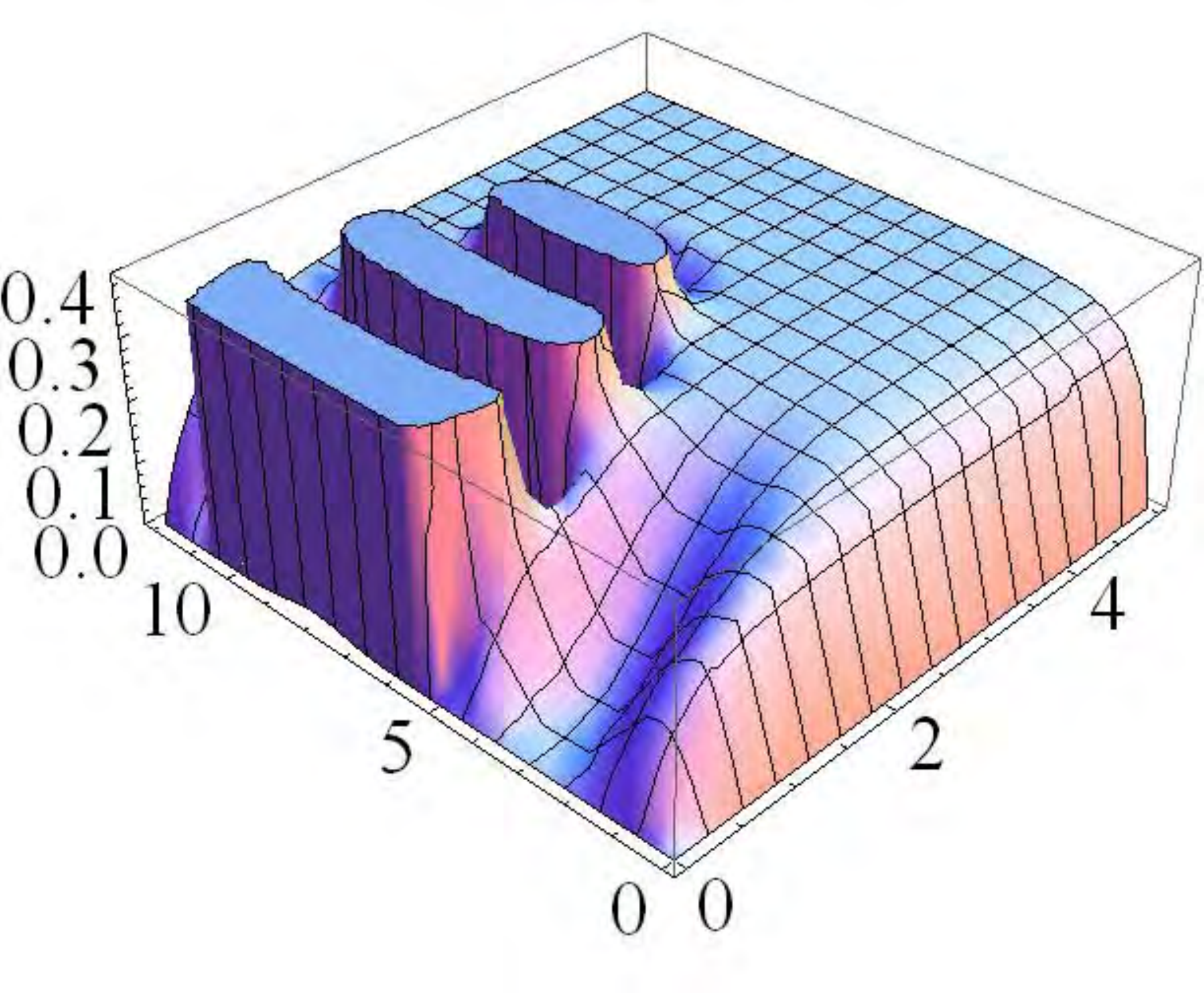}
      \put(-1.0,0.6){$\re[\xi]$}
      \put(-5.4,0.5){$\im[\xi]$}
      \put(-8.3,0.7){$\re[\xi]$}
      \put(-12.5,0.5){$\im[\xi]$}
      \put(-6.4,4.3){$\rho_{s}^{(N_f=4,\mathbb{C})}(\xi)$}
      \put(-13.8,4.3){$\rho_{s}^{(N_f=4,\mathbb{C})}(\xi)$}
    \caption{\label{fg:Complex-strong-density_Nf=4_2}
      Same as figure~\ref{fg:Complex-strong-density_Nf=4_1} but 
      for $\hm_1=\hm_2=4i$, $\hm_3=\hm_4=8i$ (left) and 
      $\hm_1=\hm_2=3i$, $\hm_3=5i$, $\hm_4=11i$ (right),
      both at $\nu=0$. Note that the scales for $\re[\xi]$ and $\im[\xi]$ are different.
      In the right figure the negative part of $\rho_s$ is cut for better readability.
    }
\end{figure}
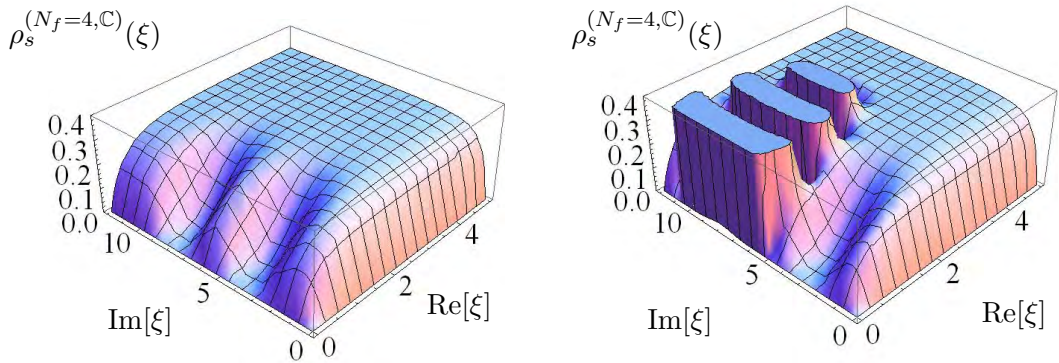

Plots of $\rho_{s}^{(N_f=4,\mathbb{C})}(\xi;\{\hm\}_4)$ are shown 
in figures~\ref{fg:Complex-strong-density_Nf=4_1} and
\ref{fg:Complex-strong-density_Nf=4_2}.  
It is seen that $N_f=4$ flavours roughly amount to two pairs of $N_f=2$ flavours, 
each of which shows either a dip or a domain of strong oscillations.

\section{The sign problem}
\label{sc:sign-problem}

\subsection{Analysis of oscillations of the Dirac spectrum}
\label{phase}

If we choose non-identical masses the spectral densities show interesting behaviour.  In this
section, we will consider in particular the case of $N_f=2$ with $\hm_1 \ne \hm_2$, and
the case of $N_f=1$ in the weak non-Hermiticity limit.  As
can be seen from a direct inspection of the expression \eqref{ZPQ} for the partition
function, the presence of non-degenerate masses can lead to ``probabilities''
which, although still real-valued, may
be negative in certain regions.  This is perhaps not unexpected.  However,
what may be surprising is that there are regions of the complex plane in which the
eigenvalue density function seems to oscillate from positive to negative
values quite dramatically. We present here a straightforward explanation. 

For simplicity, we first consider the strong limit which was discussed in
section~\ref{subsec_high_density}.  In the quenched case, the complex Dirac
eigenvalue density is essentially constant everywhere apart from close to
the real and imaginary axes, where it vanishes. As we discussed earlier in
eq.~(\ref{correction}), the
unquenched density can be written as a correction to the quenched
case.  From
eq.~(\ref{rhoCstrongNf2}), the correction in this case is 
\be
\Delta\rho_s^{(N_f=2,\mathbb{C})}(\xi;\hm_1,\hm_2) 
 = \frac{4 |\xi|^2 \hg_s(\xi^{*2},\xi^2)}{\mcK_s(\hm_1^2,\hm_2^2)}\,(-2i) \, 
  \im\Big\{ \mcK_s(\xi^2,\hm_1^2)\mcK_s(\xi^{*2},  \hm_2^2) \Big\}\,,
 \label{Deltadef}
\ee
where the weight function $\hg_s(z^*,z)$ was given in eq.~(\ref{ghats}). We next insert the explicit form of the strong kernel from eq.~(\ref{strongK}), and, after noting that a number of factors cancel, we get
\begin{multline}
  \Delta\rho_s^{(N_f=2,\mathbb{C})}(\xi;\hm_1,\hm_2) = 4 |\xi|^2 \left[ \frac{\eta_+^3}{8\pi} \right]\, 
  \hg_s(\xi^{*2},\xi^2) \frac{\exp[-\eta_-(\xi^2+\xi^{*2})]}{(\hm_1^2-\hm_2^2)\,I_{\nu}(2 \eta_+ \hm_1 \hm_2)} \\
  \times (-2i) \, \im \Big\{ (\xi^2-\hm_1^2)(\xi^{*2}-\hm_2^2)I_{\nu}(2\eta_+ \xi \hm_1)I_{\nu}(2\eta_+ \xi^* \hm_2) \Big\}\,.
\end{multline}
Here, $\eta_+$ and $\eta_-$ are as defined in eq.~(\ref{eta_hat_def}). So far
everything is exact.  However, we now make two approximations. First, we
approximate $\hg_s(\xi^{*2},\xi^2)$ by using the following argument. For the
quenched case, away from the axes, the Dirac density is essentially flat,
i.e.,  it is equal to some constant which we could determine exactly, but here
we shall simply denote it by $\rho_S$. Therefore, from eq.~(\ref{rhoCstr}),
we have 
\be
  4 |\xi|^2 \, \hg_s(\xi^{*2},\xi^2) \mcK_s(\xi^{*2},\xi^2)  \approx  \rho_S\,.
\ee
On inserting the explicit form of the strong kernel, and rearranging, we have
\bea
  4 |\xi|^2 \left[ \frac{\eta_+^3}{8\pi} \right]\, \hg_s(\xi^{*2},\xi^2) 
  \exp[-\eta_-(\xi^2+\xi^{*2})]
  & 
  \approx & \frac{\rho_S}{(\xi^2-\xi^{*2})I_{\nu}(2\eta_+|\xi|^2)}\,,
\eea
and so, away from the axes,
\begin{align}
 \Delta\rho_s^{(N_f=2,\mathbb{C})}(\xi;\hm_1,\hm_2) \approx \frac{(-2i)\,\rho_S\,\im \big\{ (\xi^2-\hm_1^2)(\xi^{*2}-\hm_2^2)I_{\nu}(2\eta_+ \xi \hm_1)I_{\nu}(2\eta_+\xi^* \hm_2) \big\}}
  {(\hm_1^2-\hm_2^2)\,I_{\nu}(2\eta_+ \hm_1 \hm_2)(\xi^2-\xi^{*2})\,I_{\nu}(2\eta_+ |\xi|^2)}\,
.
\end{align}
Our second approximation is to replace each modified Bessel function with its large-argument asymptotic limit, namely
\be
  I_{\nu}(z) \sim \frac{\e^z}{\sqrt{2\pi z}}
\ee
for $-\tfrac{\pi}{2} < \arg(z) < \tfrac{\pi}{2}$, i.e., in the right half-plane. After simplification, this gives (in the right half-plane, and sufficiently far away from the origin and from the axes)
\begin{align}
  \label{MA_delta_massive_final}
  &\Delta\rho_s^{(N_f=2,\mathbb{C})}(\xi;\hm_1,\hm_2) \notag\\ 
  & \quad \approx \frac{(-2i)\,\rho_S\,
\e^{-2\eta_+(\hm_1 \hm_2 + |\xi|^2)}}
  {(\hm_1^2-\hm_2^2)(\xi^2-\xi^{*2})} \, \im \Big\{ (\xi^2-\hm_1^2)(\xi^{*2}-\hm_2^2)\,\e^{2\eta_+(\xi\hm_1 + \xi^* \hm_2)} \Big\} \nn 
  \\
  & \quad = \frac{(-2i)\,\rho_S}{(\hm_1^2-\hm_2^2)(\xi^2-\xi^{*2})}\,\exp\Big[ 2\eta_+ \big( (\hm_1+\hm_2)x - |\xi|^2 - \hm_1 \hm_2 \big) \Big]  \notag  \\
  & \qquad \times \im \Big\{ (\xi^2-\hm_1^2)(\xi^{*2}-\hm_2^2)\Big( \cos(2\eta_+ (\hm_1-\hm_2)\hat{y}) 
  + i\sin(2\eta_+ (\hm_1-\hm_2)\hat{y}) \Big) \Big\}\,,
\end{align}
where we have written $\hat{x} \equiv \re\xi$ and $\hat{y} \equiv \im\xi$. We can regard this function as being the product of an envelope (the exponential), some polynomials in $\xi$ and $\xi^*$, and an oscillatory function of $\hat{y}$.  Let us consider each in turn.

To understand the behaviour of the exponential part, we can determine the contours in the $\xi$-plane where this part takes constant values. We can equivalently perform this analysis on the exponent. So we look for the solution of
\be
  (\hm_1+\hm_2)\hat{x} - |\xi|^2 - \hm_1 \hm_2  =  k\,,
\ee
where $k$ is some constant. Writing $|\xi|^2 = \hat{x}^2 + \hat{y}^2$, and rearranging, we have
\be
  \left( \hat{x} - \frac{\hm_1+\hm_2}{2} \right)^2 + \hat{y}^2 
  = \left( \frac{\hm_1 - \hm_2}{2} \right)^2 - k \equiv r^2\,.
\ee
This is, of course, the equation of a circle of radius $r$, centred on the point $\xi = (\hm_1 + \hm_2)/2$. The exponential function equals unity when $k=0$, i.e., when the circle passes through the points $\xi=\hm_1$ and $\xi=\hm_2$.  Crudely speaking, this marks the boundary of the region where the correction is ``significant'' when compared with the quenched case (where the density equals $\rho_S$).  Points inside this circle have $k>0$, so the circle marks the boundary of a peak (rather than a dip) in the complex plane. The maximal value of this peak is $\exp(2\eta_+ k_{\textup{max}})$, where
\be
  k_{\textup{max}} = \left( \frac{\hm_1 - \hm_2}{2} \right)^2,
\ee
so the peak grows very rapidly indeed with the difference between $\hm_1$ and $\hm_2$. Of course, our approximations are not valid at the centre of the circle since this lies on the $\hat{x}$-axis, and so this argument is merely indicative of the orders of magnitude that we might expect to observe in the density itself.

The polynomials $(\xi^2-\hm_1^2)(\xi^{*2}-\hm_2^2)$ in eq.~(\ref{MA_delta_massive_final}) result in a repulsion of eigenvalues from the points $\pm \hm_1$ and $\pm \hm_2$, similar to what we observed in the case of identical masses.

Finally, if we combine the oscillatory part of eq.~(\ref{MA_delta_massive_final}) with the $(\xi^2-\xi^{*2})^{-1}$ factor, then we get an oscillating function of $\hat{y}$ which is symmetric (even), and which has a constant ``wavelength''
\be
	\label{eq:lambda_oscillations}
  \lambda = \frac{\pi}{\eta_+\,(\hm_1-\hm_2)}\,.
\ee
In other words, this constitutes a set of ridges and troughs parallel to the real axis.

Therefore, what we observe in the total density is a circular ``peak'' of fairly dramatic oscillations, but these are almost completely suppressed elsewhere in the complex plane.  There is also some repulsion from the location of the masses, although the exponential effect causing the peak is generally more significant.  The number of oscillations visible will therefore be given by
\be
  \label{MA_num_oscillations}
  N_{\textup{oscillations}} \approx \frac{\eta_+\,(\hm_1-\hm_2)^2}{\pi}\,,
\ee
and we can say that no oscillations will be seen if, roughly speaking,
\be
  |\hm_1 - \hm_2| < \sqrt{\frac{\pi}{\eta_+}}\,.
  \label{eq:onset_of_oscillation}
\ee
These same conclusions can also be obtained directly from the group
integral eq.~(\ref{Zstrong}). On making an ansatz for the matrices $U,V\in \U(4)$ 
and keeping only the degrees of freedom that are expected to be relevant, 
a mean-field analysis reveals the same equations for the boundary and height of the
oscillations. 

For the weak non-Hermiticity limit, a similar analysis can be performed, although the situation here is slightly more complicated because we need to use not one, but two approximations for the kernel, depending on the average of the arguments.

When we have two masses, both located inside the strip that contains
the quenched eigenvalues (and both taken to be purely imaginary), then
we find that the situation is similar to the strong case above; there
is a circular region of oscillations, with the boundary of the
oscillating region passing through the two masses.  Indeed, one can
map the weak non-Hermiticity case onto the case of maximal non-Hermiticity (with $\mu=1$)  by making a simple scaling of masses and eigenvalues by $1/2\hat{\mu}$, and then the earlier results eqs.~(\ref{eq:lambda_oscillations}), (\ref{MA_num_oscillations}) and (\ref{eq:onset_of_oscillation}) for the strong case also hold true here.

If we have a single mass located within the strip at $\hat{m}=i\eta$, then we
get an elliptical region of oscillations, given in fact by the same equation
as for the corresponding situation in three-colour QCD
\cite[eq.~(41)]{Osborn:2008ab} where it was derived from an ansatz,
\be
  \frac{x^2}{\Big( \frac{2}{\sqrt{3}} (2\hat{\mu}^2-\eta) \Big)^2} 
  + \frac{\Big( y - \frac{4\hat{\mu}^2+\eta}{3} \Big)^2}
  {\Big( \frac{2}{3}(2\hat{\mu}^2-\eta) \Big)^2} = 1\,.
  \label{eq:envelope_weak_limit}
\ee
A comparison of this formula with the exact spectral density is given in figure~\ref{fg:Complex-weak-density_Nf=1___No-3}.  
Conversely, a single mass located outside the strip has virtually no effect whatsoever on the density.

The $N_f=2$ case with one of the masses considerably far outside the strip (in fact, beyond a (finite) critical point for which we can solve) is both qualitatively and quantitatively very similar to the $N_f=1$ case; we refer to Appendix \ref{heavy} for further details on the decoupling as one of the masses is taken progressively larger.  However, there is a third distinct intermediate regime where one mass is inside the strip, and the other is outside the strip, but below the critical point. Here, we find that the region of oscillations lies part-way between a circle and an ellipse.

It would be interesting to study the effects of the oscillations
analysed in this section on the chiral condensate, along the lines of
\cite{Osborn:2005ss,Osborn:2008jp}.  This is left for future work.

\subsection{Average sign factor}   \label{sc:Avsign}
\subsubsection{Introductory remarks}
In this section we study the severity of the sign problem by looking at
the average sign of the Dirac determinant. The definition of such an average is not unique, and
we will use the following guiding principles to define an average that behaves
reasonably:
\begin{itemize}
  \item The average should be bounded by 1 from above.
  \item Starting at 1 for $\mu=0$ (or degenerate masses) 
  the average should be a decreasing function of
  $\mu$ (or the mass difference).
\end{itemize}

Let us illustrate the problem of finding a proper definition using the
case of three-colour QCD (QCD3), which 
has been well studied in the RMT framework \cite{Splittorff:2006fu,
Splittorff:2007ck,Splittorff:2006vj}.  
If we denote the complex phase of $\det[{\cal D}+m]$ by
$\e^{i\theta}$, the authors proposed
to compute
\be
\big\langle \e^{2i\theta} \big\rangle_{N_f}^{\rm QCD3} \ \equiv\ 
  \left\langle\frac{\det[{\cal D}(\mu)+m]}{\det[{\cal D}(\mu)+m]^\dag}
  \right\rangle_{N_f}^{\rm QCD3}\,,
  \label{QCD3ave}
\ee
which gives the expectation value of twice the phase in the case that the
determinant is complex. Had we chosen the opposite ratio on the
right-hand side
instead, this would naively have led to $\langle \e^{-2i\theta}
\rangle_{N_f}$. However, for $N_f=1$ we have in three-colour QCD 
\be
  \left\langle \frac{\det[{\cal D}+m]^\dagger}{\det[{\cal D}+m]} 
  \right\rangle_{N_f=1}  ^{\rm QCD3}
  = \frac{\langle \det[{\cal D}+m]^\dagger\rangle_{N_f=0}^{\rm QCD3}}{\langle
  \det[{\cal D}+m] \rangle_{N_f=0}^{\rm QCD3}} = 1 
  \label{char}
\ee
because the expectation value in denominator and numerator (also called
characteristic polynomial) is equal and $\mu$-independent
\cite{Akemann:2002vy}. Obviously this would contradict the second of our
criteria above. Note that in contrast 
using the definition (\ref{QCD3ave}) the ratio
is not equal to 1, even for $N_f=1$.

Our definition of the average sign for two-colour QCD, which we will use
throughout this section, is
\begin{align}
  p^{(N_f)}(\mu;\{m\})\equiv\Big\langle\sgn\prod_{f=1}^{N_f}
  \det[D(\mu)+m_f]\Big\rangle_{||N_f||}\,,
\end{align}
where the average is computed in a sign-quenched theory denoted by
$\|N_f\|$ (i.e., only the absolute values of the determinants appear
in the measure).  We recall that in the strong limit we always need
$N_f$ to be even.  In the next two subsections we will consider
two particular cases, the weak limit with one flavour, where
\be
  p_w^{(N_f=1)}(\hmu;\hm)=\lim_{N\to\infty,\text{weak}}
  p^{(N_f=1)}(\mu;m)\,,
\label{pwdef}
\ee
and the strong limit with two flavours, where
\be
  p_s^{(N_f=2)}(\mu;\hm_1,\hm_2) = \lim_{N\to\infty,\text{strong}}
  p^{(N_f=2)}(\mu;m_1,m_2)\,.
  \label{psdef}
\ee
We shall see explicitly from these two cases that our
guiding principles are satisfied.

We finish this subsection by commenting on possible bad choices for the
average sign. Due to the obvious inequality
\be 
  \big\langle\left|\det[{\cal D}+m]\right|\big\rangle^{}_{N_f=0}\geq 
  \big\langle\det[{\cal D}+m]\big\rangle^{}_{N_f=0}
\ee
we have
\begin{align}
  \left\langle \frac{\det[{\cal D}+m]}{\left|\det[{\cal D}+m]\right|}
  \right\rangle_{N_f=1}=
  \left\langle \frac{\left|\det[{\cal D}+m]\right|}{\det[{\cal D}+m]}
  \right\rangle_{N_f=1}\geq 1\,,
\end{align}
and hence a ratio of this kind, which would have been the natural
generalisation of eq.~\eqref{QCD3ave}, can be ruled out.
A second choice that leads to an inconsistency is the following. Because the sign
problem can be turned on by detuning the masses of two originally degenerate flavours one
might consider
\be
  \left\langle\frac{\det[{\cal D}(\mu)+m_1]\det[{\cal D}(\mu)+m_2]}{
  \det[{\cal D}(\mu)+m_0]^2}
  \right\rangle_{N_f}\,,
  \label{pdef}
\ee
where in the denominator we have two degenerate masses $m_0$. 
One caveat is that the choice of $m_0$ in the denominator is 
ambiguous.  For small $m_1-m_2$ one might think that any of the choices
$m_0=m_1,m_2,(m_1+m_2)/2,\sqrt{m_1m_2}$ would be acceptable. 
However, using the latter choice  we can show that in the strong limit
this leads to a quantity that is always unity. 
Moreover, in the weak limit or for other choices of $m_0$ the quantity
in eq.~(\ref{pdef})
goes to zero more rapidly for larger topology. This contradicts our
expectation, which was confirmed in RMT studies of
three-colour QCD \cite{Bloch:2008cf},  that a larger number of exactly zero eigenvalues should push the density away from
the origin (where the oscillations are strong) and thus mitigate the average
sign problem.

\subsubsection{Average sign at low density}
\label{sec:sign_low}

Let us consider the case $N_f=1$ in the limit of weak non-Hermiticity. 
Although the sign-quenched partition function does not permit a field-theoretical interpretation 
like QCD with nonzero isospin chemical potential, it is still possible to compute it as a 
mathematical entity by RMT. In appendix \ref{app_weak} we derive
\begin{align}
  {\cal Z}_{w}^{(N_f=\|1\|,\,\nu)}(\hmu;\hm) & \equiv 
  \lim_{N\to\infty,\,\text{weak}}\big\langle \big|\det[{\cal D}(\mu)+m]\big| \big\rangle_{N_f=0}
  \notag\\
  & \sim -\hmu~\e^{\hmu^2/2} 
  \frac{ G_w(\hm^2,\hm^2) }{\hh_w(\hm^2)}\,.
  \label{eq:Z_w_|1|}
\end{align}
This result essentially follows from the quenched real spectral density. 
Combining it with eq.~(\ref{ZNf1}),
\be
  {\cal Z}_{w}^{(N_f=1,\,\nu)}(\hmu;\hm) \sim q_w(\hm^2)\,,
\nn
\ee
we obtain the average sign factor in the weak limit,
\begin{align}
  p_w^{(N_f=1)}(\hmu;\hm)
  = \frac{{\cal Z}_{w}^{(N_f=1,\,\nu)}(\hmu;\hm)}{{\cal Z}_{w}^{(N_f=\|1\|,\,\nu)}(\hmu;\hm)}
 \sim - \frac{\hh_w(\hm^2)q_w(\hm^2)}{\hmu\ \e^{\hmu^2/2}G_w(\hm^2,\hm^2)}\,.
\end{align}
The normalisation of $p_w^{(N_f=1)}(\hmu;\hm)$ can be uniquely
determined from the requirement $p_w^{(N_f=1)}(0,\hm)=1$.

\begin{figure}[t]
    \unitlength1.0cm
    \centering\!\!\!\!\!
      \includegraphics[clip=,width=6cm]{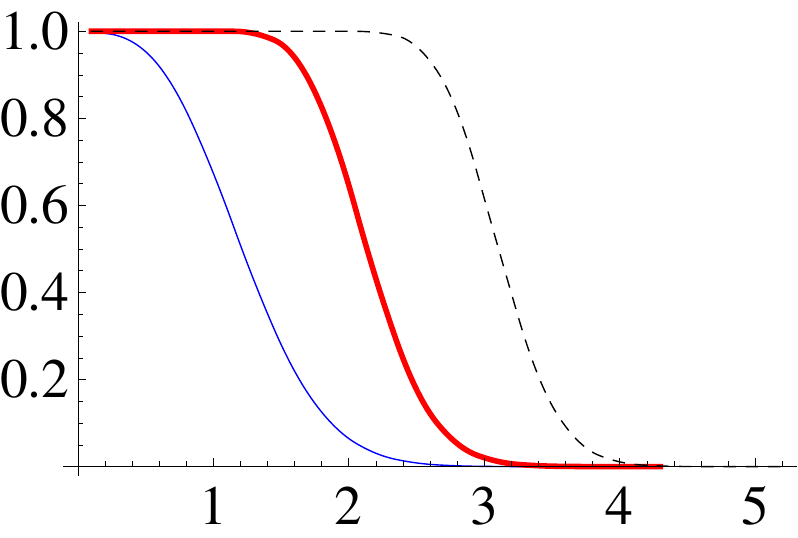} \qquad \quad 
      \includegraphics[clip=,width=6cm]{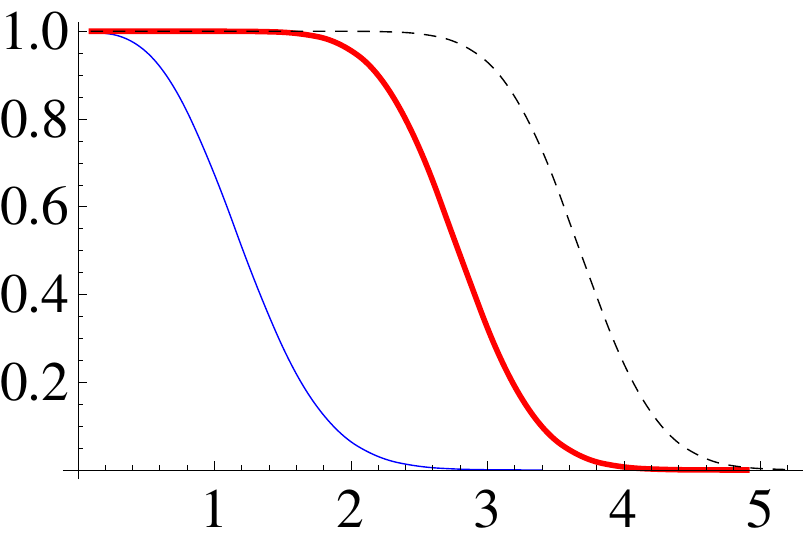}
      \put(0.3,0.4){$\hmu$}
      \put(-7.0,0.4){$\hmu$}
      \put(-6.4,4.4){$p_{w}^{(N_f=1)}(\hmu;\hm)$}
      \put(-13.7,4.4){$p_{w}^{(N_f=1)}(\hmu;\hm)$}
    \caption{\label{fg:sign_weak_Nf=1}
      Average sign for $N_f=1$ at weak non-Hermiticity. 
      Left: $\hm=0$ (blue line), $\hm=6i$ (red thick line) and $\hm=15i$ (black dashed line), all for $\nu=0$. 
      Right: $\nu=0$ (blue line), $\nu=10$ (red thick line) and $\nu=20$ (black dashed line), all for $\hm=0$.
    }
\end{figure}

\begin{figure}[t]
    \unitlength1.0cm
    \centering\!\!\!\!\!
      \includegraphics[clip=,width=8cm]{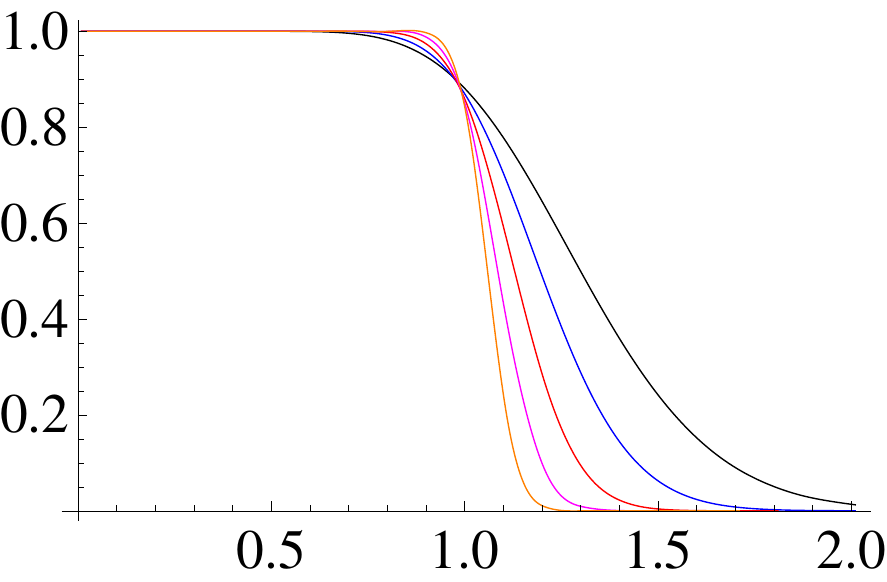}
      \put(0.3,0.4){$\displaystyle \frac{\hmu}{\sqrt{\hm/2}}$}
      \put(-10.7,4.8){$p_{w}^{(N_f=1)}(\hmu;\hm)$}
    \vspace{-.5\baselineskip}
    \caption{\label{fg:sign_weak_Nf=1_scaling}
      Average sign for $N_f=1$ and $\nu=0$ at weak non-Hermiticity. 
      The curves correspond to $\hm=4i$ (black), $\hm=8i$ (blue), $\hm=16i$ (red), $\hm=32i$ (magenta), and $\hm=64i$ (orange). 
    }
\end{figure}

In figure~\ref{fg:sign_weak_Nf=1} we depict how $p_{w}^{(N_f=1)}(\hmu;\hm)$ depends on $\hmu$ 
for several values of $\hm$ and $\nu$. We observe that the onset of the sign problem 
is delayed as $\hm$ or $\nu$ is increased, a feature which is in common with 
three-colour QCD \cite{Splittorff:2006fu,Bloch:2008cf}.
In figure~\ref{fg:sign_weak_Nf=1_scaling} we plot the average sign against 
a rescaled variable $\hmu/\sqrt{\hm/2}$. In physical units, 
it is equal to $\mu/(m_{\pi}/2)$, where $m_{\pi}$ denotes the 
mass of the Nambu-Goldstone (NG)  
bosons in the vacuum.\footnote{Note that no NG
mode appears in $N_f=1$ two-colour QCD. 
The above $m_{\pi}$ refers to the mass of the NG modes that appear in $N_f\geq 2$ two-colour QCD.} 
Interestingly, the sign problem looks almost absent for $\mu<m_{\pi}^{}/2$
while it deteriorates rapidly  
for $\mu>m_{\pi}^{}/2$. In particular, we observe a convergence of the
curves to a step function in  
the thermodynamic limit. This apparent jump of the average sign is quite intriguing, 
compared to the average phase factor in the microscopic limit of three-colour
QCD \cite{Splittorff:2006fu,Splittorff:2007ck}:  
the latter changes smoothly from $1$ to $0$ in the limit $\hm\to\infty$. 

Since the $\hmu$-dependence of ${\cal Z}_{w}^{(N_f=1,\,\nu)}$ is
explicit in eq.~\eqref{ZNf1} (being of oscillatory nature, 
see eq.~(\ref{weakQ})), it must be ${\cal
  Z}_{w}^{(N_f=\|1\|,\,\nu)}$ 
that is responsible for the apparent jump of the average sign. 
While it seems that the $N_f=\|1\|$ theory is similar to the $N_f\geq 2$ theory in that 
both undergo a transition at $\mu=m_{\pi}/2$, 
see, e.g., \cite{Kogut:2000ek}, a comparison of the order of the
transition needs further investigation.

Finally we comment on a lattice simulation using one flavour of staggered
fermions in the adjoint representation of $\SU(2)$
\cite{Hands2000,Hands:2001ee}. This theory belongs to the same symmetry class as
$N_f=1$ two-colour QCD with fundamental fermions.  In \cite{Hands2000,Hands:2001ee}
this theory was simulated by two different algorithms, 
Hybrid-Monte-Carlo (HMC) and Two-Step-Multi-Boson (TSMB).
The HMC algorithm is non-ergodic in this case and actually simulates a
different theory, namely the one-flavour sign-quenched theory.  The
authors of \cite{Hands2000,Hands:2001ee} report that the sign-quenched theory
simulated by HMC seems to undergo a phase transition at
$\mu=m_{\pi}/2$, while the correct one-flavour theory simulated by TSMB
exhibits no singularity at all. These results are consistent with
ours. However, the observables considered here and in \cite{Hands2000,Hands:2001ee}
are different so that a direct comparison is difficult.

\subsubsection{Average sign at high density}

Let us now consider the case $N_f=2$ in the limit of strong non-Hermiticity. 
This theory is free from the sign problem if and only if the masses are degenerate. 
Here we are interested in the severity of the sign problem for $|\hm_1-\hm_2|$ nonzero.
In appendix \ref{app_strong} we derive
\begin{align}
  {\cal Z}^{(N_f=\|2\|,\,\nu)}_{s}(\mu;\hm_1,\hm_2) & \equiv 
  \lim_{N\to\infty,\,\text{strong}}\Big\langle 
  \big|\det[{\cal D}(\mu)+m_1]\det[{\cal D}(\mu)+m_2]\big| \Big\rangle_{N_f=0}
  \notag\\
  & \sim 
  \frac{f_{s}(\hm_1^2,\hm_2^2)}{\ab{\hm_1^2-\hm_2^2}}\,,
  \label{eq:Z_s_|2|}
\end{align}
where\footnote{In eq.~(\ref{eq:fs}) the squared masses are assumed to be nonpositive since
this is the case relevant for two-colour QCD.  If one or both of the
arguments of $f_s$ are positive we obtain a more complicated
expression for $f_s(x_1,x_2)=\lim_{N\to\infty}
(x_1x_2)^{\nu/2}f_N(x_1,x_2)$.}
\begin{multline}
  f_{s}(\hm_1^2,\hm^2_2) = {\cal K}_{s}(\hm_1^2,\hm^2_2)\sgn(\hm^2_1-\hm^2_2)
  \\
  +
  \int\limits_{\mathbb{R}}dx
  \int\limits_{\mathbb{R}}dx'~
  \hh_s(x)\hh_s(x')  \Big[
    \sgn(\hm_1^2-x)\sgn(\hm_2^2 -x')-\sgn(\hm_2^2 -x)\sgn(\hm_1^2 -x')
  \Big]
  \\
  \times  \Big[
    {\cal K}_s(\hm_1^2,x){\cal K}_s(\hm^2_2,x')-\frac{1}{2}{\cal K}_s(\hm^2_1,\hm^2_2){\cal K}_s(x,x')
  \Big]
  \label{eq:fs}
\end{multline}
with the kernel in the strong limit from eq.~(\ref{strongK}),
\be
  {\cal K}_s(u,v) \equiv  \frac{\eta_+^3}{8\pi}
  (u - v)\,\e^{-\eta_-(u+v)} 
  I_{\nu}\left( 2 \eta_+\sqrt{uv} \right)\,.
\ee
These results are obtained from the two-point correlation function of real eigenvalues in the strong limit. 
Combining eq.~\eqref{eq:Z_s_|2|} with eq.~(\ref{ZNf2}),
\be
  {\cal Z}_{s}^{(N_f=2,\,\nu)}(\mu;\hm_1,\hm_2) \sim \frac{{\cal K}_s(\hm_1^2,\hm_2^2)}{\hm_1^2-\hm_2^2}\,,
\nn
\ee
we obtain
\be
  p_s^{(N_f=2)}(\mu;\hm_1,\hm_2) 
  = \frac{{\cal Z}_{s}^{(N_f=2,\,\nu)}(\mu;\hm_1,\hm_2)}{{\cal Z}_{s}^{(N_f=\|2\|,\,\nu)}(\mu;\hm_1,\hm_2)}
  = \sgn (\hm_1^2-\hm_2^2) \frac{{\cal K}_s(\hm_1^2,\hm_2^2)}{f_s(\hm_1^2,\hm_2^2)}\,.
\ee
This expression satisfies the normalisation condition $\displaystyle \lim_{\hm_2\to\hm_1}p_s^{(N_f=2)}(\mu;\hm_1,\hm_2)=1$. 
By explicit calculation a scaling law is verified,
\be
  p_s^{(N_f=2)}(\mu;\hm_1,\hm_2) = p_s^{(N_f=2)}(1;\sqrt{2\eta_+}\hm_1,\sqrt{2\eta_+}\hm_2)\,,
\ee 
hence we can assume $\mu=1$ without loss of generality.

\begin{figure}[t]
    \unitlength1.0cm
    \centering
      \includegraphics[clip=,width=6cm]{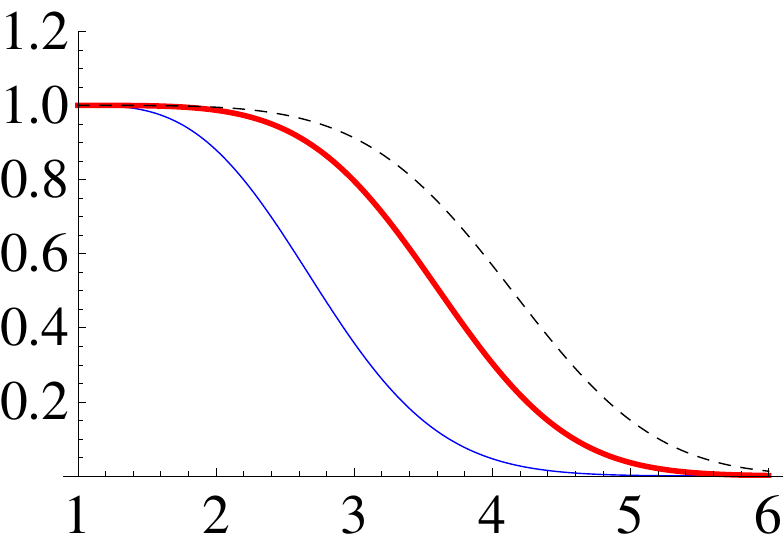} \qquad \quad 
      \includegraphics[clip=,width=6cm]{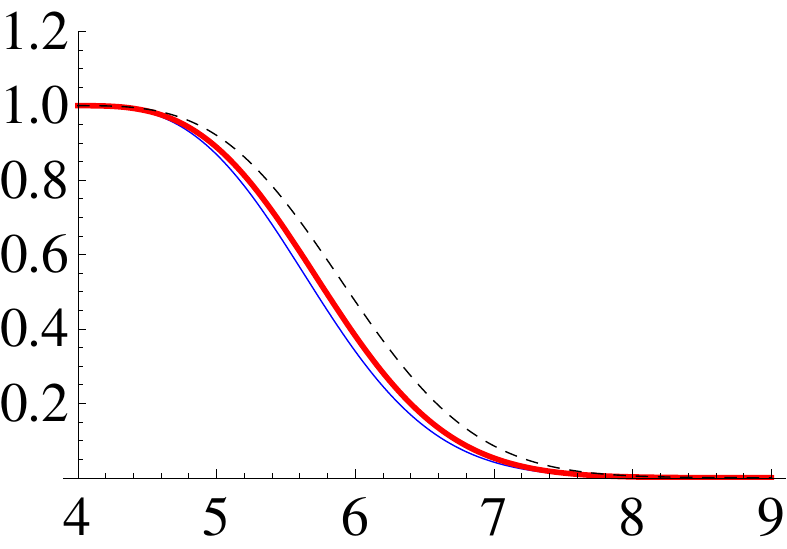}
      \put(0.3,0.4){$|\hm_2|$}
      \put(-7.0,0.4){$|\hm_2|$}
      \put(-6.4,4.4){$p_{s}^{(N_f=2)}(1;\hm_1=4i,\hm_2)$}
      \put(-14.0,4.4){$p_{s}^{(N_f=2)}(1;\hm_1=i,\hm_2)$}
    \caption{\label{fg:sign_strong_Nf=2}
      Average sign factor for $N_f=2$ at maximal non-Hermiticity 
      ($\hmu=1$) with $\hm_1=i$ (left) and $\hm_1=4i$ (right), 
      at $\nu=0$ (blue line), $\nu=10$ (red thick line),
      and $\nu=20$ (black dashed line).
    }
\end{figure}

In figure~\ref{fg:sign_strong_Nf=2} we plot $p_s^{(N_f=2)}(1;\hm_1,\hm_2)$ 
as a function of $\hm_2$. The results indicate that larger topology mitigates the sign 
problem but becomes less effective for heavier masses. 
This phenomenon is easily understood if one recalls that a large $\nu$ 
that causes a depletion of eigenvalues near the origin can smoothen the strongly 
oscillating domain of the spectrum if the domain is close 
to the origin; since the domain moves away from the origin for heavier masses, 
the smoothing by $\nu$ becomes less effective.

As a result, the average sign turns out to be essentially independent of topology 
for larger masses,
where the sign problem becomes severe  
for $|\delta\hm|\gtrsim 2.5$, as can also be seen from  
figure~\ref{fg:sign_strong_Nf=2}.  In physical units we have, using
eq.~\eqref{strongscale}, 
\be
  \big|\delta m_{\text{phys}} \big| \gtrsim \frac{4.5}{\sqrt{V_4\Delta^2}}\,.
\ee
We remark that 
the oscillations of the complex spectral density start if 
$|\delta \hm|\gtrsim \sqrt{2\pi}\approx 2.5$ 
as seen from eq.~\eqref{eq:onset_of_oscillation} at $\hmu=1$. 
It is reasonable that these two values of $|\delta \hm|$ are close to each other.

\section{Conclusions}\label{cons}

In this paper we have solved a random two-matrix model as a mathematical model 
for unquenched two-colour QCD with non-vanishing chemical potential.
Our model has very interesting features. First of all it describes 
two very different physical situations, at low and at high density, in the
approximation of a static chiral Lagrangian. 
Despite the sign problem the unquenched random two-matrix model can be solved
exactly, and the Dirac operator has very distinct real, imaginary, and
complex eigenvalue densities.
The sign problem can be switched off
or on at fixed chemical potential by tuning, say, the masses of two quark flavours to
be degenerate or distinct. 

We have given many explicit examples for the various spectral densities for
$N_f=1$, $2$ and $4$ flavours and have shown how they behave quantitatively differently
when the sign problem is switched on. 

We have also introduced a measure for the severity of the sign problem and
have analysed the envelope, amplitude, number, and frequency of the
oscillations of the spectral density once it becomes negative.

It would be very interesting to compare the predictions we have made with ab
initio lattice computations, in both settings of low and high
density. There are some remarks to be made for this symmetry class, however. At
zero chemical potential two-colour QCD has a substantial probability to have
small or zero Dirac operator eigenvalues, which is in contrast to other
QCD (-like) theories. Therefore it may be very difficult to reach small enough
masses (that make the inversion of the Dirac determinant difficult) in order
to see the effect of unquenching in the low-lying spectrum. 

The chemical potential introduces an extra repulsion of the spectral density
from the origin, which might actually help in this situation. The sign
problem on the other hand could be kept under control if the non-degeneracy of
the quark masses is kept sufficiently small.

\acknowledgments 

We would like to thank  K. Splittorff for fruitful discussions 
and a critical reading of the manuscript, and J. Bloch for helping us in eliminating typos in some
equations.  
The following financial support is  acknowledged: GA thanks the
Niels Bohr Foundation as well as the Niels Bohr Institute and International
Academy where part of this work was done, 
TK is supported by the Japan Society
for the Promotion of Science for Young Scientists,  MJP by an EPSRC doctoral
training grant, and TW by DFG through SFB/TR-55.

\appendix

\section{The weakly non-Hermitian limit of the real eigenvalue density}\label{weakrealR}

\subsection{Initial steps}

In this appendix, we consider the large-$N$ limit at weak non-Hermiticity of
the real eigenvalue density for the quenched case, which was not analytically
determined in section 4.4 of \cite{Akemann:2009fc}. The real density of Wishart
eigenvalues for finite matrix of ``size'' $N$ (where $N$ is even) and
topology $\nu$ is given by (see section 4.2 of \cite{Akemann:2009fc}) 
\be
  \label{WL_R_finite_N}
  R_{1,N}^{\mathbb{R}}(x) = - G_N(x,x;\mu)\,,
\ee
where
\begin{align}
  \label{GN_definition}
  G_N(x,x';\mu) 
  & \equiv - \int_{-\infty}^{\infty} dy \, \sgn(x'-y)\,h(x')h(y){\cal K}_{N}(x,y)
  \\
  & = - \int_{-\infty}^{\infty} dy \, \sgn(x'-y)\,
  |x'y|^{\nu/2}\,
  \e^{\eta_-(x'+y)}\,2K_{\frac{\nu}{2}}(\eta_+|x'|)\,2K_{\frac{\nu}{2}}(\eta_+|y|)
  {\cal K}_{N}(x,y) \notag
\end{align}
and the quenched kernel at finite $N$ under the integral is given by
\begin{multline}
  {\cal K}_{N}(x,y) \equiv \frac{\eta_-}{8\pi(4\mu^2\eta_+)^{\nu+1}} 
  \sum_{j=0}^{N-2} \left( \frac{\eta_-}{\eta_+}
  \right)^{2j} \frac{(j+1)!}{(j+\nu)!} 
  \\
  \times \left\{ L_{j+1}^{\nu}\left(
  \frac{y}{4\mu^2\eta_-} \right)L_{j}^{\nu}\left( \frac{x}{4\mu^2\eta_-} \right)
  - L_{j+1}^{\nu}\left(
  \frac{x}{4\mu^2\eta_-} \right)L_{j}^{\nu}\left( \frac{y}{4\mu^2\eta_-} \right)
  \right\} .
  \label{kernel2N}
\end{multline}
For completeness we also give here the even skew-orthogonal polynomials since they appear
in their limiting form in eq.~(\ref{weakQ}),
\be
  q_{2k}(z)\equiv (4\mu^2\eta_{-})^{2k} (2k)!\ L_{2k}^\nu\Big( \frac{z}{4\mu^2\eta_-}\Big)\,.
  \label{q2N}
\ee
We now define the weakly non-Hermitian large-$N$ limit by
\bea
  \label{WL_R_weak_limit}
  G_w (x,x';\hat{\mu}) & \equiv & \lim_{N \rightarrow \infty} \frac{1}{4N} \, \frac{x^{\nu/2}}{(x')^{\nu/2}} \, G_N \left(\frac{x}{4N},\frac{x'}{4N};\frac{\hat{\mu}}{\sqrt{2N}} \right).
\eea
The problem is that the limit in eq.~(\ref{WL_R_weak_limit}) and the integral in eq.~(\ref{GN_definition}) do not commute. Using the fact that
\be
  \int_{-\infty}^{\infty} dy\,\sgn(x'-y)\,f(y) = \left( \int_{-\infty}^0 dy + 2 \int_0^{x'} dy - \int_0^{\infty} dy \right) f(y)
\ee
we can correspondingly write $G_w(x,x';\hat{\mu})$ as the sum of three parts,
\be
  \label{WL_R1_W_split}
  G_w(x,x';\hat{\mu}) = - \big\{ A(x,x';\hat{\mu}) + 2 B(x,x';\hat{\mu}) - C(x,x';\hat{\mu}) \big\}\,.
\ee
For $A$ and $B$, it is possible to interchange the limit and the integral, and so we have simply
\bea
  A(x,x';\hat{\mu}) & = & (-i)^{\nu} \, \frac{\hat{h}_w(x')}{[\sgn(x')]^{\nu/2}} \int_{-\infty}^0 dy \,  
  \mcK_w(x,y) \hat{h}_w(y)\,, 
  \label{WL_A_formula_convergent} 
  \\
  B(x,x';\hat{\mu}) & = & \frac{\hat{h}_w(x')}{[\sgn(x')]^{\nu}} \, \int_{0}^{x'} dy \,  
  \mcK_w(x,y) \hat{h}_w(y)\,, 
  \label{WL_B_formula_convergent}
\eea
where the scaled weight function and the weak kernel are given by eqs.~(\ref{hhdef}) and (\ref{weakK}), respectively,
\bea
  \hat{h}_w(x) &=& \e^{x/8\hat{\mu}^2}\,2K_{\frac{\nu}{2}}\left( \frac{|x|}{8\hat{\mu}^2} \right), 
  \nn 
  \\ 
  \mcK_w(x,y) &=& \frac{1}{256\pi\hat{\mu}^2} \int_0^1 ds\,s^2
  \e^{-2\hat{\mu}^2 s^2}
  \,\left\{ \sqrt{x}J_{\nu+1}(s\sqrt{x})J_{\nu}(s\sqrt{y}) 
  - \sqrt{y}J_{\nu+1}(s\sqrt{y})J_{\nu}(s\sqrt{x}) \right\}. 
  \nn
\eea
The main problem is then reduced to determining $C(x,x';\hat{\mu})$, where the non-commutativity issue remains.

Let us next perform some fairly straightforward manipulation of $C(x,x';\hat{\mu})$. We can immediately take $N \rightarrow \infty$ for all the $N$-dependent factors that lie outside both the integral and the sum. Therefore
\be
  \label{WL_C_in_terms_of_D}
  C(x,x';\hat{\mu}) = \frac{1}{32\pi\hat{\mu}^2}\, \frac{\hat{h}_w(x')}{[\sgn(x')]^{\nu/2}} \, x^{\nu/2} \lim_{N \rightarrow \infty} D_N(x;\hat{\mu})\,,
\ee
where
\begin{align}
  &D_N(x;\hat{\mu}) \equiv \frac{1}{(4N)^{2+\nu}}\, \int_0^{\infty}dy\,
  \exp\left(\frac{(1-\frac{\hat{\mu}^2}{2N})y}{8\hat{\mu}^2}\right)K_{\frac{\nu}{2}} 
  \left(\frac{(1+\frac{\hat{\mu}^2}{2N})y}{8\hat{\mu}^2}\right)\,|y|^{\nu/2} 
  \\
  &\quad\times \sum_{j=0}^{N-2} \left( \frac{1-\frac{\hat{\mu}^2}{2N}}{1+\frac{\hat{\mu}^2}{2N}} \right)^{2j}
  \frac{j!}{(j+\nu)!} 
   \,\Bigg\{ 
x \, L_{j}^{\nu+1}\left( \frac{x}{4N(1-\frac{\hat{\mu}^2}{2N})} \right)L_{j}^{\nu}
  \left( \frac{y}{4N(1-\frac{\hat{\mu}^2}{2N})} \right) - (x
  \leftrightarrow y) \Bigg\}\,. \notag
\end{align}
Since $N$ is finite, we can switch the integral and the sum. We will also split the two terms in the curly brackets, writing $D_N(x;\hat{\mu})=D_N^+(x;\hat{\mu}) - D_N^-(x;\hat{\mu})$, where
\be
  \label{WL_D_N_plus}
  D_N^+(x;\hat{\mu}) \equiv \frac{1}{(4N)^{2+\nu}} \sum_{j=0}^{N-2} 
  \left( \frac{1-\frac{\hat{\mu}^2}{2N}}{1+\frac{\hat{\mu}^2}{2N}} \right)^{2j} 
  \frac{j!}{(j+\nu)!} \,x\,L_{j}^{\nu+1}\left( \frac{x}{4N(1-\frac{\hat{\mu}^2}{2N})} \right) P(j,N,\hat{\mu})
\ee
with
\be
  \label{WL_P_def}
  P(j,N,\hat{\mu}) \equiv \int_0^{\infty}dy\,\exp\left(\frac{(1-\frac{\hat{\mu}^2}{2N})y}{8\hat{\mu}^2}\right)
  K_{\frac{\nu}{2}}\left(\frac{(1+\frac{\hat{\mu}^2}{2N})y}{8\hat{\mu}^2}\right) L_{j}^{\nu}
  \left( \frac{y}{4N(1-\frac{\hat{\mu}^2}{2N})} \right)\,y^{\nu/2}
\ee
and
\be
  \label{WL_D_N_minus}
  D_N^-(x;\hat{\mu}) \equiv \frac{1}{(4N)^{2+\nu}}\, \sum_{j=0}^{N-2} 
  \left( \frac{1-\frac{\hat{\mu}^2}{2N}}{1+\frac{\hat{\mu}^2}{2N}} \right)^{2j} 
  \, \frac{j!}{(j+\nu)!}  \,L_{j}^{\nu}\left( \frac{x}{4N(1-\frac{\hat{\mu}^2}{2N})} \right) \, Q(j,N,\hat{\mu})
\ee
with
\be
  \label{WL_Q_def}
  Q(j,N,\hat{\mu}) \equiv \int_0^{\infty}dy\,\exp\left(\frac{(1-\frac{\hat{\mu}^2}{2N})y}{8\hat{\mu}^2}\right)
  K_{\frac{\nu}{2}}\left(\frac{(1+\frac{\hat{\mu}^2}{2N})y}{8\hat{\mu}^2}\right) \, L_{j}^{\nu+1}
  \left( \frac{y}{4N(1-\frac{\hat{\mu}^2}{2N})} \right)\,y^{\nu/2+1}\,.
\ee

\subsection[Solution for limit of $D_N(x;\hat{\mu})$]{\boldmath Solution for limit of $D_N(x;\hat{\mu})$}

Our plan is to try to evaluate the large-$N$ limits of eqs.~(\ref{WL_D_N_plus}) and (\ref{WL_D_N_minus}) in a manner similar to the way we calculated the weak kernel in section 4.4 of \cite{Akemann:2009fc}. This involves taking the limit for large $j$ and $N$, with $j=tN$ and $t \in [0,1]$, of each of the factors inside the sums in eqs.~(\ref{WL_D_N_plus}) and (\ref{WL_D_N_minus}), and replacing the sums  with integrals. So, in the present case, our primary task will be to determine appropriate large-$N$ limits of $P(j,N,\hat{\mu})$ and $Q(j,N,\hat{\mu})$.

It turns out that for $D_N^+(x;\hat{\mu})$, this method works in a relatively straightforward manner. However, for $D_N^-(x;\hat{\mu})$, we also find an extra contribution that does not involve an integral. This arises from a different large-$N$ scaling of $Q(j,N,\hat{\mu})$.

\subsubsection[Determination of limits of $P(j,N,\hat{\mu})$ and
$D_N^+(x;\hat{\mu})$]{\boldmath Determination of limits of $P(j,N,\hat{\mu})$ and $D_N^+(x;\hat{\mu})$}

Let us start with eq.~(\ref{WL_P_def}). First, we use \cite[eq.~(8.432.3)]{Gradshteyn:2007} to replace the modified Bessel function of the second kind with an integral, giving
\be
  P(j,N,\hat{\mu})  =  \frac{\sqrt{\pi}}{\Gamma\left( \frac{\nu+1}{2} \right)} 
  \, \left( \frac{ 1+\frac{\hat{\mu}^2}{2N}}{16\hat{\mu}^2} \right)^{\nu/2} \, \int_0^{\infty} dy
  \, \int_1^{\infty} du\,(u^2-1)^{(\nu-1)/2}\,\e^{-by}\,L_j^{\nu}\left( \frac{y}{c} \right)\, y^{\nu}
\ee
with
\be
  b \equiv \frac{1}{8}\left( \frac{u-1}{\hat{\mu}^2} + \frac{u+1}{2N} \right)\qquad\text{and}\qquad
  c \equiv 4N \left( 1 - \frac{\hat{\mu}^2}{2N} \right).
\ee
After writing the Laguerre function explicitly as a polynomial, we interchange the order of integrations and perform the integral over $y$. The resulting sum can be recombined into a binomial as follows,
\begin{align}
  &P(j,N,\hat{\mu}) =  \frac{\sqrt{\pi}}{\Gamma\left( \frac{\nu+1}{2} \right)} 
  \, \left( \frac{ 1+\frac{\hat{\mu}^2}{2N}}{16\hat{\mu}^2} \right)^{\nu/2} \, 
  \frac{(-1)^j\, (j+\nu)!}{j!\,c^j}\, \int_1^{\infty}\,du\,(u^2-1)^{(\nu-1)/2}\,\frac{(1-bc)^j}{b^{j+\nu+1}}  \nn 
  \\
  &= \frac{2^{\nu+3}\hat{\mu}^{\nu+2}\sqrt{\pi}}{\Gamma\left(\frac{\nu+1}{2}\right)} \, \frac{(j+\nu)!}{j!} 
  \, \frac{\left( 1+\frac{\hat{\mu}^2}{2N} \right)^{\nu/2}}{\left( 1-\frac{\hat{\mu}^2}{2N} \right)^j} 
  \, \int_1^{\infty}\,du\,\frac{(u+1)^{(\nu-1)/2}}{(u-1)^{(\nu+3)/2}}\,\frac{\left( 1 - \frac{\hat{\mu}^2}{N\,(u-1)} 
  - \frac{\hat{\mu}^4(u+1)}{4N^2(u-1)}\right) ^j}{\left( 1 + \frac{\hat{\mu}^2}{2N}\,\frac{(u+1)}{(u-1)} \right)^{j+\nu+1}}\,.
\end{align}
We now wish to set $j=tN$, and take the large-$N$ limit whilst keeping $t$ fixed. In fact, $j$ should remain an integer, so we will set $j=[tN]$, where the square brackets denote the integer part.

In order to take the limit, it is necessary to split $P(j,N,\hat{\mu})$ into two parts, $P_1(j,N,\hat{\mu})$, where the integral runs over $1 \le u < 1+\hat{\mu}^2/N$, and $P_2(j,N,\hat{\mu})$, where the integral is over the range $1+\hat{\mu}^2/N \le u < \infty$.  For $P_1(j,N,\hat{\mu})$, we write $u+1 = 2 + \mcO(N^{-1})$, and then make a change of variables to
\be
  s = 1 - \frac{\hat{\mu}^2}{N(u-1)}
\ee
with the result that (dropping the $\mcO(N^{-1})$ which will vanish when the limit is taken)
\be
  \label{WL_P1_integral_over_s}
  \int_1^{1+\hat{\mu}^2/N} \, du\, \cdots  =  \frac{(-1)^j \, 2^{(\nu-1)/2} \, N^{(\nu+1)/2}}{\hat{\mu}^{\nu+1}}
  \, \int_0^{\infty}\,ds\,\frac{s^j\,(1+s)^{(\nu-1)/2}}{(2+s)^{j+\nu+1}}\,\left( 1 - \frac{\hat{\mu}^2 (1-s)}{2N\,s} \right)^j.
\ee
It is important to note that the overall sign depends on whether $j$ is an even or an odd integer.  This implies that no single limit of $P_1(j,N,\hat{\mu})$ will exist. To evaluate the integral on the right-hand side of eq.~(\ref{WL_P1_integral_over_s}), we change variables to $u = j/s$, leading to
\be
  \int_0^{\infty}\,ds\,\cdots  =  \frac{1}{j^{(\nu+1)/2}}\,\int_0^{\infty}\,du\,\Bigg\{ j^{(\nu+3)/2} 
  \, \frac{u^{(\nu-1)/2}\,(u+j)^{(\nu-1)/2}}{(2u+j)^{\nu+1}\,\left( 1 + \frac{2u}{j} \right)^j}
  \, \left( 1-\frac{\hat{\mu}^2 (u-j)}{2N\,j} \right)^j \Bigg\}\,.
\ee
In fact, we are only interested in the large-$N$ limit of this integral, with $j = [tN]$ and $t$ fixed. As $N \rightarrow \infty$, the integrand in the curly brackets converges to an integrable function in a manner that allows us to interchange the limit and the integral. Therefore
\bea
  \lim_{\substack{N \rightarrow \infty \\ j=[tN]}} j^{(\nu+1)/2} \int_0^{\infty}\,ds\,\cdots 
  = e^{\hat{\mu}^2 t}\,\int_0^{\infty}\,du\,u^{(\nu-1)/2}\,e^{-2u} 
  = \frac{\Gamma\left(\frac{\nu+1}{2}\right)}{2^{(\nu+1)/2}}\,e^{\hat{\mu}^2 t}\,.
\eea
Combining everything, we have
\be
  \lim_{\substack{N \rightarrow \infty \\ j=[tN]}} \, \frac{P_1(j,N,\hat{\mu})}{N^{\nu}} 
    =   (-1)^j \, 2^{\nu+2} \, \sqrt{\pi} \, \hat{\mu} \, t^{(\nu-1)/2}\,\e^{\hat{\mu}^2 t}\,.
\ee
Note that we have been somewhat imprecise with our notation; there are actually two limits, depending on whether $j$ steps through the even or the odd integers.

We now turn to consider the limit of $P_2(j,N,\hat{\mu})$, for which we can write
\bea
  \int_{1+\hat{\mu}^2/N}^{\infty}\, du \, \cdots & = & \int_{1}^{\infty}\,du\,\Theta\left[u-\left(1+\frac{\hat{\mu}^2}{N}\right)\right] \, \cdots\,,
\eea
where $\Theta(x)$ is the Heaviside step function. We can then use the Monotone Convergence Theorem to show that
\bea
  \lim_{\substack{N \rightarrow \infty \\ j=[tN]}} \frac{P_2(j,N,\hat{\mu})}{N^{\nu}} 
  & = & \frac{2^{\nu+3}\sqrt{\pi}\hat{\mu}^{\nu+2}t^{\nu}}{\Gamma\left( \frac{\nu+1}{2} \right)}\,\int_1^{\infty}\,du 
  \frac{(u+1)^{(\nu-1)/2}}{(u-1)^{(\nu+3)/2}}  \,\exp\left[ -\,\frac{2\hat{\mu}^2 t}{u-1} \right] \nn 
  \\
  & = & \frac{(2\hat{\mu})^{\nu+2}\sqrt{\pi}}{\Gamma\left( \frac{\nu+1}{2} \right)}\,t^{\nu}\,
  \e^{\hat{\mu}^2 t}\,E_{\frac{1-\nu}{2}}(\hat{\mu}^2 t)\,,
\eea
where we used a simple change of variables $s = (u+1)/(u-1)$ in the last step, and $E_n(x)$ is the exponential integral defined in eq.~(\ref{Edef}). 
Our formula for the limit of $P(j,N,\hat{\mu})$ is therefore
\bea\label{WL_P_limit_general}
  \lim_{\substack{N \rightarrow \infty \\ j=[tN]}} \frac{P(j,N,\hat{\mu})}{N^{\nu}} 
  & = & 2^{\nu+2}\,\sqrt{\pi}\,\e^{\hat{\mu}^2 t}\,\left\{ \frac{\hat{\mu}^{\nu+2}\,t^{\nu}\,
  E_{\frac{1-\nu}{2}}(\hat{\mu}^2 t)}{\Gamma\left( \frac{\nu+1}{2} \right)} + (-1)^j\,\hat{\mu}\,t^{(\nu-1)/2} \right\},
\eea
where it should be understood that there are different limits depending on whether $j$ takes even or odd values.

Let us now consider the limit of $D_N^+(x;\hat{\mu})$.  Because of the two different limits of $P(j,N,\hat{\mu})$, technically we should now split the sum in eq.~(\ref{WL_D_N_plus}) into odd and even parts, and treat each part separately.  However, because of the way that everything combines, it is completely equivalent if we merely use the ``average'' limit over even and odd $j$,
\be\label{WL_P_lim_av_using_k}
  \Bigg\langle \lim_{\substack{N \rightarrow \infty \\ j=[tN]}}  \frac{P(j,N,\hat{\mu})}{N^{\nu}} \Bigg\rangle 
  = \frac{2^{\nu+2}\,\sqrt{\pi}}{\Gamma\left( \frac{\nu+1}{2} \right)}\,
  \e^{\hat{\mu}^2 t}\,\hat{\mu}^{\nu+2}\,t^{\nu}\,E_{\frac{1-\nu}{2}}(\hat{\mu}^2 t)\,,
\ee
where it should be understood that this is the average limit over a small interval around $t$ which includes (an infinite number of) both even and odd terms in $j$.

We can now proceed as in \cite{Akemann:2009fc},
\bea
  \lim_{N \rightarrow \infty} \sum_{j=0}^{N-2}\frac{1}{N}  & = & \int_0^1\,dt\,, \label{WL_lim_sum} \\
  \lim_{\substack{N \rightarrow \infty \\ j=[tN]}} \left( \frac{1-\frac{\hat{\mu}^2}{2N}}{1+\frac{\hat{\mu}^2}{2N}} \right)^{2j} & = & \exp[-2\hat{\mu}^2 t]\,,  \\
  \lim_{\substack{N \rightarrow \infty \\ j=[tN]}} \frac{j!}{(j+\nu)!}\,N^{\nu} & = & t^{-\nu}\,,  \\
  \lim_{\substack{N \rightarrow \infty \\ j=[tN]}} \frac{1}{N^{\nu+1}}\,L_{j}^{\nu+1}\left( \frac{x}{4N(1-\frac{\hat{\mu}^2}{2N})} \right) 
  & = & \lim_{N   \rightarrow \infty} \frac{1}{N^{\nu+1}}\,L_{[tN]}^{\nu+1}\left( \frac{x}{4N} \right) \nn\\
  & = & \left(\frac{2\sqrt{t}}{\sqrt{x}}\right)^{\nu+1}\, J_{\nu+1}(\sqrt{x t})\,.
\eea
Hence
\be
  \label{WL_D_N_plus_limit}
  \lim_{N \rightarrow \infty} D_N^+(x;\hat{\mu})  =   \frac{\sqrt{\pi}}{2\Gamma\left(\frac{\nu+1}{2}\right)}
  \, \frac{\sqrt{x}}{x^{\nu/2}}\,\hat{\mu}^{\nu+2}\,\int_0^1\,dt\,\e^{-\hat{\mu}^2 t}\,t^{(\nu+1)/2}
  \, E_{\frac{1-\nu}{2}}(\hat{\mu}^2 t)\,J_{\nu+1}(\sqrt{x t})\,.
\ee

\subsubsection[Determination of limits of $Q(j,N,\hat{\mu})$ and
$D_N^-(x;\hat{\mu})$]{\boldmath Determination of limits of $Q(j,N,\hat{\mu})$
and $D_N^-(x;\hat{\mu})$} 

For $Q(j,N,\hat{\mu})$, we find that we expect the limit
\be
  \lim_{\substack{N \rightarrow \infty \\ j=[tN]}} \frac{Q(j,N,\hat{\mu})}{N^{\nu+2}}
\ee
to exist.  The key point is that there are \textit{two} powers of $N$ difference compared with the $P(j,N,\hat{\mu})$ case, which we can see as follows. Using the recurrence relation \cite[eq.~(8.971.4)]{Gradshteyn:2007}
\be
  z\,L_j^{\nu+1}(z)   =  (j+1)\Big\{ L_j^{\nu}(z) - L_{j+1}^{\nu}(z) \Big\} + L_j^{\nu}(z)
\ee
we can show that
\be
  \label{WL_Q_in_terms_of_P_exact}
  \frac{Q(j,N,\hat{\mu})}{N}   =   4\left( 1-\frac{\hat{\mu}^2}{2N} \right) 
  \left[ (j+1) \Big\{ P(j,N,\hat{\mu}) - P(j+1,N,\hat{\mu}) \Big\} +\nu\,P(j,N,\hat{\mu}) \right].
\ee
Since $j+1 \propto N$ at large $N$, there are two powers of $N$ difference between the leading orders of $P$ and $Q$, and not one, as might na\"ively have been expected.

\paragraph{Leading order contribution:}

We use eqs.~(\ref{WL_P_limit_general}) and (\ref{WL_Q_in_terms_of_P_exact}) to show that
\be
  \lim_{\substack{N \rightarrow \infty \\ j=[tN]}} \frac{Q(j,N,\hat{\mu})}{N^{\nu+2}} 
   = (-1)^j\,2^{\nu+5}\,\sqrt{\pi}\,\hat{\mu}\,t^{(\nu+1)/2}\,\e^{\hat{\mu}^2 t}\,,
\ee
where the sign depends on whether $j$ is even or odd. Note that the other terms from eqs.~(\ref{WL_P_limit_general}) and (\ref{WL_Q_in_terms_of_P_exact}) are of smaller order, and so will vanish in the large-$N$ limit.

Now let us consider the limit of $D_N^-(x;\hat{\mu})$. The presence of the $N^{\nu+2}$ scaling for $Q$ is the crucial difference compared with the $P$ case, which scaled only with $N^{\nu}$.  Although the $N^{\nu+2}$ here can be matched with the $N^{\nu+2}$ prefactor in eq.~(\ref{WL_D_N_minus}), this means that there is then no residual $N^{-1}$ to make $dt$ when we take the large-$N$ limit (see eq.~(\ref{WL_lim_sum})), and so we will not get an integral.

Because adjacent even and odd terms have opposite signs, however, we can use the fact that, for a continuous function $f(t)$ on the interval $0 \le t \le 1$,
\be
  \lim_{N \rightarrow \infty} \sum_{j=0}^{N/2-1}\left\{ f\left( \frac{2j+1}{N} \right) - f\left( \frac{2j}{N} \right) \right\}
   =  \frac{1}{2}\{f(1)-f(0)\}
\ee
to take the large-$N$ limit of the sum in eq.~(\ref{WL_D_N_minus}). This results in a contribution of
\be\label{WL_D_N_minus_limit_1}
  \lim_{\substack{N \rightarrow \infty \\ \textup{Regime 1}}} D_N^-(x;\hat{\mu}) 
  = \frac{\sqrt{\pi}}{x^{\nu/2}}\,\hat{\mu}\, \left[ \e^{-\hat{\mu}^2 t}\,\sqrt{t}\,J_{\nu}(\sqrt{x t}) \right]_{t=0}^{t=1} 
  = \frac{\sqrt{\pi}\,\hat{\mu}\,\e^{-\hat{\mu}^2}}{x^{\nu/2}}\,J_{\nu}(\sqrt{x})\,.
\ee

\paragraph{Next-to-leading order contribution:}

There will also be a contribution from the next-to-leading order term of
$Q$. Here, we \textit{can} average even and odd terms (as we did for
$P(j,N,\hat{\mu})$ in eq.~(\ref{WL_P_limit_general})), and this average gives
zero for the leading order.
So the effective limit (over a small neighbourhood of $t$) is 
here (from eq.~(\ref{WL_Q_in_terms_of_P_exact}))
\begin{align}
  \label{WL_Q_lim_av}
  \Bigg\langle \lim_{\substack{N \rightarrow \infty \\ j=[tN]}} \frac{Q(j,N,\hat{\mu})}{N^{\nu+1}} \Bigg\rangle 
  & = 4\left(-t\,\frac{\partial}{\partial t}\,+ \nu \right) \Bigg\langle \lim_{\substack{N \rightarrow \infty \\ j=[tN]}} 
  \frac{P(j,N,\hat{\mu})}{N^{\nu}} \Bigg\rangle \nn 
  \\
  & = \frac{\sqrt{\pi}}{\Gamma\left(\frac{\nu+1}{2}\right)}\,(2\hat{\mu})^{4+\nu}t^{1+\nu}\e^{\hat{\mu}^2 t}
  \, \left\{ E_{\frac{-1-\nu}{2}}(\hat{\mu}^2 t) - E_{\frac{1-\nu}{2}}(\hat{\mu}^2 t) \right\},
\end{align}
where we used
\be
  \frac{d}{dx}\,E_n(x)  = -\,E_{n-1}(x)\,.
\ee
Hence
\begin{align}
  \label{WL_D_N_minus_limit_2}
  \lim_{\substack{N \rightarrow \infty \\ \textup{Regime 2}}} D_N^-(x;\hat{\mu}) 
  = \frac{\sqrt{\pi}\,\hat{\mu}^{\nu+4}}{\Gamma\left(\frac{\nu+1}{2}\right)}\,\frac{1}{x^{\nu/2}}
\int_0^1\,dt\,\e^{-\hat{\mu}^2 t}\,t^{1+\nu/2}\,
  \left\{ E_{\frac{-1-\nu}{2}}(\hat{\mu}^2 t) -
  E_{\frac{1-\nu}{2}}(\hat{\mu}^2 t) \right\}J_{\nu}(\sqrt{x t})\,.
\end{align}
We should add that lower orders will not contribute to the integral when we take the large-$N$ limit of the sum in eq.~(\ref{WL_D_N_minus}).

\subsubsection{Final formula}

Pulling everything together and simplifying, we have
\begin{align}
  \label{WL_C_formula_final}
  C(x,x';\hat{\mu}) = \frac{1}{32\sqrt{\pi}}\, \frac{\hat{h}_w(x')}{[\sgn(x')]^{\nu/2}}
  \bigg\{ - \frac{1}{\hat{\mu}}\,\e^{-\hat{\mu}^2}\,J_{\nu}(\sqrt{x}) 
  + \frac{\hat{\mu}^{\nu}}{\Gamma\left(\frac{\nu+1}{2}\right)}\,\int_0^1\,dt\,\e^{-\hat{\mu}^2 t}t^{(\nu+1)/2} \Bigg.  
  \nn
  \\
  \times \left[ \frac{\sqrt{x}}{2}\,E_{\frac{1-\nu}{2}}(\hat{\mu}^2 t)\,J_{\nu+1}(\sqrt{x t}) 
  - \hat{\mu}^2\,\sqrt{t}\left( E_{\frac{-1-\nu}{2}}(\hat{\mu}^2 t) - E_{\frac{1-\nu}{2}}(\hat{\mu}^2 t) \right) J_{\nu}(\sqrt{x t}) \right] \bigg\}\,. 
\end{align}
We then substitute this (after an easy change of variables), together with eqs.~(\ref{WL_A_formula_convergent}) and (\ref{WL_B_formula_convergent}), into eq.~(\ref{WL_R1_W_split}) to give eq.~(\ref{Gdef}).  The final result, although not obtained in a mathematically rigorous manner, is well supported by numerical checks. Furthermore, we have also verified analytically that, on taking the Hermitian limit $\hat{\mu}\rightarrow 0$, our result for the density shows complete agreement with \cite{Verbaarschot:1994ia}.

\section{Decoupling of heavy flavours}\label{heavy}

Looking at the RMT partition function 
(\ref{ZPQ}) -- and in fact for any QCD-like theory with or without chemical
potential -- it is expected that the microscopic spectral density would reduce
to that for a fewer number of flavours when one or more of the quark masses were sent to infinity,\footnote{We recall that the masses are purely imaginary.}
\begin{gather}
  \rho_{w}^{(N_f)}(\xi)\xrightarrow[]{|\hm_1|\to\infty}\rho_{w}^{(N_f-1)}(\xi)\xrightarrow[]{|\hm_2|\to\infty}
  \rho_{w}^{(N_f-2)}(\xi)\;\cdots\,,
  \\
  \rho_{s}^{(N_f)}(\xi)\xrightarrow[]{|\hm_1|,~|\hm_2|\to\infty}\rho_{s}^{(N_f-2)}(\xi)
  \xrightarrow[]{|\hm_3|,~|\hm_4|\to\infty}
  \rho_{s}^{(N_f-4)}(\xi)\;\cdots\,.
\end{gather}
We note that an even number of masses must be sent to infinity in the strong limit. 
The decoupling can easily be seen by numerical methods.

We do not give a full proof. Instead we wish to illustrate the decoupling for
\be
  \rho_{w}^{(N_f=2,\mathbb{C})}(\xi) \xrightarrow[]{|\hm_1|\to\infty} 
  \rho_{w}^{(N_f=1,\mathbb{C})}(\xi) \,.
\ee
To show this, we use that the weak kernel ${\cal K}_w$ obeys
\be
  {\cal K}_w(v,\hm^2) = -{\cal K}_w(\hm^2,v) \simeq f_w(\hm^2)q_w(v)
  \quad \text{for } -\hm^2\gg 1\,,
  \label{eq:K_w_factorize}
\ee
where
\be
  f_w(\hm^2) \equiv \frac{   (2i)^{\nu}   }{256\pi\hmu^2}\,\e^{-\frac{3}{2}\hmu^2}I_{\nu+1}(|\hm|)\,.
\ee
Thus, in the limit $-\hm_1^2 \gg 1$, eq.~\eqref{rhoCweakNf2} yields
\begin{align}
  \rho_{w}^{(N_f=2,\mathbb{C})}(\xi;\hm_1,\hm_2)
  & \simeq \rho_{w}^{(0,\mathbb{C})}(\xi)
  + 4 |\xi|^2 \hg_w(\xi^{*2},\xi^2) \nn\\
&\quad\times
  \frac{{\cal K}_w(\xi^2,\hm_2^2)\{q_w(\xi^{*2})f_w(\hm_1^2)\} - 
  \{q_w(\xi^2)f_w(\hm_1^2)\}{\cal K}_w(\xi^{*2},\hm_2^2)}{-f_w(\hm_1^2)q_w(\hm_2^2)}
  \notag\\
  & = \rho_{w}^{(0,\mathbb{C})}(\xi)
  + 4 |\xi|^2 \hg_w(\xi^{*2},\xi^2) 
  \frac{-{\cal K}_w(\xi^2,\hm_2^2) q_w(\xi^{*2}) \!+\! 
  q_w(\xi^2){\cal K}_w(\xi^{*2},\hm_2^2)}{q_w(\hm_2^2)}
  \notag\\
  & = \rho_{w}^{(N_f=1,\mathbb{C})}(\xi;\hm_2)\,,
\end{align}
as seen from eq.~\eqref{rhoCweakNf1}. These steps can be generalised to higher $N_f$ in a straightforward manner.

The decoupling occurs in the sectors of real/imaginary eigenvalues as well, as seen 
in figures~\ref{fg:Weak-density_Nf=1_for_real_and_imaginary} and \ref{fg:Weak-density_Nf=2_for_real_and_imaginary}. 
For illustration, let us prove
\be
  \rho_w^{(N_f=2,\,(i)\mathbb{R})}(\xi;\hm_1,\hm_2)   \xrightarrow[]{|\hm_1|\to\infty}  
  \rho_w^{(N_f=1,\,(i)\mathbb{R})}(\xi;\hm_2)   \,.
\ee
With eq.~\eqref{eq:K_w_factorize} it can be proven that
\be
  G_w(x,x') \xrightarrow[]{x \to -\infty} f_w(x)Q_w(x')\,.
  \label{eq:G_w_factorize}
\ee
Using eqs.~\eqref{eq:K_w_factorize} and \eqref{eq:G_w_factorize} in eq.~\eqref{rhoRweakNf2}, 
we find, in the limit $-\hm_1^2\gg 1$, that
\begin{align}
  &\rho_{w}^{(N_f=2,\,(i)\mathbb{R})}(\xi;\hm_1,\hm_2) 
  \nn
  \\
  &\quad\simeq
  \rho_{w}^{(0,\,(i)\mathbb{R})}(\xi)
  \ +\ 2|\xi| 
  \frac{{\cal K}_w(\xi^2,\hm_2^2)  \{ f_w(\hm_1^2)Q_w(\xi^2) \} -
  \{q_w(\xi^2)f_w(\hm_1^2)\} G_w(\hm_2^2,\xi^2)}{-f_w(\hm_1^2)q_w(\hm_2^2)}
  \notag\\
  &\quad=
  \rho_{w}^{(0,\,(i)\mathbb{R})}(\xi)
  \ +\ 2|\xi| 
  \frac{{\cal K}_w(\xi^2,\hm_2^2)  Q_w(\xi^2) -
  q_w(\xi^2) G_w(\hm_2^2,\xi^2)}{-q_w(\hm_2^2)}
  \notag\\
  &\quad=
  \rho_{w}^{(N_f=1,\,(i)\mathbb{R})}(\xi;\hm_2)\,,
\end{align}
as seen from eq.~\eqref{rhoRweakNf1}. Moreover 
using eqs.~\eqref{eq:K_w_factorize} and \eqref{eq:G_w_factorize} once again 
we can easily prove 
$\rho_{w}^{(N_f=1,\,(i)\mathbb{R})}(\xi;\hm_2)\to \rho_{w}^{(0,\,(i)\mathbb{R})}(\xi)$ for $-\hm_2^2\gg 1$.

\section{Sign-quenched partition functions}\label{ZsignQ}

In two-colour QCD the fermion determinant has no complex phase but only a sign, so we shall 
introduce a ``sign-quenched'' partition function, in which $\det[{\cal D}(\mu)+\hm]$ is 
replaced by $|\det[{\cal D}(\mu)+\hm]|$, as a direct counterpart of the phase-quenched 
partition function in three-colour QCD. The purpose of this appendix is 
to derive analytical expressions for such sign-quenched partition functions, first for finite $N$, and 
later in the limits of both weak and strong non-Hermiticity. The results in this appendix 
will be used in the main text for the analysis of the sign problem.

\subsection[Main result for finite $N$]{\boldmath Main result for finite $N$}

Let us begin with a formula for the probability measure, with
$z_j\equiv x_j+iy_j$, 
\begin{multline}
  {\cal Z}_N^{(N_f,\nu)}(\mu;\{m\}) = c_N^{}
  \prod_{f=1}^{N_f}\mm_f^\nu 
  \prod_{k=1}^{N} \int\limits_{\mathbb{C}} d^2z_k~
  w(z_k)\ab{\Delta_N(\{z\})} 
  \prod_{j=1}^{N}\prod_{f=1}^{N_f}(\mm_f^2-z_j)
  \label{eq:Z_measure_def}
  \\
  \qquad \times \sum_{n=0}^{[N/2]}
  \frac{1}{n!(N-2n)!}
  \prod_{i=1}^{n}\delta^2(z_{2i-1}-z_{2i}^*)
  \, 
  \delta(y_{2n+1})\cdots\delta(y_N)\,,
\end{multline}
which is a minor modification of \cite[eq.~(2.9)]{Akemann:2009fc}. 
Whilst being equal to eq.~\eqref{Zgen} for even $N$, this expression is more convenient 
for our current purpose. The weight function $w(z)$ is defined in \cite[eq.~(2.12)]{Akemann:2009fc},
and below we only use that $w(z)=w(z^*)$ for $z\in\mathbb{C}$, and $w(z)=h(z)$ for $z\in\mathbb{R}$, 
with $h(z)$ defined in eq.~\eqref{wch}. The prefactor $c_N^{}$ reads
\be
  c^{}_N \equiv (\mu\text{-independent factor})
  \times (2\mu)^{-N(N+\nu)}\eta_{+}^{-N(N+\nu-1)/2}\,,
  \label{eq:c_N_def}
\ee
and we refer to \cite[eq.~(3.46)]{Akemann:2009fc} for more details.

We also define a ``partially sign-quenched'' partition function,
which we denote by
${\cal Z}^{(\|N_f'\|+N_f,\nu)}_{N}(\mu;\{\tilde{\mm}\},\{\mm\})$,
as being identical to ${\cal Z}^{(N_f'+N_f,\nu)}_{N}
$ 
in eq.~\eqref{eq:Z_measure_def} except that the fermion determinant is 
replaced by
\be
  \prod_{j=1}^{N}\left[
  \prod_{h=1}^{N_f'}|\tilde{\mm}_h|^\nu \ab{\tilde{\mm}_{h}^2-z_j}
  \prod_{f=1}^{N_f}\mm_f^\nu (\mm_f^2-z_j)\right]\,.
\ee

After these prerequisites, we can now state our main result at finite $N$,
\be
  R^{(N_f,\mathbb{R})}_{k,N}(x_1,\ldots,x_k)
  = 
  \frac{c_N^{}}{c_{N-k}^{}}
  \ab{\Delta_{k}(\{x\})}
  \prod_{i=1}^{k}
    \frac{w(x_i)}{|x_i|^{\nu/2}}
    \prod_{f=1}^{N_f}(\mm_f^2-x_i)
  \frac{{\cal Z}^{(\|k\|+N_f,\nu)}_{N-k}(\mu;\{\sqrt{x}\},\{\mm\})}
  {{\cal Z}_{N}^{(N_f,\nu)}(\mu;\{\mm\})}\,.
  \label{eq:R_Z_sign}
\ee
Below we give a proof of this formula for $k=1$. The generalization to $k>1$
is then straightforward.

\subsection[Proof: the $k=1$ case]{\boldmath Proof: the $k=1$ case}
By definition, we have
\begin{multline}
  R^{(N_f)}_{1,N}(z) 
   = \frac{c^{}_N }{{\cal Z}_{N}^{(N_f,\nu)}(\mu;\{\mm\})}
  \prod_{f=1}^{N_f}\mm_f^\nu 
  \prod_{k=1}^{N} \int\limits_{\mathbb{C}} d^2z_k~
  w(z_k)\ab{\Delta_N(\{z\})} 
  \prod_{j=1}^{N}\prod_{f=1}^{N_f}(\mm_f^2-z_j)
  \\
  \times 
  \sum_{n=0}^{[N/2]} \frac{1}{n!(N-2n)!}
  \prod_{i=1}^{n}\delta^2(z_{2i-1}-z_{2i}^*)
  \,
  \delta(y_{2n+1})\cdots\delta(y_N)
  \sum_{\ell=1}^{N}\delta^2(z-z_\ell)\,.
\end{multline}
For $2n+1\leq \ell\leq N$, $\delta^2(z-z_\ell)$ in the last sum will yield $\delta(y)$ 
(with $z\equiv x+iy$), hence
\allowdisplaybreaks{
\begin{align}
  \delta(y) R^{(N_f,\mathbb{R})}_{1,N}(x) 
  & = \frac{c^{}_N }{{\cal Z}_{N}^{(N_f,\nu)}(\mu;\{\mm\})}
  \prod_{f=1}^{N_f}\mm_f^\nu 
  \prod_{k=1}^{N} \int\limits_{\mathbb{C}} d^2z_k~
  w(z_k)\ab{\Delta_N(\{z\})} 
  \prod_{j=1}^{N}\prod_{f=1}^{N_f}(\mm_f^2-z_j)
  \nn
  \\
  & \quad \times 
  \sum_{n=0}^{[N/2]} \frac{1}{n!(N-2n)!}
  \prod_{i=1}^{n}\delta^2(z_{2i-1}-z_{2i}^*)
  \,
  \delta(y_{2n+1})\cdots\delta(y_N)
  \sum_{\ell=2n+1}^{N}\delta^2(z-z_\ell)
  \notag\\
  & = \frac{c^{}_N }{{\cal Z}_{N}^{(N_f,\nu)}(\mu;\{\mm\})}
  \prod_{f=1}^{N_f}\mm_f^\nu 
  \prod_{k=1}^{N} \int\limits_{\mathbb{C}} d^2z_k~
  w(z_k)\ab{\Delta_N(\{z\})} 
  \prod_{j=1}^{N}\prod_{f=1}^{N_f}(\mm_f^2-z_j)
  \nn
  \\
  & \quad \times 
  \sum_{n=0}^{[N/2]} \frac{N-2n}{n!(N-2n)!}
  \prod_{i=1}^{n}\delta^2(z_{2i-1}-z_{2i}^*)
  \,
  \delta(y_{2n+1})\cdots\delta(y_N) \, \delta^2(z-z_N)
  \notag\\
  & = \frac{c^{}_N }{{\cal Z}_{N}^{(N_f,\nu)}(\mu;\{\mm\})}
  w(x) \prod_{f=1}^{N_f}(\mm_f^2-x) 
  \prod_{f=1}^{N_f}\mm_f^\nu 
  \nn
  \\
  & \quad \times 
  \prod_{k=1}^{N-1} \int\limits_{\mathbb{C}} d^2z_k~
  w(z_k)\left(\ab{\Delta_{N-1}(\{z\})} \prod_{\ell=1}^{N-1}\ab{x-z_\ell} \right)
  \prod_{j=1}^{N-1}\prod_{f=1}^{N_f}(\mm_f^2-z_j)
  \nn
  \\
  & \quad \times \sum_{n=0}^{[N/2]}
  \frac{1}{n!(N-1-2n)!}
  \prod_{i=1}^{n}\delta^2(z_{2i-1}-z_{2i}^*)
  \delta(y_{2n+1})\cdots\delta(y_{N-1}) \delta(y)\,.
\end{align}
}
Therefore
\begin{align}
   R^{(N_f,\mathbb{R})}_{1,N}(x) = \frac{c_N^{}}{c_{N-1}^{}}\frac{w(x)}{|x|^{\nu/2}}
   \prod_{f=1}^{N_f}(\mm_f^2-x)
   \frac{{\cal Z}_{N-1}^{(\|1\|+N_f,\nu)}(\mu;\sqrt{x},\{\mm\})}
   {{\cal Z}_{N}^{(N_f,\nu)}(\mu;\{\mm\})}\,.
   \label{eq:R_Z_sign_k=1}
\end{align}

\subsection[The large-$N$ limit of weak non-Hermiticity]{\boldmath The large-$N$ limit of weak non-Hermiticity}
\label{app_weak}

Here we shall compute ${\cal Z}^{(N_f=\|1\|,\nu)}(\mu;\{\mm\})=\gl{|\det[{\cal D}(\mu)+m]|}$ 
in the limit of weak non-Hermiticity defined in eq.~(\ref{weaklim}),
\begin{align}
  \hmu^2 & \equiv  2N\mu^2\,, \nn\\
  \hm_{f} & \equiv  2\sqrt{N}\mm_f\,, \nn\\
  \xi & \equiv  2\sqrt{N}\La \,.
\nn
\end{align}
Putting $N_f=0$ and $x=m^2\in\mathbb{R}$ in eq.~\eqref{eq:R_Z_sign_k=1}, we find
\be
\lim_{N\to\infty}
  {\cal Z}_{N-1}^{(N_f=\|1\|,\,\nu)}(\mu;m) \sim \frac{c^{}_{N-1}}{c_N^{}}
  \frac{|m|^{\nu}}{w(m^2)} R^{(N_f=0,\mathbb{R})}_{1,N}(m^2)\,.
\ee
In the weak limit, it follows from eqs.~\eqref{eq:weak_mappings} and \eqref{rhoRweakQ} that
\begin{align}
  \lim_{N\to\infty}
  R^{(N_f=0,\mathbb{R})}_{1,N}(m^2) &\sim R^{(N_f=0,\mathbb{R})}_{w}(\hm^2)
  = -G_w(\hm^2,\hm^2) \,,\\
  \lim_{N\to\infty}  \frac{|m|^\nu}{w(m^2)}&\sim \frac{1}{\hh_w(\hm^2)}\,.
\end{align}
From eq.~\eqref{eq:c_N_def}, we have
\begin{align}
\lim_{N\to\infty}
  \frac{c^{}_{N-1}}{c^{}_N}
  & \sim \frac{(2\mu)^{-(N-1)(N-1+\nu)}\eta_{+}^{-(N-1)(N+\nu-2)/2}}
  {(2\mu)^{-N(N+\nu)}\eta_{+}^{-N(N+\nu-1)/2}}
  = (2\mu)^{2N+\nu-1} \eta_{+}^{N-1+\nu/2}
  \\
  & \sim \sqrt{2N}\mu ~\e^{N\mu^2} = \hmu~\e^{\hmu^2/2} \,.
\end{align}
Collecting all results, we finally obtain eq.~(\ref{eq:Z_w_|1|}),
\be
  {\cal Z}_{w}^{(N_f=\|1\|,\,\nu)}(\hmu;\hm) \sim -\hmu~\e^{\hmu^2/2} 
  \frac{ G_w(\hm^2,\hm^2) }{\hh_w(\hm^2)}\,.
\nn
\ee

\subsection[The large-$N$ limit of strong non-Hermiticity]{\boldmath The large-$N$ limit of strong non-Hermiticity}
\label{app_strong}

Below we show how to derive an analytical expression in the limit of strong
non-Hermiticity for 
${\cal Z}_N^{(N_f=\|2\|,\,\nu)}(\mu;\mm_1,\mm_2)=
\gl{|\det[{\cal D}(\mu)+m_1]\det[{\cal D}(\mu)+m_2]|}$, 
 where $N$ goes to infinity with $\mu$ and $m_f$ fixed 
($\hm_f=m_f$). 

Putting $k=2$, $N_f=0$, $x_1=m_1^2$, and $x_2=m_2^2$ (with $m_1^2,\,m_2^2 \in\mathbb{R}$) 
in eq.~\eqref{eq:R_Z_sign} and using $w(x)=h(x)$ for $x\in\mathbb{R}$, we find%
\be
  {\cal Z}^{(N_f=\|2\|,\,\nu)}_{N-2}(\mu;\mm_1,\mm_2) \sim 
  \frac{c^{}_{N-2}}{c_{N}^{}} \frac{1}{\ab{m_1^2-m_2^2}}  \frac{|m_1m_2|^\nu}{h(m_1^2)h(m_2^2)}
  R^{(N_f=0,\mathbb{R})}_{2,N}(m_1^2,m_2^2)\,.
  \label{eq:Z-R_strong_Nf=2}
\ee
Next, let us recall that according to \cite{Sommers-Wieczorek2008,Akemann:2009fc} the real two-point function reads
\be
  R^{(N_f=0,\mathbb{R})}_{2,N}(x_1,x_2)
  = h(x_1)h(x_2)f_N(x_1,x_2)\,,
  \label{eq:R_strong_Nf=2}
\ee
where
\begin{align}
  f_N(x_1,x_2) \equiv {\cal K}_N(x_1,x_2)\sgn(x_1-&x_2)
  \notag\\
  +
  \int_{\mathbb{R}}dx
  \int_{\mathbb{R}}dx'\,
  h(x)h(x')&\Big[
    \sgn(x_1-x)\sgn(x_2-x')-\sgn(x_2-x)\sgn(x_1-x')
  \Big]
  \notag\\
  \times &\Big[
    {\cal K}_N(x_1,x){\cal K}_N(x_2,x')-\frac{1}{2}{\cal K}_N(x_1,x_2){\cal K}_N(x,x')
  \Big]\,.
\end{align}
We can plug eq.~\eqref{eq:R_strong_Nf=2} into eq.~\eqref{eq:Z-R_strong_Nf=2} to get
eq.~(\ref{eq:Z_s_|2|}),
\be
  {\cal Z}^{(N_f=\|2\|,\,\nu)}_{s}(\mu;\hm_1,\hm_2) \sim 
  \frac{f_{s}(\hm_1^2,\hm_2^2)}{\ab{\hm_1^2-\hm_2^2}} \,,
\nn
\ee
where $f_s(x_1,x_2)$ is defined as the large-$N$ limit of $(x_1x_2)^{\nu/2}f_N(x_1,x_2)$, with no additional rescaling 
of parameters involved, and is explicitly given in eq.~\eqref{eq:fs} 
for nonpositive arguments. 
In eq.~\eqref{eq:Z_s_|2|} we omitted a $\mu$-dependent prefactor on the right-hand side, 
as our primary interest in the strong limit is in the mass dependence 
of the quantities considered.

\providecommand{\href}[2]{#2}\begingroup\raggedright\endgroup

\bibliographystyle{JHEP}
\end{document}